\documentclass[12pt]{iopart}

\pdfoutput=1

\usepackage{epsfig}

\newcommand{\be}{\begin{equation}}
\newcommand{\ee}{\end{equation}}
\newcommand{\bea}{\begin{eqnarray}}
\newcommand{\eea}{\end{eqnarray}}
\newcommand{\ba}[1]{\begin{array}{*{#1}{c}}}
\newcommand{\ea}{\end{array}}

\newcommand{\pr}{{\cal P}}
\newcommand{\pd}[2]{\frac{\partial #1}{\partial #2}}
\newcommand{\la}{\left\langle}
\newcommand{\ra}{\right\rangle}
\newcommand{\lb}{\left}
\newcommand{\rb}{\right}
\newcommand{\half}{\frac{1}{2}}
\newcommand{\Oh}{{\cal O}}

\newcommand{\Lop}{{\cal L}}
\newcommand{\Lod}{{\cal L}^\dagger}
\newcommand{\Up}{\Upsilon}
\newcommand{\n}{\rho}
\newcommand{\nl}{\hat{\rho}}
\newcommand{\nt}{\tilde{\rho}}
\newcommand{\na}{\bar{\rho}}
\newcommand{\nL}{{\nu_L}}
\newcommand{\K}{K}
\renewcommand{\k}{\kappa}
\newcommand{\kt}{\tilde{\kappa}}
\newcommand{\phix}{\phi^\ast}
\newcommand{\rhox}{\rho^\ast}
\newcommand{\phih}{\hat{\phi}}
\newcommand{\ag}{A}
\newcommand{\bg}{B}
\newcommand{\zc}{Z_c}
\newcommand{\Ub}{\tilde{U}_b}
\newcommand{\xb}{\bar{x}}
\newcommand{\tsw}{\tau_{sw}}
\newcommand{\test}{\tau_{est}}
\newcommand{\st}{\tilde{s}}
\newcommand{\Q}{{\bar{Q}}} 
\renewcommand{\P}{P}    
\newcommand{\chit}{\tilde{\chi}}
\newcommand{\gamt}{\tilde{\gamma}}
\newcommand{\Ut}{\tilde{U}_b}
\newcommand{\Nt}{\tilde{N}}
\newcommand{\zetat}{\tilde{\zeta}}
\newcommand{\dtau}{\delta\tau}
\newcommand{\A}{{\cal A}}

\newcommand{\Mz}{{\cal M}_0}
\newcommand{\Mo}{{\cal M}_1}
\newcommand{\Mt}{{\cal M}_2}
\newcommand{\Mmo}{{\cal M}_{-1}}

\newcommand{\eps}{\epsilon}

\newcommand{\good}{Good et al \cite{GOOD} }

\begin{document}

\title{Asexual Evolution Waves: Fluctuations  and Universality}

\author{Daniel S. Fisher}

\address{McCullough Laboratory, Department of Applied Physics, Stanford University,  CA 94305-4045, USA}

\begin{abstract} 

In large asexual populations, multiple beneficial mutations arise in the population,  compete, interfere with each other, and accumulate on the same genome, before any of them fix.  The resulting dynamics, although studied by many authors, is still not fully understood, fundamentally  because the effects of fluctuations due to the small numbers of the fittest individuals are large even in enormous populations.  
In this paper, branching processes and various asymptotic methods for analyzing the stochastic dynamics are further developed and used to obtain information on fluctuations, time dependence, and the distributions of sizes of subpopulations,  jumps in the mean fitness, and other properties.   The focus is on the behavior of a broad class of models: those with a distribution of selective advantages of available beneficial mutations that  falls off more rapidly than exponentially. For such distributions, many aspects of the dynamics are universal --- quantitatively so for extremely large populations.     On the most important time scale that controls coalescent properties and fluctuations of the speed, the dynamics is reduced to a simple stochastic model that couples the peak and the high-fitness ``nose" of the fitness distribution. 
Extensions to other models and distributions of available mutations are discussed briefly.   


\end{abstract} 


 
  \maketitle


\section{Introduction}

Classical population genetics primarily focussed on the dynamics and statistics of mutations  that exist or arise in a population under the approximation that their eventual fate is primarily determined by their own contribution (or sometimes also  those of  a few other alleles)  to the fitness of the organism.  But in large populations many new beneficial mutations  arise each generation: these compete with each other and further mutations  occur in  the descendants of mutant individuals, before the fate of the earlier mutations is apparent. Especially as microbial populations are usually large enough for these effects to be important, one would like to understand the dynamics and statistics of this multiple mutation regime. But 
even the simplest models of the effects of multiple beneficial mutations  and  the resulting  population dynamics have proved to be surprisingly --- at least to this statistical physicist --- difficult to analyze. (For a recent review, see reference \cite{park:etal:2010}, and for earlier work, see, e.g. references \cite{BARTON95, FELSREV}. )

The simplest model  is a large asexual population of very similar organisms descended very recently from a common ancestor which evolve in a fixed environment with the only ``ecological" interactions among the organisms  acting to keep the population approximately constant (because, e.g., of a limited supply of a primary nutrient or predation) with no dependence of the interactions on the genotypic differences.   The ``fitness" is then solely a function of an individual's genome: simply  the difference between the birth and death rates with that genome.   The ``mean-field" interactions between the organisms act to make the net growth rate of a clonal sub-population equal to the difference between the fitness of its genome and the average fitness of the population.  As long as this does not change by too much, only the net growth rate matters: one can assume, e.g., that the death rate is constant and the birth rate equal to the death rate plus the fitness relative to the mean.  

If the ``fitness  landscape" --- the fitness as a function of the genome  --- in the vicinity of the genomes of the (assumed almost identical) existing organisms has a great number of uphill directions accessible by single mutations, then  the evolutionary dynamics is primarily determined by the statistical properties of the landscape.  If these do not change much as the population evolves, then the most crucial feature of the landscape is the distribution of the   numbers and selective advantages, $s$, of potentially beneficial mutations available.  Multiplied by the corresponding mutation rates, these can be lumped together into a $\mu(s)ds$,  defined as the rate per individual for beneficial mutations in an interval $(s,s+ds)$.   Although this model is often framed in terms of an assumption that the effects of the mutations on the fitness are additive, it is actually far more general obtaining as long as the fitness landscape is static with statistically-uniform properties and there are no genotype-dependent interactions or other properties of the organisms that affect their birth or death rates. 

The effects of stochastic fluctuations in the number of births and deaths in a subpopulation are small when the subpopulation is large: one would thus expect that the dynamics in the limit of very large populations would be essentially deterministic.  But this is fundamentally wrong except close to a fitness peak: indeed, it quickly leads to absurd results.  The crucial feature of evolutionary dynamics is that even in enormous populations,  a single individual that is fitter than all the others can take over the population in a time that is only logarithmic in the population size.  Thus rare events --- in particular mutations that produce individuals fitter than any that already exist --- play an essential role in the dynamics.  One can analyze  these by branching process methods, as we do herein, but analyzing  the interplay between the  stochastic rare events and the almost-deterministic but non-linear (due to the fixed population size constraint)  dynamics of the bulk of the population is not straightforward.  And average quantities are usually very bad characterizations of the typical behavior.  This is true even in the presence of several helpful small parameters that are related to the very large populations and  low beneficial mutation rates.  

The most basic question is the average rate of increase of the mean fitness of the population, $v$.  In the deterministic approximation, this will increase exponentially fast and there will be no steady state.  In contrast, for any finite $N$ there will be a statistical steady state with a finite average speed.  The most basic task is to understand the properties of this steady state.

Many authors --- particularly statistical physicists --- have studied this model over the past decade, \cite{park:etal:2010,tsimring:etal:1996, gerrish:lenski:1998,rouzine:etal:2003,wilke:2004,joyce:etal:2008,brunet:etal:2008,DFG,OH,GOOD,desai:etal:2007,fogle:etal:2008,park:krug:2007,rouzine:etal:2008,sniegowski:gerrish:2010,SCHIFFELS} often focussing on the simplest case of $\mu(s)=\Ub \delta(s-\st)$: i.e. a single fitness increment.  This corresponds most simply to a landscape that looks like a uniform staircase with equal sized steps of height $s$. But if one is interested in the diversity within the evolving population, one has to remember that the landscape is actually very high-dimensional with a large number of possible mutations for each step: $\Ub$ is  the total rate for all of these.  \cite{DerridaExactLong,DerridaSelLett,DWF}

At this stage, many of the properties of the statistical steady state of the staircase model are pretty well understood. \cite{DFG, park:etal:2010,GOOD} But most of the methods used are somewhat tricky and rely on various Ansatz's and heuristic arguments. They are hard to generalize to a distribution of mutational effects, although these are the focus of ``clonal interference" analyses \cite{gerrish:lenski:1998,DFG,sniegowski:gerrish:2010} for which competition between mutations of different selective advantages is assumed to dominate the dynamics, and recent work by \good.   One of the  purposes of this paper is technical: to analyze the general model with a distribution  of mutational effects by several methods, reproducing in a more controlled way many of the existing results, elucidating some of the difficulties and correcting some errors,  establishing which aspects are universal, and --- it is hoped --- providing means for analyzing other models and phenomena.   
But the primary goal of this paper is to move away from  average properties and explicitly  focus on fluctuations: the primary new results concern these.  The basic strategy is to first  allow the population size to vary and then use results from this to study the fixed-population-size model of primary interest. 

\subsubsection{Outline:}  In the remainder of this introduction the model is defined and its heuristic behavior then discussed  in Sec.(2) whose  first subsection is a review of some previous work.  Analysis of the branching process that generates the fluctuations and the dynamical consequences of these are   carried out in Sec.(3)  and Sec. (4), which also discusses distributions.  Technical details from these sections are relegated to appendices.  Section (5) focusses on the dynamics on the most important time-scale: this is controlled by  a simple effective model. Sections (6) and (7)  introduce other possible approaches and their pros and cons.  The last section summarizes the results and their limitations,  and discusses possible extensions and future directions.

\subsection{Model and fitness distribution}

As long as one is not interested in individuals, all that matters is the dynamics of the fitness distribution of the population: we denote by  $\n(x,t)dx$ the number of individuals with fitness in the interval  $(x,x+dx)$ at time $t$.  We refer to $\rho(\chi)$ as the ``fitness distribution" although its integral is the total population, $N(t)\equiv \int dx \n(x,t)$ , rather than unity: dividing it by $N(t)$ yields the instantaneous probability density of the fitness.  The growth rate of a population with fitness $x$ is 
\be  
\chi\equiv x-\Up(t)    \label{GROWTHRATE}
\ee 
with $\Up(t)$ adjusted to keep the total population constant:  $\Up(t)$ must thus be  (up to small fluctuations) equal to the mean fitness of the population at time $t$, 
\be
\bar{x}(t)\equiv \int dx \,  x\n(x,t).
\ee
To understand the behavior, it is useful to consider relaxing the fixed population size constraint: this we will do in various ways. In the last section we briefly discuss different forms of the dependence of the growth rate on $x$, but largely focus on the simple conventional form, Eq.(\ref{GROWTHRATE}).

Although the argument about individuals with the highest fitness suggests that one can not make a continuous approximation to $\n$, in fact one can as long as the width of the fitness distribution is small compared to the mean birth and death rates --- which we will assume.  But in making this continuous population size approximation, it is essential to include the right form of the stochastic noise that represents the fluctuations in numbers of births and deaths: these must be of order $\sqrt{\n}$.   This approximation is then exactly equivalent to the ``diffusion approximation" often used in population genetics. \cite{GILLBOOK}

The dynamics of the general model we consider is given by a stochastic integro-differential equation of the form 
\be
\frac{\partial \n}{\partial t}=[\Lop_0 - \Up(t)] \n + \sqrt{2\n}\, \eta_{BD}(x,t)
\ee
where  the linear operator, $\Lop_0$, includes the effects of selection and mutations,   $\eta_{BD}$ is gaussian white noise (with the Ito convention) which, when  multiplied by $\sqrt{2\n}$, represents the close-to-gaussian fluctuations in the difference between births and deaths, and $\Up(t)$ can be adjusted to keep the total population, $\int \n(x)dx$,  constant at $N$. Time is measured in units of   the inverse death or birth rate which are approximately equal and roughly constant as long as the total increase in fitness is small. 
We consider  a general distribution of mutations,  $\mu(s)ds$.  Then with the linear dependence of the selection on $x$, we have 
\be
 [\Lop_0 \n](x) = x + \int dy \mu(x-y) [\n(y) -\n(x)]\ . \label{LOPDEF}
 \ee
 [For derivation of the basic equation for the stochastic dynamics, see appendices of \good.]

We expect that  in a large steadily evolving population with no depletion of the supply of beneficial mutations, the mean fitness of the population will increase steadily, so with $N$ fixed,    $\Up(t)\approx vt$, with fluctuations around this. The simplest quantities of interest are  how the mean speed, $v$, depends on $N$ and on $\mu(s)$.
As values of $s$ in the tail of the distribution will play an important role, but many quantities will only depend logarithmically on $\mu$, it is useful to define the --- typically large ---  
\be
\Lambda(s)\equiv \log[1/\mu(s)] \ .
\ee
As discussed in \cite{DFG} --- henceforth DFG --- and \good the behavior depends qualitatively on whether $\Lambda(s)$
is convex up or down: i.e. whether it increases faster or more slowly than linearly for large $s$.   We will primarily focus on the former: the {\it short-tailed case} for which $\mu(s)$ decreases faster than exponentially: many properties of the short-tailed case are universal in the limit of large populations sizes, as we shall see.  The latter, long-tailed, case is discussed  in the last section --- along with an argument that it is unlikely to be relevant for continuous evolution which is our primary interest here.

Note that if  $\Up(t)$ were fixed independent of the stochastic fluctuations in the populations, the dynamics  would be simply a branching process which --- at least in principle --- can be analyzed fully.  But, as with all branching processes, at long times the population will either go extinct or grow without bound.  We are thus faced with a choice: either to assume the speed is essentially constant  and keep the total population fixed by  some heuristic  means, or to try to analyze the full problem with the population size fixed, or to change the model in some way so that the population size fluctuates but neither goes extinct nor diverges.  \cite{OH}
Most authors have taken the heuristic approach and we consider that first.

\section{Heuristics}

\subsection{Staircase model}

 The basic heuristics for the dynamics of the fitness distribution of the  staircase model has been worked out in a variety of ways.  \cite{DFG,rouzine:etal:2008,park:etal:2010} We focus solely on populations large enough that there are many new beneficial mutations each generation, $N\Ub>>1$: the multiple mutations regime.  This subsection reviews the heuristic arguments of DFG.  
 
 At any given time, the high-fitness tail of the distribution of the population extends out to a ``nose" at fitness $Q$  above the mean:  beyond this the total population is so small that it is very  likely to quickly die out in the absence of mutations feeding it.  \cite{tsimring:etal:1996}  We refer to $Q$ as the {\it lead} of the nose. At $x= \xb+Q$ there is a population large enough that it has established --- i.e. very unlikely to die out from fluctuations --- and growing exponentially at rate $Q$: this we call  the lead population.   The condition that this subpopulation  has established  is that its size is larger than roughly $1/Q$. 

  After it is has established, the lead population grows exponentially and after some  time, $\test$, will itself have produced enough beneficial mutations to further advance the lead by $\st$. Thus the average speed of the nose is simply. 
\be
v= \frac{\st}{\la \test \ra} 
\ee
Self-consistency requires that  the average speeds of the mean and  the nose be the same.

A time $t$ after a lead population is established, it will have grown by a factor of  $e^{Qt-vt^2/2}= e^{(Q-vt/2)t}$ because its growth rate slows down as $Q-vt$ due to the advance of the mean.    The lead population produces further beneficial  mutations at rate $\Ub$ per individual.  The probability that a new mutant will establish is roughly $Q$. \cite{GILLBOOK} Thus the time to establish a new lead population is when of order $1/Q$ new mutants have been produced.  The average time for the lead population to grow large enough to produce mutations  that will establish the next lead population is thus
\be
\test\approx \frac{\log(\st/\Ub)}{\Q-\half\st} 
\ee
with $\Q$ the mean lead and  $\Q-\half\st$  the average growth rate of the lead population between when it establishes and when the next population establishes, by which time its growth rate will have decreased to $\Q-\st$.    Note that $\st$ rather than $\Q$ appears in the logarithm because of multiple mutations that establish soon enough to contribute to the new lead population. \cite{rouzine:etal:2008}


The populations near the nose comprise only a tiny fraction of the population. Indeed, when it first establishes,  the lead population has size only $\sim \frac{1}{Q}$.   If further mutations into the lead population can be ignored, then a time $t$ after it is established this subpopulation's  size  is  roughly  $\exp(Qt-\half vt^2)/Q$ which  peaks after time $Q/v$ at  $ \frac{1}{Q}\exp(Q^2/2v)$. At this point it is the largest subpopulation: its peak size is thus a simple estimate for the total population size.  Given the sloppiness in this argument, the best we can hope for at this stage is logarithmic accuracy: 
\be
\log(N\Q)\approx \frac{\Q^2}{2v}  \ .
\ee
The time it takes for the lead population to grow until it becomes  the largest subpopulation  is roughly analogous to the (half) sweep time  for a single mutation in a small population, thus we call it --- rather loosely --- the ``sweep-time":
\be
\tsw\approx Q/v \ .
\ee
This is the time delay between the dynamics of the nose and the consequences of these for  
the bulk of  the distribution: it is thus a particularly important time scale. 
 
From the simple arguments above, one obtains 
\be
v\approx \st^2 \lb\{ \half [2\log(N\st)-\log(\st/\Ub)] + \half\sqrt{[2\log(N\st)-\log(\st/\Ub)] ^2-\log^2(\st/\Ub)}\rb\} \ .
\ee
When $\log(N\st)$ is substantially larger than $\log(\st/\Ub)$, the regime on which we will focus, then  the basic result of DFG follows:
\be
v\approx \frac{\st^2 [2\log(N\st) -\log(\st/\Ub)]}{\log^2(\st/\Ub)}
\ee
As we shall see, this result turns out to be true much more generally than the staircase model,  \cite{DFG,GOOD} provided the appropriate effective $\st$ and $\Ub$ are used. 

\begin{figure}
\includegraphics[width=6.5in]{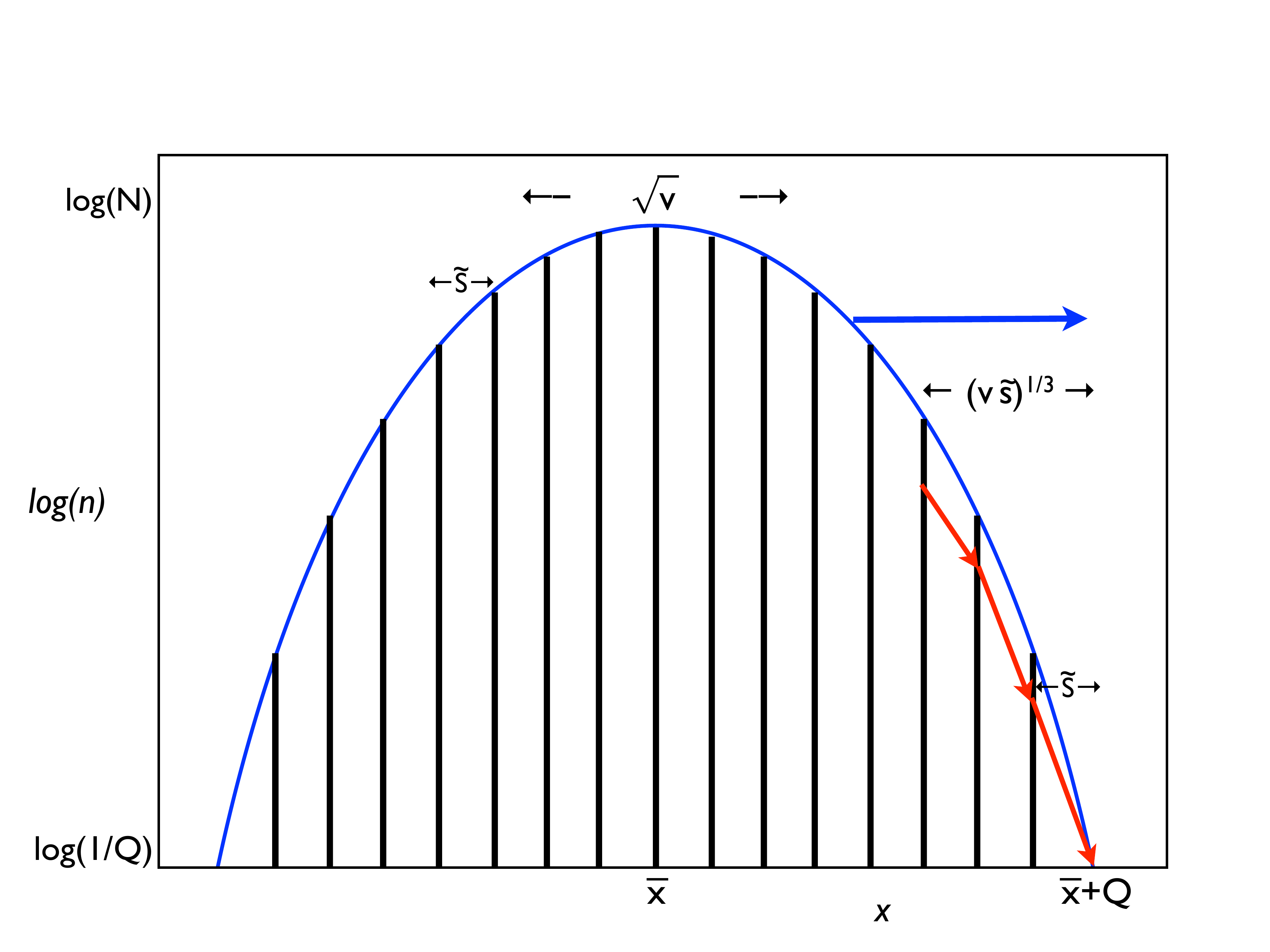}
\caption{Schematic of the  fitness distribution of a steadily evolving population in the huge population, high speed, regime for the staircase model. The sizes, $n$, of the subpopulations as a function of their fitness, $x$, are shown on a logarithmic scale by vertical bars.  Mutations from the fittest subpopulation will establish a new lead population that is  fitter by a fixed amount $\st$, as indicated by the rightmost red arrow.  This increases the fitness of the nose of the distribution, which leads the mean, $\bar{x}$, by $Q$. After a  time delay of $\tsw=\frac{Q}{v}$, the former lead population will have risen to be the largest subpopulation thereby driving the dynamics of the mean.    The envelope of the fitness distribution (blue) is close to a gaussian cutoff at $\bar{x}+Q$: this envelope moves at average speed $v$ equal to the rate of advance of the nose. 

The red arrows indicate  a series of mutations  that  could lead to the lineage from which they arose taking over the population: this is unlikely to occur except from a fitness of order $(v\st)^{1/3}$ or less from the nose.   This and several of the other important fitness scales are indicated. 
  In order of decreasing magnitude: the lead, $Q$, the standard deviation of the fitness distribution, $\sqrt{v}$,  $(v\st)^{1/3}$,   and the fitness increment, $\st$.  
  
     \label{FIG1} }
\end{figure}

The gaussian time-dependence of a sub-population with any fixed fitness implies that the the fitness distribution is a sum of sharp peaks with a close-to-gaussian envelope:
\be
\n(x,t)\sim \sum_k  N \delta(x-ks) e^{-[x-\Up(t)]^2/2v}  \label{FITDISTSUM}
\ee
with $\Up\approx \bar{x}$. 
This is illustrated schematically in \fref{FIG1}. 

{\subsubsection{Large parameters and asymptotic simplifications:} The crucial simplification that enables both heuristic and systematic  analyses to be valid is the separation, at any {\it  fixed} time,  between the strongly fluctuating subpopulations  near the nose  and the subpopulations near the mean fitness  which dominate the  total population size.   For the former, non-linearities needed to keep the population fixed do not matter, while for the latter, stochastic fluctuations from births and deaths do not matter.   In between, neither fluctuations nor non-linearities matter: the established subpopulations grow essentially deterministically at a gradually slowing down rate (due to the advancing $\Up(t)$), but they are still small enough that they do not contribute substantially to the total population.  This is thus a classic case for which one can hope that asymptotic methods should be good. Near the nose one can use branching process methods; near the peak of the population distribution one can fix the total population from the deterministically growing subpopulations; and in between both approximations should be good.  Thus, at least  in principle, one should be able to develop separately asymptotic expansions for the nose and for the dynamics of the mean and then match them together in the intermediate regime.  

The basic large parameter that enables the nose and the bulk of the fitness distribution to be treated separately is the large population size, more precisely 
\be 
L\equiv \log(N\st).
\ee
   However as the beneficial mutation rate is usually small as well (except perhaps in some viral populations) the parameter 
   \be
   \ell\equiv \log(\st/\Ub) 
   \ee
 is also substantially large than unity.  Here $\Ub$ is the effective mutation rate of the predominant beneficial mutations that drive the evolution: these have selective advantage near to $\st$.  The dependence of $\st$ and $\Ub$ on $\mu(s)$ will be analyzed below.

An important  quantity is the ratio between the typical time for the nose to advance, $\test$, and the time for the lead population to become the largest population: the sweep time $\tsw$.  The ratio of these, which we denote $q$, is the typical number of mutations above the mean that are needed to establish the nose. If the distribution of strengths of the predominant mutations is narrow, then $q\approx \Q/\st$.  When the population is not too large --- $\log(N\st)$ only a small multiple of $\log(\st/\Ub)$) --- or if there is a long tail to the distribution of beneficial mutations, $\mu(s)$,  $q$ will be small and will fluctuate (as we discuss later).  But in very large populations, especially with high mutation rates such as occur in mutator populations (so that  $\log(\st/\Ub)$ is not large),  $q$ can  become substantial: $1/q$ is then a useful small parameter.   The self-consistency condition between the assumed speed of the mean and the speed of the nose yields for the staircase model, 
\be
q\approx \frac{2\log(N\st)}{\log(\st\Ub)} =\frac{2L}{\ell}   
\ee
(or more precisely, $q\approx(L+\sqrt{L^2-\ell L})/\ell$, but neglecting the difference in the regimes of interest is consistent with our other approximations).
The condition for the population to be in the multiple mutations regime is that $N\Ub\gg1$, equivalent to $L>\ell$ so that $q
\ge 2$.  
For $q$ to be truly large requires $L\gg\ell$: much larger populations. 
  Thus asymptotic large-population-size behavior will be approached very slowly. 
 Nevertheless, the qualitative, and many of the quantitative, features are correctly captured by the behavior in the large $q$ limit.  
 
 The behavior further simplifies if the population is large enough that  $v/\st^2\approx q/\ell\approx 2L/\ell^2$ is not too small:  the mean fitness then increases relatively steadily because many subpopulations contribute substantially to it at any given time. This can be seen by summing over the subpopulations  from Eq.(\ref{FITDISTSUM}): for $v\gg \st^2$ the sum can be well approximated by an integral. In practice, factors of $\pi$ help and even for $v\sim \st^2 /5$,  the instantaneous speed,  $\frac{d\bar{x}}{dt}\approx
 \frac{d\Up(t)}{dt}$, does not fluctuate much on time scales of order $\test=\st/v$.  But we shall see later that when $v> \st^2$, small variations in the timing of the establishments are important on longer time scales than $\test$ and these  yield  fluctuations of the speed that are of order the average speed.    


 We call the limit $L\gg\ell^2$, for which $v\gg \st^2$, the {\it huge population size}, or {\it high speed},  regime. 
The regime in which the population is not large enough for the mean fitness to increase smoothly on time scales of the establishments of new lead populations, but still large enough that multiple mutations accumulate before any fix,  we call the  {\it modest population size} or {\it modest speed} regime.  This obtains for  $\ell<L<\ell^2$ for which   $v<\st^2$. 

\subsection{Instability and fluctuations of the nose} 

A crucial aspect of the dynamics --- and the root cause of the difficulties in analyzing the model --- is that the nose is unstable. If  a fluctuation causes the nose to advance faster than average, then $Q$ will increase so that  the lead population will grow at a faster rate and the next population  be established sooner, causing $Q$ to increase further.  Similarly, if $Q$ decreases, the slower growth of the lead population will cause it to decrease further.  On time scales longer than the time between establishments, $\test$, but shorter than the sweep  time, $\tsw$,  the heuristic arguments given above imply that
\be 
\frac{dQ}{dt}\approx  \epsilon (Q-\Q) + \ {\rm fluctuations}
\ee
with $\Q(v)$ the mean lead of the nose at mean speed $v$ and $\epsilon$ the growth rate of  fluctuations in $Q$.     A time $\tsw$ later, when the former lead population has grown to be the largest subpopulation,   the fluctuations of the nose would  affect the total population size  so that $\Up(t)$ has to be adjusted to keep $N$ fixed: the nose fluctuations therby drive the fluctuations of the whole fitness distribution.   For the constant $s$ model with beneficial  mutation rate $\Ub$, a simple calculation yields the eigenvalue, $\eps$, of the instability of the nose: if $Q$ increases due to a fluctuation, the speed will increase proportionally.  Thus at fixed speed of the mean of the population, the dynamics of the lead is 
\be
\la \frac{dQ}{dt}\ra \approx  \frac{sQ}{\log(\st/\Ub)} -v \approx v\frac{Q-\bar{Q}}{\bar{Q}}\approx \frac{ Q-\bar{Q}}{\tsw}
\ee
therefore
\be
\eps\approx \frac{1}{\tsw}  
\ee
(where we have here neglected differences between $Q$ and $Q-\half\st$). 

For large populations the sweep time is the most important time scale for the dynamics. 
That the time scale of the nose instability, $1/\eps$, is similar, is the cause of many of the difficulties in the analysis: there is a time delay but no separation of time scales between the fluctuations of the nose and the needed adjustments of the mean to keep the population fixed. [In the last section we will discuss models for which there is such a separation of time scales.]

\subsection{Predominant mutations with short-tailed $\mu(s)$}

With a distribution of the strengths of the selective advantages,  a heuristic argument for the relative contributions of the various $s$ can be made in terms of their contributions to establish a new lead population with lead $Q$.  The rate of mutations to $Q$ from a fitness distribution $\n(\chi)$ is $\int ds \mu(s) \n(Q-s)$.  If the envelope of the fitness distribution has a similar shape near its nose to that of the staircase model (see \fref{FIG1}),  then we expect it to rise roughly exponentially for some range of $\chi$ behind the nose: $\rho(\chi)\sim e^{\gamt(Q-\chi)}$ with $\gamt\approx - \frac{d\log\n}{d\chi}|_Q$ its log-slope which in the staircase model is $\gamt\approx \frac{\Q}{v}$.  This suggests that the weighted effects of the mutations of different strengths will involve 
\be
\int ds\, \mu(s) e^{\gamt s} \ . \label{MUWEIGHT}
\ee
This is reasonable if $\mu(s)$ decays slower than exponentially --- the short-tailed case --- for which the integral of $\mu(s)\n(Q-s)$ will be dominated by $\chi$ near $Q$.
Furthermore,  the exponential weighting, Eq.(\ref{MUWEIGHT}, suggests that there will be a relatively narrow range near to some $s$, the {\it predominant} $s$ that we denote $\st$, that dominate the evolutionary dynamics: we shall see that this is  indeed correct. \cite{DFG,GOOD}    The predominant $s$ is determined by the maximum of the integral in Eq.(\ref{MUWEIGHT}) and is thus related to $\gamt$ by 
\be
-\, \frac{d\log\mu}{ds}|_{\st} \approx\gamt \ .
\ee 

In the {\it long-tailed case} with $\mu(s)$ decaying more slowly than exponentially, the exponentially weighted integral over $s$ is divergent so that the downwards curvature of $\log\n(\chi)$ must be taken into account: the behavior, even for very large populations, is then somewhat different as we discuss in the last section. \cite{DFG,GOOD}

\section{Branching Process at Fixed Speed}

For general $\mu(s)$, and even for the staircase model with $\mu=U\delta(s-\st)$, if one is interested in fluctuations, distributions of sizes of subpopulations, dynamics, etc., analyzing the  behavior near the nose is subtle even though formally it can be done exactly.  [\good also carry out a branching process analysis: the approach we take here 
is somewhat different although some of it parallels theirs closely as discussed at the end of this section.]
We first give the general branching process analysis in the approximation of fixed speed --- allowing the total population size, $N(t)$, to fluctuate around some typical value that is of order the fixed population size of interest. We then use this to derive various results that can be ``matched" on to analysis of variations of the mean fitness that are needed to  keep the population size fixed --- compensating for the delayed effects of the nose fluctuations that occurred a time $\tsw$ earlier. 

  It is convenient to go to the moving frame and define
\be
\chi=x-\Up(t)  \ \ \ \  {\rm with} \ \ \ \   \Up=vt  
\ee
with the associated linear operator
\be
\Lop_v\equiv \Lop_0  + v \pd{}{\chi} \ ,
\ee
the second term coming from the shift to the moving frame and $\Lop$ defined in Eq.(\ref{LOPDEF}). 
To compute any quantities of interest at time $T$, one can use the generating function  
\be
Z_{T}(\{\zeta\})   
\equiv  \left\langle \exp\lb[- \int d\chi \,\zeta(\chi)\n(\chi,T)\rb] \right\rangle \ ,
\ee
where the average is over the stochastic fluctuations from some initial time to  time $T$ and, if the initial conditions are not fixed, also over the distribution of these.  

 With the branching process dynamics --- either for discrete integer numbers of individuals in subpopulations or for the continuous $\n$ (diffusion) approximation we use --- the generating function at time $T$ can be related to that at $T-dt$  by averaging over the fluctuations in the $dt$ time interval.  This can be seen as follows: With the continuous-population-size stochastic dynamics of $\n(\chi,t)$, the noise term in $\n(\chi,T)-\n(\chi,T-dt)$ is proportional to $\sqrt{dt}$.   Expand the generating function to order $dt$ and then average over the noise in the $dt$ time  interval and reexponentiate.  
Second order in the noise times $\sqrt{dt}$ yields a term $\zeta^2\n$.  And first order in the rest yields a linear term from the deterministic part of the dynamics: this has the form $\int d\chi \zeta(\chi)\Lop_v \n(\chi,T-dt)$. The latter can be rewritten in terms of the  {\it adjoint operator}, $\Lod_v$, acting on $\zeta$ to yield an exponent of the generating function of the {\it same form} but in terms of the fitness distribution at time $T-dt$ and $\zeta(\chi)$ replaced by a function $\phi(\chi)$.  
 
 By iteration one thereby obtains the generating function at time $T$ in terms of the generating function at any earlier time, $t<T$:
 \be
Z_{T} = Z_{t}\lb(\{\phi(\chi,t)\}\rb)
\ee
with $\phi$ obeying the backwards-in-time equation:
\be
-\, \pd{\phi}{t}=\Lod_v \phi -\phi^2  \label{PHIDYN}
\ee
with ``initial" --- i.e. final --- condition 
\be
\phi(\chi,T)=\zeta(\chi)    \ .
\ee
The adjoint operator, $\Lod_v$,  involves input from where the mutations {\it go} --- i.e. from $\chi+s\to \chi$ ---  thus
\be
\Lod_v \phi = -v\pd{\phi}{\chi} + \chi \phi  +  \int ds\,  \mu(s) [\phi(\chi+s) -\phi(\chi)]   \ . \label{LOD}
\ee

The dynamics can be integrated back to the initial time where the generating function is found simply in terms of the initial fitness distribution, $\n(\chi,0)$ (or  more generally averages of this):
\be 
Z_{T} = Z_0(\{\phi(\chi,0)\})= \left\langle \exp\lb[-\int d\chi \, \phi(\chi,0)\n(\chi,0)\rb] \right\rangle
 \ee
 Note that the backwards equation is analogous to using the method of characteristics to solve the Laplace transform of the multi-(or infinite-)dimensional Fokker Planck equation for the joint distribution of $\{\n(\chi,T)\}$. If one is  interested in quantities at more than one time, then one can generalize the exponent of the generating function to $- \int^{T}_0 dt \int d\chi \, \theta(\chi,t) \n(\chi,t)$ with the one-time case  equivalent to $\theta(\chi,t)=\delta(t-T)\zeta(\chi)$. The equation for $-\pd{\phi}{t}$, Eq.(\ref{PHIDYN}), then has an additional term: $+\theta(\chi,t)$. 

The key quantities will turn out to be the fixed point,  $\phix$, of the branching process equation, $\Lod_v \phi -\phi^2=0$  and the lowest eigenfunction, $\psi(\chi)$, of the linearized operator about this fixed point which  has eigenvalue $-\epsilon$ with $\epsilon>0$ corresponding to the fixed point being stable. As we shall see, this eigenvalue is controlled by the dynamics of the nose with $\epsilon$   the mean growth rate of deviations of the lead, $Q(t)$, from its average value in steady state, as discussed above.   The other --- trivial ---  fixed point, $\phi(\chi)=0$ for all $\chi$, is unstable: $\phix$ is a global attractor.

Note that integer-population-size models can be analyzed similarly: the only difference is that $-d\phi/dt$ has a different non-linear term which i is periodic in $\phi\to\phi + 2\pi i$ because the population density, $\n(\chi)$, is a series of $\delta$ functions with integer weights. The detailed form of the non-linear term depends on the distribution of numbers of offspring, etc.,  but for small $\phi$, is proportional to $\phi^2$.  All of the universal features are dominated by this small $\phi$ behavior as long as the range of selective advantages or disadvantages of  the population is small compared to the birth and death rates at all times: i.e. the conditions under which the continuous (diffusion) approximation is good.

\subsubsection{Fixation probability:}

Because of the instability of the branching process, at long times $N(t)\equiv\int d \chi\, \n(\chi,t) $ will either be zero or very large and diverging with $t$.  This is reflected in the backwards-in-time convergence of $\phi$ to $\phix$ for almost all ``initial" conditions  $\zeta(\chi)$.
The fixed point, $\phix$, of the backwards time equation for $\phi$, is simply related to the fixation probability of an ``individual" with fitness relative to the mean $\chi$.   More precisely,  given $\n(\chi,t=0)$, 
\be
\pr[\lim_{t\to \infty}N(t)<\infty] = \lim_{t\to \infty}  \la \exp\left[-\zeta N(t))\right]\ra= \la \exp(-\int d \chi\, \n(\chi,0) \phix(\chi) \ra 
\ee
for any positive $\zeta$. 
Thus  for small $\n(\chi,0)$,  
\be
\pr[N(\infty)=\infty]\approx   \int d \chi\, \n(\chi,0) \phix(\chi).
\ee
  
\subsubsection{Information on the distribution of $N(t)$  from the generating function:}\label{BRDDIST}  

If initially $\n(\chi,0)$ is typical of the steady-state traveling wave, then we expect that for  large $t$ the distribution of $N(t)$ will be very broad as its value is roughly determined by the time at which the nose goes unstable after which  $N(t)$ will either grow or shrink in a roughly deterministic manner.  For a {\it broad distribution}  the generating function has a simple interpretation for positive $\zeta$:   
\be
\la \exp(-N\zeta)\ra \approx \pr[N<1/\zeta]
\ee 
as the contributions to the generating function  from almost all $N>\zeta$ are very small, while those from almost all $N<\zeta$ are approximately unity.  Even when the distribution is not so broad, this is a useful approximation and the inverse of the value of $\zeta$ at which the generating function is $1/e$ (or some other $\Oh(1)$ value) can --- at least in some sense --- be considered a {\it typical } value of $N$.  [Note that we use $\zeta$ here as the generating function variable:  this corresponds to $\zeta(\chi)$ being independent of $\chi$.] 

\subsubsection{Pathologies of averages of fitness distributions:}  

For branching process generally, and especially for faster-than-exponentially unstable ones as in the present case, averages of population sizes are very poor characterizations of their typical values.   Averages of $\n(\chi,t)$ are given by the linear term in the expansion of the generating function in powers of $\zeta(\chi)$: the results are, of course, the same as obtained by directly averaging the equation for $\frac{d\n}{dt}$ --- i.e. the deterministic approximation.  By use of Laplace transforms, the dynamics of the average $\la\n(\chi,t)\ra$ can be analyzed as carried out in  \ref{AVDYNSAPP}.    The mutation spectrum enters the results via its Laplace transform:
\be 
\Mz(\lambda)\equiv  \int ds\, \mu(s)\lb [e^{\lambda s} -1\rb] 
\ee
(with the mutation-out term subtracted so that $\Mz(0)=0$). In particular, the time dependence involves $\Mz(\lambda+t)$, Eq.(\ref{RHOHAT}).

{\it Without mutations} 
an initial gaussian $\n(\chi,0)\approx N(0)\exp(-\chi^2/2v)/\sqrt{2\pi v}$ would give $\la\n(\chi,t)\ra$ independent of time.  But  with a truncated gaussian $\n(\chi,0)$ cutoff at a  nose with fitness $\chi=Q$ above the mean --- i.e. $\n(\chi,0)\approx N(0)\Theta(Q-\chi)\exp(-\chi^2/2v)/\sqrt{2\pi v}$ --- the behavior after a time $\tsw=Q/v$ is dominated by descendants of the lead population.  Beyond  this time the population would decay as $\exp(-v(t-\tsw)^2/2)$ in the absence of mutations.   

{\it With  mutations}, the long-time  behavior of the average fitness distribution is very bad. For truncated gaussian initial conditions, the effects of the mutations on the total population size are small initially but grow rapidly for $t > t_M \approx \log[1/\mu(s_M)]/s_M$  with  $s_M$ a characteristic $s$ that dominates  $\Mz(\lambda=t_M)$. 
 Beyond time $t_M$, the mutational term grows exponentially or faster in time as does  the position of the peak  of  the distribution, $\la\n(\chi)\ra$.  
Concomitantly, the average total population size grows as a double exponential of time. 

\subsubsection{Mutation spectrum, bounds,  and dominant range of  $s$:}

In general, the typical population will grow less rapidly than the mean.  In  \ref{AVDYNSAPP}   it is shown that this results in a bound for the (average) lead:
\be
\Q>\chit \equiv \gamt v-\Mz(\gamt) 
\ee
where $\gamt$ is the value of $\lambda$ at which $\Mo\equiv d\Mz/d\lambda =v$: i.e.
\be
 \Mo(\gamt)\equiv \frac{d\Mz}{d\lambda}(\gamt) = \int ds\, \mu(s) s \, e^{\gamt s}=v \ . \label{EXPWEIGHTING}
\ee

As we shall see, for short-tailed $\mu(s)$ for which $\Mz(\lambda)$ is finite for all $\lambda$,  $\Q$ is close to $\chit$ and the contributions of the spectrum of $s$ to the speed are  close to those given by their weight in the condition $\Mo(\gamt) =v$, i.e. weighted by $e^{\gamt s}$. This  is close to the condition anticipated from heuristic arguments given above which correctly concluded that the values of $s$ that dominate for short-tailed $\mu(s)$ are centered around the value $\st$, at which $\exp(\gamt s) \mu(s)$ is maximum. The quantity $\gamt$ is close to the slope of a typical $\log\n(\chi)$ at $\Q$ that appeared in the heuristic argument, justifying the notation used there.   With the beneficial mutation rate small, as we are assuming, 
\be
\gamt\st\gg 1  .
\ee
    A pretty good approximation for short-tailed distributions is to replace  $\mu(s)$ by $\Ut\delta(s-\st)$ with 
\be
\Ut\equiv \Mz(\gamt) e^{-\gamt \st} \ :
\ee
this is close to the  ``predominant-$s$" approximation of DFG.  Unless the curvature of 
$\Lambda(s)\equiv - \log \mu(s)$
 is anomalously small, the range of $s$ around $\st$ that substantially impact $v$ is small compared to $\st$, being of order $1/\sqrt{d^2\Lambda/ds^2|_{\st}}$ which is typically much less than $\st$ for $\gamt\st\gg 1$.  

For example,  if the distribution of beneficial mutations  is a half-gaussian: $\mu(s)=2U/\sqrt{2\pi\sigma^2} \exp(-s^2/2\sigma^2)$, $\gamt\approx \sqrt{2\ell_0}/\sigma$ and $\st\approx \sqrt{2\ell_0}\sigma$, with $\ell_0=\log(\sigma/U)$. But the range of $s$ that dominate are within $\sigma$ of $\st$ since $d^2\Lambda/ds^2=1/\sigma^2$. The effective mutation rate to the predominant-range mutations is $\Ub \sim U^2/\sigma$  (so that  $\ell\equiv\log(\st/\Ub)=2\ell_0$) which can be a tiny fraction of the total beneficial mutation rate, $U$. In terms of $\Ub$, the range of $s$ that dominates is narrower than $\st$ by a factor of $1/\sqrt{\log(\st/\Ub)}$.

\subsubsection{Formal steady state $\n(\chi)$:}  

Formally, there exists a steady state solution to the average equation analyzed above, i.e., a fixed point $\rhox(\chi)$ satisfying $\Lop_v\rhox=0$ as found in \ref{FORMALAPP}.
 The function $\rhox(\chi)$ is well-behaved for $\chi$ less than some value --- which turns out to be very close to $\Q$ --- beyond which it becomes negative and starts to oscillate, while continuing to decay rapidly, as  $\chi$ increases. 
But a truncated $\rhox$ restricted to the range $\chi<\Q$, i.e.
\be
\rhox_\Q(\chi)\equiv \Theta(Q-\chi)\rhox(\chi),
\ee
 is close to a {\it typical} $\n(\chi)$ with the constant-$N$ constraint. Nevertheless,  although $\rhox_\Q(\chi)$ is essentially what many authors have analyzed as the ``average" distribution, it is at best only a pseudo-average --- closer to a logarithmically weighted average, $\exp[\la\log\n\ra]$ --- and very far from the actual average of $\n(\chi)$ which is dominated by rare fluctuations as we shall show in Sec. (\ref{AVFITSEC}).

\subsection{Fixation probability at fixed speed}\label{FIXPROBSEC}   

The fixed point, $\phix(\chi)$, of the backwards in time equation for $\phi(\chi,T)$ obeys
\be
\Lod_v \phix -\phi^{\ast 2}=0
\ee
with $\Lod_v \phi$ given by Eq.(\ref{LOD}).
The fixed point, shown  on a logarithmic scale in red in \fref{FIG2},  has some simple features and some rather more subtle ones.  

\begin{figure}
\includegraphics[width=6.0in]{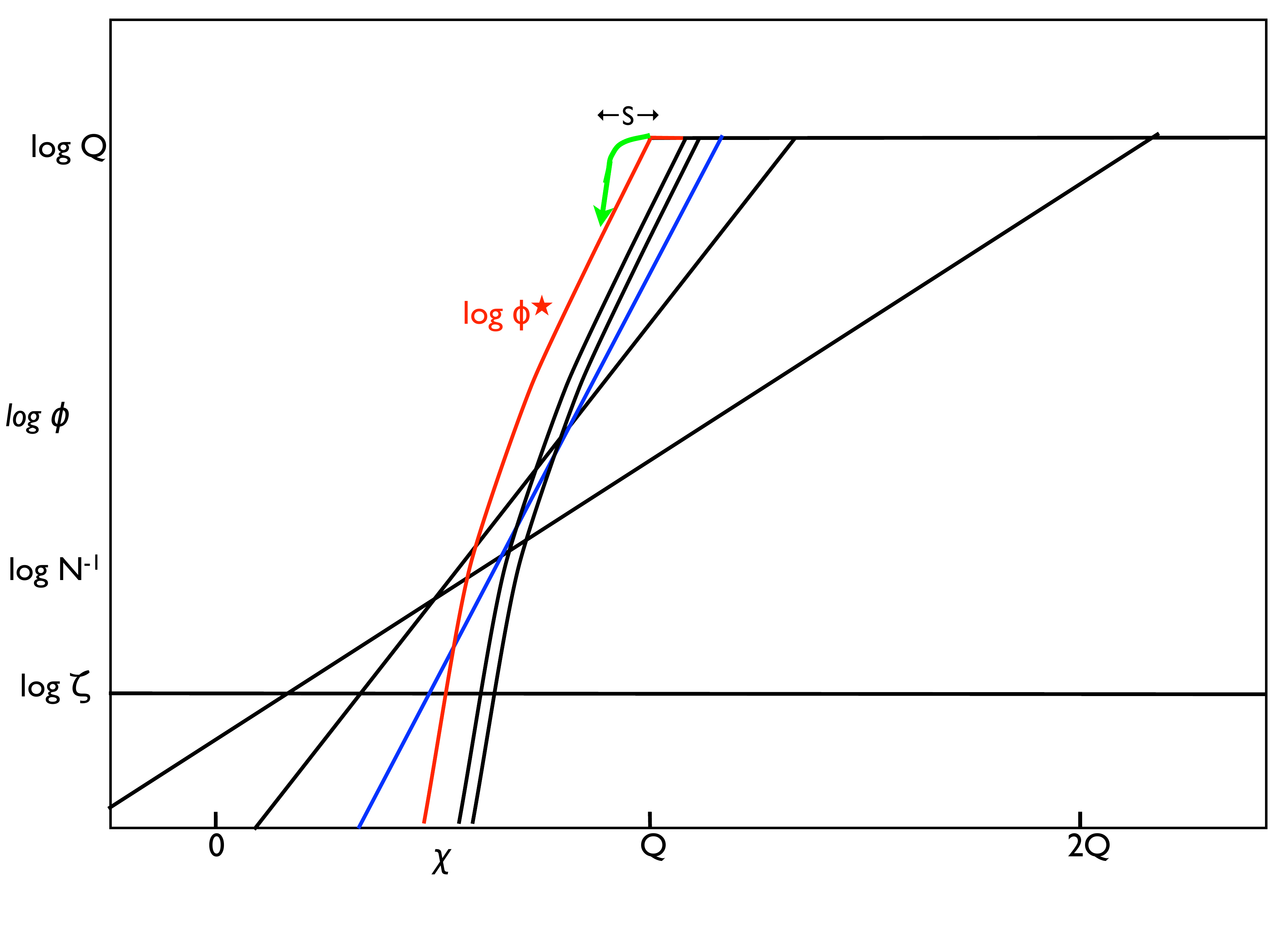}
\caption{Schematic of the  backwards-time dynamics of  the function $\phi(\chi,t)$ that determines how the size of the population at time $T$ depends on the fitness distribution at earlier times, $t<T$, with the speed of the mean of the fitness distribution fixed at $v$.   Note the logarithmic scale for $\phi$. The dynamics are shown in the moving frame with $\chi=x-vt$. The generating function $\la e^{-\zeta N(T)}\ra$ is obtained by starting at $t=T$ with $\phi=\zeta$ (horizontal line) and integrating backwards which results in $\log\phi$ becoming steeper with decreasing $t$ until $t=T-\tsw$ (blue line).  For $\chi$ larger than a sharp shoulder (of width $\frac{v}{\Q}$) whose position is denoted  $P(T-t)$, $\phi$ saturates at $\phi\approx \chi$ which is almost horizontal on this logarithmic scale. For $T-t>\tsw$, $\phi$ rapidly attains its asymptotic shape which then moves slowly with $P(T-t)$ converging exponentially to $Q$ and $\phi$ to its fixed point, $\phix(\chi)$ (red).  The green arrow indicates where the most important mutations that advance the nose of the fitness distribution come {\it from}: these cause the slope of $\log \phi$ near the shoulder to saturate at  $\gamt\approx\tsw$.  
    
The value of $\zeta$ shown is much less than the inverse of the typical $N$, $\Nt$: the generating function then gives information on the large $N$ tail of its distribution. At fixed population size, such  fluctuations drive fluctuations of the speed of the mean fitness.  For $\zeta>1/\Nt$, the blue curve would be to the left of the red curve and the final approach to the fixed point would be from $P<Q$. 
  \label{FIG2} }
\end{figure}

{\it High fitness limit}: At large $\chi$, the dominant balance is between the $\chi\phi$ and $\phi^2$ terms so that $\phi\approx \chi$,   the fixation probability of individuals that are substantially ahead of the average lead: this rarely occurs but when it does, the fixation probability is just the establishment probability as the future population will be almost entirely descendants of any such anomalously fit mutant that establishes and starts to grow deterministically.  This is a consequence of the instability of the nose: if $Q$ is substantially ahead of the typical lead, it will grow even faster giving the rest of the population even less chance to produce mutations quickly enough to catch up.  

{\it Linear regime:} For $\chi$ substantially below some $\Q$  we expect that there is little chance that individuals will produce enough mutants to reach the lead and have a chance at taking over the population.  Thus $\phi$ should be very small.  As linear behavior in the dynamics of $\phi$  corresponds to deterministic behavior in the dynamics of $\n(\chi)$, we expect this to obtain except near to the nose.  Thus the simple --- and correct --- conjecture is that non-linearities in $\phix$ will be unimportant except close to and above the value of $\chi$ that roughly corresponds to the average lead of the nose, $\Q$. 

{\it Shoulder}: Near to $\Q$ we expect a crossover between the fluctuation-dominated and deterministic regimes: i.e. from saturated to linear behavior of $\phi$.   In the saturation regime the effects of mutations are small and we expect this to hold for at least some range below the shoulder at $\Q$ as the effects of the mutations from $\chi+s>\Q$ will be modest until $\phi(\chi)$ has dropped substantially below $\Q$.   If we can ignore both the mutations and  the difference between $\chi$ and $\Q$ in this shoulder regime, then we have simply 
\be
\phix \approx \frac{\Q \exp(\Q(\chi-\Q)/v) }{1+ \exp(\Q(\chi-\Q)/v)}
\ee
which satisfies the simple equation, $-v d\phix/d\chi + Q\phix - \phi^{*2} = 0$ (which is linear substantially below $\Q$). Note that for convenience we have chosen the mid-point of the crossover to define $\Q$.   

{\it Multiple scales:} The scale over which the crossover occurs is simply $v/\Q$:  the amount the mean fitness increases in a time that is the inverse of the growth rate of the lead population (which is also the typical time for new mutants that advance the lead to escape the fluctuation regime and become established).    Note that this fitness scale is $\ll \Q$ for large populations, and $\ll \st$ for small mutation rates.   Thus we have at least four scales of fitnesses: $\Q, \sqrt{v}$ -- the width of the population distribution ---, $\st$, and $v/\Q$ with $\Q \gg \sqrt{v}\, \&\, \st \gg v/\Q$.  The inverses of these correspond, respectively, to the time for the nose population to grow by a factor of $e$, the time for the mean fitness to increase by the standard deviation of the fitness distribution, the range of times over which most of the mutants that advance the nose by one step are established,  and  the time, $\tsw$, for the nose population to rise to be the largest population. For  huge  populations for which $\sqrt{v}\gg \st$,  we will see that there is generally a  fifth scale, of order $(v\st)^{\frac{1}{3}}$ which is  intermediate between $\sqrt{v}$ and $\gg \st$. 
 Finally, except for the staircase model, there is a sixth scale, between $v/\Q$ and $\st$, that is the range of $s$ around $\st$ that contribute substantially to the dynamics, as discussed earlier.   Several of the most significant fitness scales are illustrated in \fref{FIG1}.

\subsubsection{Well-behaved solution:}  If the effects of mutations could be ignored for $\chi$ well below $\Q$, the only solution in this regime would be simply $\exp[\chi^2/2v]$ which is clearly nonsense   and it can be checked that the effects of the mutational term cannot save it.   But this observation suggests how $\Q$ will be determined.  With only beneficial mutations, the fixed point equation can be integrated down from large $\chi$, starting with the general well-behaved solution in that regime: $\phi\approx  \chi  - A \exp(-\chi^2/2v)$, with $A$ an arbitrary constant. This is obtained  by linearizing around the $\phi\approx \chi$ limiting behavior and ignoring the effects of mutations which can be justified and treated perturbatively in this regime.   The value of $A$ will determine $Q$, defined roughly as the point at which the solution substantially deviates from $\phi\approx \chi$ so that it matches onto the solution in the shoulder regime derived above.  For small $A$, the behavior in the linear regime below the shoulder will be dominated by the badly behaved upside-down gaussian solution. In contrast, for large $A$, it can be seen that $\phi$ will go negative and diverge to $-\infty$.  Adjusting $A$ to find a well-behaved solution  is thus like solving an eigenvalue problem by shooting from one end and adjusting a freely chosen coefficient to find the well-behaved solution (or solutions) at the other end.   Thus we need to find a solution that is well-behaved in the linear regime.     The detailed analysis is carried out in \ref{WELLBEHAPP}.

{\it Log-slope approximation:} The most salient features in the linear regime are well captured by a WKB-like approximation.  Writing
\be
\phi=\Q\exp\lb[-\int_\chi^{\Q} \gamma(\chi') d\chi'\rb]
\ee
and assuming that the log-slope 
\be
\gamma(\chi)\equiv \pd{\log\phi}{\chi} 
\ee
 is not too rapidly varying with $\chi$,  yields 
 \be
\phi(\chi+s)\approx \phi(\chi)\exp[s\gamma(\chi)]
\ee
 so that the mutational term is approximately $\Mz[(\gamma(\chi)]\phi(\chi)$. 
 
 The linearized fixed point equation in the log-slope approximation  simply becomes:
\be
v\gamma-\Mz(\gamma)\approx \chi \label{CHIOFGAM}
\ee
with $\chi$ as a function of $\gamma$ obtained by integration.  
The well-behaved solution is that for which $\gamma\to \infty$ as $\chi\to -\infty$: the corresponding $\phi(\chi)$ is manifestly less than the ``generic" badly-behaved upside-down gaussian solution which corresponds to $\gamma=\chi/v$.

The WKB solution cannot be correct for $\chi>\chit=v\gamt-\Mz(\gamt)$ at which $\chi(\gamma)$ from Eq.(\ref{CHIOFGAM}) is maximum.  An expansion around the WKB approximation suggests that it breaks down when
\be
\chit-\chi \sim \left[\Mt(\gamt)\right]^{\frac{1}{3}} \approx (v\st)^{\frac{1}{3}} 
\ee
with $\Mt(\gamma)\equiv\frac{d^2\Mz}{d\gamma^2}$. For huge populations for which $v\gg \st^2$, this is a distance of more than $\st$ from $\chit$ and the behavior changes over a substantial range of $\chi$.  But for more modest population sizes with smaller $v$, the scale $\st$ is less than $(v\st)^{\frac{1}{3}}$  and higher order terms make the saddle point approximation break down for $\chit-\chi$ of order $\st$: the scale at which the discreteness of the typical jumps, and corresponding fine-structure in $\phix(\chi)$, occurs.

\subsubsection{Moderate speeds, $v\ll \st^2$:} 

To find the fixed point, $\phix$, we are faced with matching the linear solution(s) to the non-linear solution in the region just below $\Q$.   For small $v/\st^2$, a simple argument leads to the correct behavior.   With $\Q$ not much different from $\chit$ and  $\Q-\chi$ substantially less than  $\st$, the corrections to the shoulder solution from mutations are not large. Specifically, the contributions to the fixed point equation from mutations are well approximated by the input from the non-linear region: $\Q\int_{\Q-\chi}^\infty \mu(s)ds$. 
As shown in  \ref{MODAPP}, this input is only important for $\chi$ near $\Q-\st$ and the condition that the solution matches onto the well-behaved linear solution yields 
\be
\Q\approx v\gamt+\frac{1}{2}\st +  {\cal O}\lb[\frac{v}{\st}\log(\st^2/v)\rb] \approx \chit + \frac{1}{2}\st + {\cal O}\lb[\frac{v}{\st}\log(\st^2/v)\rb]   \ .
\ee 
(Note that the $\Mz(\gamt)$ in $\chit$ is of order $v/\st$ and thus smaller than the neglected correction terms.)
The observation that the behavior for all $\chi<\Q$ is controlled by the behavior of the last mutational step below $\Q$ corresponds exactly with the approximations made in DFG for the staircase model of ignoring the mutational input and fluctuations for all but the lead population.   One important note, however, is that in this regime with $v/\st^2$ small, there will be large oscillations in $N(t)$ at fixed $v$ caused by the close-to-periodic  occurrence of the predominant mutations that control the dynamics of the lead.  Correspondingly, there will be oscillations in $\gamma(\chi)$,  the log-slope of $\phi$, with period $\st$ --- but their magnitude is small compared to the non-oscillatory part and decays more rapidly as $\chi$ decreases. The corresponding oscillations in $\phi(\chi)$ are, however, not small, being similar to those in $N(t)$.

\subsubsection{High speeds, $v\gg\st^2$:} 

For huge populations with $v\gg \st^2$, the saddle point approximation breaks down much further than $\st$ below $\Q$. But (in contrast to the moderate speed regime) the behavior below $\Q$ is now smooth and one can expand in derivatives.  
It is convenient to pull out the rapidly varying factor, $e^{\gamt\chi}$, and $\Q$, defining
\be
f\equiv \frac{\phi e^{-\gamt(\Q-\chi)}}{\Q}
\ee
so that the linearized fixed point equation becomes 
\bea
0 & = & -v \pd{f}{\chi} + (\chi-\chit)f +\int ds \mu(s)e^{\gamt s}[f(\chi+s)-f(\chi) \\  
& = & b^3\pd{^2f}{\chi^2} +(\chi-\chit)f + {\cal O}\lb[v\st^2 \pd{^3f}{\chi^3}\rb]
\eea
with the fitness scale determined by the second derivative term
\be
b=    \left[\half \frac{d^2\Mz}{d\lambda^2}(\gamt)\right]^{\frac{1}{3}}  \approx \left(\frac{v\st}{2}\right)^\frac{1}{3} \ .
\ee
This is Airy's equation with the condition that the solution is well behaved for $\chit-\chi$ large mandating the Ai solution.  Matching this onto the shoulder solution and the mutational effects near the shoulder indicate that the non-linearities become important only very close to the first negative zero of the Airy function which occurs when its argument is $-a_1$. Analysis in  \ref{FASTAPP}  yields 
\be
\Q \approx v\gamt-\frac{v}{\st} +a_1 (v\st/2)^\frac{1}{3} + \Oh(1/\gamt)  
\ee
with the correction term of order the width of the shoulder region which is much smaller than $\st$. 


\subsection{Sharp shoulder approximation:} 

In the limit in which the width of the crossover from linear to non-linear is much smaller than the other relevant scales --- particularly $\st$ --- we can treat the shoulder approximately as a sharp boundary at $\Q$ with the linear approximation taken all the way up to $\Q$: this is valid with $\log(\st/\Ub)$ large.   The inferred value of $\phix$ at $\chi=\Q$ should be simply  $\Q$ as would be obtained from extrapolation of the linear region of the shoulder solution.   Using this, along with the appropriate conditions at $\chi=\Q$, we can  find  all the properties of $\phix$ discussed above --- and much more --- far more directly.  The approximate fixed point equation on the interval $(-\infty,\Q)$ is 
\be
\Lod_v \phix + \Q \int_{\Q-\chi}^\infty \mu(s)ds =0
\ee
with the second term the input from the non-linear saturation regime (ignoring the difference between $\phix\approx \chi$ and $\phix\approx \Q$ which is readily justified since $\Q\gg\st$).   
This can be solved formally by Laplace transforms as carried out in \ref{SHOULDAPP}.  The condition for a well-behaved solution that determines $\Q$ is a ``solvability" condition as occurs in many problems: it enforces the condition that the ``projection" onto the badly behaved solution is zero.   Since the linear operator is the adjoint of the operator that determines  the  formal fixed point of the average fitness distribution, $\rhox$, it is not surprising that the condition can be written in terms of the truncated, well-behaved,  form of this:
\be
\int d\chi \, \rhox_{\Q} (\chi) \Omega(\chi) = v\rhox_{\Q}(\Q)  \label{RHOQCOND}
\ee
where $\Omega(\chi)=\int_{\Q-\chi}^\infty ds \mu(s)$.
The integral over $\chi$ extends only up to $\Q$ --- beyond which $\rhox_\Q$ is zero --- and $\rhox_{\Q}$ must have a non-zero value at this end-point.  The condition Eq. (\ref{RHOQCOND})  is simply that the total rate of mutations to fitnesses higher than the lead is ``typically" just the right amount needed to advance the nose at speed $v$ --- similar  to what one might expect from the heuristic argument for the stochastic dynamics of the nose.  

The results quoted above that are derived  in the Appendix by matching together the various regimes, can be obtained directly and far more efficiently from this solvability condition. Furthermore, as discussed in the last section, the sharp shoulder approximation is valid in many circumstances and can be used to quickly derive results in very different regimes than those analyzed here.  In particular, Eq.(\ref{RHOQCOND}) is  applicable for obtaining $\phix$ even  if $\mu(s)$ decays more slowly than exponentially for large $s$. 
   
 \subsection{Fixation probability and determination of $N$}
 
 The fixed point of the branching process does not provide any direct information about the $N$ that corresponds to the assumed average speed, $v$ in the fixed population size model.  A natural approximation is to use the interpretation of $\phix(\chi)$ as the fixation probability of individuals with fitness $\chi$ above the mean and use the fact that the descendants of one individual will eventually takeover the population.   Thus the average fixation probability should be unity so that $\int d\chi \phix(\chi)\n(\chi) = \Oh(1) $
which fixes $N$ as a function of $v$ up to a factor of order unity, \cite{OH} hence $v(N)$ much more accurately because of the dependence of $v$ on $\log N$. 
But it is not clear what $\n(\chi)$ should be used: as we shall see later, the average is certainly not correct.   A natural guess is to use the same $\n$ that yielded the condition for $\Q(v)$: i.e. the formal fixed point truncated at $\Q$ leading to the conjecture that
\be
\int d\chi \phix(\chi)\rhox_\Q(\chi)  = \Oh(1) \ . \label{PHIRHOINT}
\ee
As we shall show, this integral is dominated by a narrow range just below $\Q$. 
 
It is not at all clear at this stage whether fluctuations in the speed induced by fluctuations of the nose and/or fluctuations in $\n(\chi,t)$, will invalidate the conjecture Eq.(\ref{PHIRHOINT}). We thus must analyze the fluctuations,  which we do in the next section.
 
 \subsection{Comparison with results of Good et al}
 
 \good have analyzed the fixation probability at  fixed speed and found some of the same results discussed above, with a few differences.  They work directly with the self-consistent equation for the fixation probability, which they denote $w(\chi)$ (denoting by their $x$ our $\chi$). This is readily derived from the observation that for the descendants of an individual not to fix, none of its descendants produced by mutation can fix. With mutations occurring continuously in time and populations approximated as continuous, this yields $\Lod w -w^2$. Thus $w=\phix$.  
 
The first three properties of $\phix$ listed in Sec. \ref{FIXPROBSEC} were obtained by Good et al.   But they did not analyze correctly the behavior of $\phix$ for $\chi<\Q$: they approximate it by the badly-behaved upside-down gaussian solution, $\phix \approx \Q\exp[(\chi^2-\Q^2)/2v]]$ for $0<\chi<\Q$, and zero for $ \chi<0$.    If one uses the condition Eq.(\ref{PHIRHOINT}) to fix $N$ --- the most natural condition in the fixation probability framework --- then this approximation for $\phix$ would imply that the distribution of fitnesses of the individuals whose descendants take over the population would be uniform on $(0,\Q)$ in contrast to the sharply peaked distribution just below $\Q$ that actually obtains as we show below.  However,  they primarily use a condition involving the fixation probability of new mutants, \cite{OH}
\be
\int d\chi\int ds\, \phi(\chi)s\,\mu(s)\n(\chi-s)=\Oh(v),
\ee
 which does not suffer from this pathology.  They show, via comparisons with simulations, that this leads to good results in the population-size regimes that are quite far from the asymptotic behaviors on which we focus in this paper. 

 For the fitness distribution, $\n(\chi)$,  Good et al make a simple gaussian approximation which is good except in the Airy regime near the nose that obtains in the high-speed regime: while this should affect factors inside the logarithm in the relationship between $\log N$ and $v$, this is not a large effect.  
An exception for which the gaussian approximation for $\n$ is not good, is if 
  deleterious mutations play a substantial role: we also have  not analyzed this case (but see reference  \cite{RATCHETSHRNEH}).   

With long-tailed distributions, which we do not consider in detail here, the condition  for the weighting of $s$ that determines $\st$ and  the predominant range around it is different than the simple exponential weighting of Eq.(\ref{EXPWEIGHTING}) which is non-sensical in this case. Good et al's analysis of the weighting includes the effects of the downward curvature of the fitness distribution from which the mutations come: although this is implicit in our general expressions that determine $\Q$ and hence $\phix$, such as  Eq. (\ref{RHOQCOND}), we have not analyzed this regime in detail.

\section{Dynamics and Distribution of $N(t)$ at Fixed Speed}

From the branching process methods used above, we can analyze the 
  fluctuations of $N(t)$ at fixed $v$:  these we can then  use to infer the fluctuations of $v(t)$ and other quantities at fixed $N$.   We also consider distributions of other quantities, particularly the effective population size of the nose and its temporal fluctuations. 
  Of primary interest is the typical, rather than average, behavior. 
    

We are interested in the distribution of $N(T)$ given  the distribution of $\n(\chi,t=0)$.  We thus compute the generating function:
\be
Z_T(\zeta)\equiv \la e^{-N(T)\zeta} \ra
\ee
as a function of $\zeta$ (here independent of $\chi$ as $N=\int d\chi \n(\chi))$.  Although this generating function --- simply the Laplace transform of the the distribution of $N(T)$ --- can only be inverted  by continuing $\zeta$ into the complex plane, as we shall see much information can be inferred from the behavior for real positive $\zeta$.  The relevant $\zeta$ of interest are those of order the inverse of a typical population size, call this $\Nt$, which we want to be of order the actual population size when it is held fixed.

Defining $t=T-\tau$, the backward time dynamics of $\phi$ are given by 
$\pd{\phi}{\tau}=\Lod_v \phi -\phi^2$
with initial condition $
\phi(\chi,\tau=0)=\zeta$ independent  of $\chi$
with $\zeta$ very small.  The dynamics  are illustrated schematically in \fref{FIG2}. Initially, the log-slope of $\phi$ will be small enough that the mutational terms can be neglected.  In the linear-regime, we then have
\be
\phi(\chi,\tau)\approx e^{\chi\tau-v\tau^2/2} \zeta
\ee
which is valid for $\chi<\P(\tau)$ with 
\be
\P(\tau)\approx \frac{1}{\tau}\log\lb(\frac{\P(\tau)}{\zeta}\rb) +\frac{v\tau}{2} :  \label{PAPPROX}
\ee
$\P(\tau)$ is the time-dependent position of the shoulder,  the value of $\chi$ above which $\phi$ saturates at $\phi\approx \chi$.   As the log-slope in the linear regime is simply $\tau$, the mutational term will start to become important when $\tau\approx \gamt$.  This is close to, as could have been anticipated, the sweep time $\tsw$ after which the total population size is affected by fluctuations of the nose.  Concretely, the subpopulation in the lead at time $T-\tsw$ will be the largest population at time $T$ while  up until this time the dynamics will be dominated by the almost deterministic growth of the already established subpopulations and the stochastic behavior of the advance of the nose via newly established subpopulations will not yet affect the total population size.  This suggests that for $T<\tsw$, the nonlinearities in $\pd{\phi}{\tau}$ which reflect the fluctuations, should not much affect  $\int \phi(\chi,0) \rho(\chi,0)$: i.e. that with $\zeta\sim1/\Nt$ the shoulder $\P(\tau)$,  should, for $\tau<\tsw$, be larger than $\Q$  in the regime where $\rho(\chi)$ is close to zero.    

With $\Q\approx v\gamt$, there is one particular value of $\zeta$, $\zetat\approx \Q e^{-\Q^2/2v}$ at which $\P(\tau=\gamt)\approx \Q$ and for this value the important parts of $\phi$ --- the position of the shoulder and log-slope just below it --- are quite close to the fixed point $\phix$ already at $\tau\approx \gamt$.  The inverse of $\zetat$ is a first approximation to the typical population size: $\Nt \sim \frac{1}{\zetat}\sim
e^{\Q^2/2v}/\Q$ as from the heuristic arguments.

The dynamics of $\phi(\chi,T-\tau)$ under the full linear operator can be analyzed generally by Laplace transforms and approximately by the log-slope approximation.  This must then be matched to the moving shoulder at $\P(\tau)$ to determine its dynamics self-consistently, as carried out in \ref{DYNAPP}.  The approximation of ignoring mutations up until $\tau\approx \tsw$ is shown to be rather good and  after that time the most important effect of  mutations 
is to shape and move the shoulder region centered at $\P(\tau)$.  At $\tau\approx \tsw$, the log-slope of $\phi$ at $\P(\tsw)$ is close to $\gamt$ so that $\phi$ has a shape similar to $\phix$ --- above, through, and  to some distance below the shoulder --- but it is shifted by $\P(\tsw)-\Q$.  Below the shoulder the log-slope continues to increase for some time after $\tsw$ and $\phi$ drops concomitantly, but is prevented from decreasing too far by the mutations from higher $\chi$ which quickly stabilize $\phi$ at close to the same shape as $\phix$.    Thus on a time scale (asymptotically) much shorter than $\tsw$, $\phi$ converges from its form at $\tau =\tsw$ to one of a family of functions, $\{\phi_\P(\chi)\}$, parametrized by the position of the shoulder, $\P$.  As long as $\P$ is relatively close to $\Q$, the shape of all the $\phi_\P$  are very similar, even quantitatively, to $\phix$.  

But for $\P\ne \Q$, the mutational input from the saturation regime to just below the shoulder is not quite balanced by the combined effects of the growth in the rest frame and the motion at speed $v$ which acts to decrease $\phi$.  This imbalance results in $\P(\tau)$  slowly changing.  For $|\P-\Q|\ll \Q$ --- as it certainly will be unless $\zeta \Nt$ is either huge or tiny --- 
the dynamics of $\P$ after $\phi$ has converged to some $\phi_\P$ is simple: 
\be
 \frac{d\P}{d\tau} \approx \eps (\Q-\P) 
 \ee
 with $\eps\approx v/\Q\approx1/\tsw$. The stability of the  fixed point, $\phix$, backwards in time simply reflects the instability of the nose and is thus controlled by the same eigenvalue $\eps$.    
 
We could have directly analyzed the convergence of $\phi$ to $\phix$ by linearizing around $\phix$. The dynamics of the deviation from the fixed point, $\Delta\phi \equiv \phi-\phix$, is $\pd{\Delta\phi}{\tau}\approx (\Lod_v -2\phix)\Delta\phi$.  The eigenvalues of $\Lod_v-2\phix$ are all negative, with the least negative being $-\eps$. The corresponding eigenfunction, $\psi(\chi)$, must be   
\be
\pd{\phi_\P}{\P}|_{\P=\Q}.
\ee  
The other eigenvalues, which also control the approach to the whole family of $\phi_\P$'s, 
are (asymptotically) much larger than $\eps$, of order $\st$ for moderate speeds, and of order $(v\st)^{\frac{1}{3}}$ for high speeds.

At the special value of $\zeta$, $\zetat$, the initial stage of the dynamics will take $\P(\tsw)$ to $\Q$ and then, much more quickly, take  $\phi(\chi,\tsw)$ to $\phix$.   This special behavior corresponds to vanishing amplitude of the slow transient (i.e. of the  eigenfunction, $\psi$).  For larger $\zeta$, $\P(\tau)$ will overshoot $\Q$ and then approach it from below, while for $\zeta<\zetat$, $\P$ will approach $\Q$ from above as shown in \fref{FIG2}.  For a wide range of $\zeta$, because of its logarithmic dependence on $\zeta$ from Eq.(\ref{PAPPROX}),
$\P(\tsw)$ will be close to $\Q$, deviating from it as
 \be
 \P(\tsw)\approx \Q + \frac{\log(\zetat/\zeta)}{\tsw} \ .
\ee
Since $\tsw\gg \frac{1}{\st}$, $\P(\tsw)$ deviates from $\Q$ by only a small fraction of $\st$ unless $\zetat/\zeta$ is very large or very small.   

\subsubsection{Distribution of $N(T)$:}

We are now in a position to understand the statistics of the fluctuations in population size at fixed speed.   A  crucial simplification arises from the simple form of the slow transient  to the fixed point $\phix$.  After a time $\tsw$, $\phi$ will be mostly determined by $\P(\tsw)$  independent of the earlier history, with   $\P(\tau)$ then decaying exponentially from $\P(\tsw)$ to $\Q$.  The $\phi_\P(\chi)$ that obtain after $\tsw$ are close to $\chi$ for $\chi>\P$, have a narrow shoulder region of width $v/\P\approx v/\Q$, and below the shoulder drop exponentially as $\phi\approx \P\exp[\gamt (\chi-\P)]$ with steepening log-slope at lower $\chi$.  This exponential behavior that is only weakly dependent of $\P$, means that for any $\n(\chi)$ with support that does not extend beyond $\P$, 
\be
\int d\chi \phi_\P(\chi) \n(\chi) \approx e^{\gamt (\Q-\P)} \int d\chi \phix(\chi)\n(\chi) \label{PHIP}
\ee
as long as the latter is dominated by the regime in which $\phix$ has log-slope close to $\gamt$.   As both  $\n(\chi)$ and $\phix$ are log-concave downwards, this will indeed be the case.  Furthermore we expect that the actual $\n(\chi)$ will not usually  extend beyond $\Q$ --- the typical lead of the nose. Thus at least for $\P>\Q$, which will turn out to be the important regime,  the approximation of Eq.(\ref{PHIP}) should be good.  

For the generating function for $N(T)$ with $T>\tsw$,  the dynamics of $\P$ yield
\bea
\P(\tau=T)-\Q &\approx& [\P(\tsw)-\Q]\exp[-\eps(T-\tsw)] \\
& \approx& \frac{1}{\tsw}\log(\zetat/\zeta)\exp[-\eps(T-\tsw)] 
\eea
so that the needed quantity at time zero is 
\be
\int d\chi \phi(\chi,0)\n(\chi,0) \approx \int d\chi \phix(\chi)\n(\chi,0) \lb(\frac{\zeta}{\zetat}\rb)^{\alpha(T-\tsw)} 
\ee
with the exponent
\be
\alpha(t)\approx e^{-\eps t} :  \label{ALPHA}
\ee
since $\eps \approx1/ \tsw$,  $\alpha(\tsw)\approx 1/e$.  

We have seen that the distribution of $N(t+\tsw)$ depends on 
\be
\nL(t) \equiv\int d\chi\, \phix(\chi)\n(\chi,t)
\ee 
--- in particular $\la \exp(-\nL(t))\ra$ --- from a  time $\tsw$ earlier.  
But what about the distribution of $\nL$ itself?     Since this is --- in the branching process --- the sum over all the individuals at one time of the probability that each fixes, we expect that it will typically be of order unity.  Indeed, since in the branching process the fate of each individual is independent, the quantity that arises in the generating function calculations 
 \be
\la e^{-\int \phix \n}\ra = 1- \pr[{\rm at\  least\ one \ fixes}]
\ee 
which should be of order unity.  

The quantity $\nL(t)$ plays a special role: its distribution at one time is essentially determined by its distribution at earlier times. This can be seen from its generating function, $\la\exp(-\theta\tilde{\nL}\ra)$, for which  we are interested in $\theta=\Oh(1)$ (in contrast to $\zeta =\Oh(1/\Nt)$ in the generating function for $N$).  We thus need to analyze the backwards time dynamics starting from $\phi=\theta\phix$.  For $\chi>\Q$, $\phi$ will rapidly approach $\phi\approx \chi$, while moving to higher $\chi$ at speed $v$.   This will quickly produce a shoulder at some position, $\P_\theta$, which for $\theta<1$ is
$\P_\theta\approx \Q+\frac{v}{\Q}\log(1/\theta)$. Between $\Q$ and $\P_\theta$ the log-slope of $\phi$ will become close to $\gamt\approx \frac{\Q}{v}$. Thus after a time of order $\log(1/\theta)/\Q\ll  \test $ we will have $\phi\approx \phi_{\P_\theta}$.  From then on, the dynamics is as above with exponential convergence of $\P(\tau)$ to $\Q$.   The result
\be
\la e^{-\theta\nL(t)} \ra \approx \la e^{-\theta^{\alpha(\tau)}\nL(t-\tau)} \ra \ ,
\ee
with $\alpha(\tau)$ from Eq.(\ref{ALPHA}), then follows.  At fixed $\nL(t-\tau)$, this
is a the Laplace transform of a (one-sided) Levy distribution with parameter $\alpha(\tau)$.  There is thus a long tail of the distribution of $\nL$ of the form 
\be
\pr[d\nL] \sim \frac{d\nL}{\nL^{1+\alpha}} 
\ee
with the typical value 
\be
\nL(t) \sim [\nL(t-\tau)]^{1/\alpha(\tau)}
\ee
from the scaling with $\theta$.   This is a consequence of the exponential growth of fluctuations of the lead which corresponds to double exponential growth of  $\nL$. 

Note, however,  that the simple result for the generating function that led to the Levy distribution is not valid for $\theta$ substantially greater than unity as the approach to $\phi_P$ is then controlled by the non-linearities and $\P_\theta$ is no longer a simple function of $\theta$. This means that the distribution of anomalously small $\nL$ is not given correctly by the above analysis.  But this does not matter much:  the probability,is in any case very small  in this regime as can be seen from the Levy distribution.

The quantity $\frac{1}{\Q}\nL(t)$ is the effective  number of individuals in the lead population at time $t$ with the factor of $1/\Q$ the average number of individuals with fitness greater than $\Q$ that is needed for one to establish and eventually be likely to fix.   In the staircase model, the number in the lead population  starts small and grows exponentially before the next higher fitness population establishes and advances the lead.  It has been shown previously \cite{DFG, DWF} that when normalized by its typical value, the
 number in each lead population has a Levy distribution with exponent $\alpha=1-\frac{1}{q}$ with  the scale given by the number in the previous lead population. When iterated through a series of steps this gives essentially the results above.   The advantage of weighting with $\phix$ to define $\nL$, is that the effective number in the nose is defined at all times --- essentially by their collective  potential  to takeover the population. Some  such weighting is of course always needed  for a continuous distribution $\mu(s)$ for which there is no obvious --- and good --- definition of the lead population.

The instability of the front that is responsible for the exponentially broadening Levy distributions will be cutoff, as analyzed in Sec.(5),  by  changes in the speed of the mean to keep the total population size fixed.  As this will have occurred no longer ago than a time $\tsw$ before the lead population of interest is established, and the exponent $\alpha(\tsw)=1/e$,  we expect that at fixed $N$, $\nL$  will have  a multiplicative width of order unity  
with the needed quantities, in particular $\la e^{-\theta\nL}\ra$, that determines the distributions at later times being relatively well-behaved.  But the resulting distributions will not, as we shall see, be Levy distributions.

We now return to the distribution of $N(T)$ at fixed speed. Given $\nL$  a time $\tau>\tsw$ earlier, we have that 
\be
\la e^{-\zeta N(T)} \ra \approx \la \exp\lb[-\nL(T-\tau)(\zeta/\zetat)^{\alpha(\tau-\tsw)}\rb]\ra \ .
\ee
At fixed $\nL(T-\tau)$, this
is a Levy distribution with parameter $\alpha$ which, as for $\nL$ itself, yields a long tail of the distribution of $N(T)$: for $N\gg \Nt \approx 1/\zetat$,
\be
\pr[dN] \sim \frac{dN}{N^{1+\alpha}}  \ .
\ee
Note that this distribution has infinite mean for any $\alpha<1$ (as obtains here).  Thus, as could have been anticipated from the analysis of the dynamics of the average population at fixed $v$, the mean is a very poor characterization of the distribution of $N$ at times more than $\tsw$ after typical steady-state conditions.   

By  time $T\approx 2 \tsw$ later than a time with typical $\nL$, the effects of the fluctuations of the nose will have caused the distribution of $N(T)$ to have a width of order its typical value.   But as many properties depend on logarithmically on $N\Q$, the important result is that the width of the distribution of $\log(N/\Nt)$ is still quite narrow even for several times $\tsw$.  The distribution of $\log(N\zetat)$ for the Levy distribution can be computed from its characteristic function. Since the factor of $\nL$ just sets the scale, its effects can be factored out: for simplicity we set it to unity and obtain
\be
\la e^{-\beta \log(N\zetat)} \ra = \la (N\zetat)^{-\beta}\ra= \frac{\Gamma(1+\beta/\alpha)}{\Gamma(1+\beta)}  \label{GAMMADIST} 
\ee
whence cumulants can be found by expanding the logarithm of this in powers of $\beta$.  
One thus finds that, given $\nL=1$ at time zero, 
\bea
\la \log(N(T)\zetat) \ra  &=& [\exp(T/\tsw) -1]\gamma_E   \\
{\rm var}[\log(N(T)\zetat)] &=& \frac{\pi^2}{6} [\exp( 2T/\tsw)-1]
\eea
with $\gamma_E$ Euler's constant.  

\subsubsection{Typical $N$ and determining the speed, $v(N)$:} 
At long times the instability of the nose causes it to either run away, resulting in very large and rapidly growing populations, or fail to keep up with the advance of the mean (at speed $v$), causing $N$ to shrink rapidly and the population to go extinct.  Which of these fates is more likely depends on the lead of the nose or, essentially equivalently, the value of $N$ a time $\tsw$ later: the population scale $\Nt$ is, up to a multiplicative factor of order unity,   the population size above which the population is  likely to diverge and below which it is likely to go extinct.  Note that for long enough times that $\alpha$ is relatively small, 
$Z(\zeta)= \la e^{-\zeta N} \ra \approx \pr\lb[N<\frac{1}{\zeta}\rb]$ from the
general result  for broad distributions of Sec.\ref{BRDDIST}.   Thus $\log[1/\zetat]$ is roughly the median of $\log N$ (up to differences between $1/e$ and $1/2$, the latter being the probability that $N$ is below its median value).    As we shall see, the fluctuations are sufficiently small on time scales of order $\tsw$, that $\Nt(v)$ is within a multiplicative factor of order unity of the value of $N$ in the fixed population-size model that would evolve at average speed $v$. 

In the modest speed regime, with $v/\st^2$ small, there will be large oscillations of the population distribution due to the roughly periodic advance of the nose.  This will be true even with a distribution of fitness increments: a large --- $\Oh(1)$ --- fluctuation of the population in the nose will occur on average once per time $\tsw$ and the descendants of the particular early-establishing mutation that caused it will contribute a substantial fraction of the population for a time at least of order $\tsw$ with the fitness of these descendants  increasing by close to evenly spaced establishments.  These close-to-periodic oscillations are reflected in the roughly periodic (in $\P-\chi$) structure of the log-slope of both $\phix$ and the slow transients $\phi_\P$, superimposed on top of their average log-slope of $\gamt$ near their shoulder.  These yield oscillations in $\int \phi_\P \n$ which determines the distribution of $N(T)$.  It  is  useful to consider  $v(t)$ having an oscillatory component --- as it will at fixed $N$ --- that suppresses these oscillations.  As they are fast on the time scale of $\tsw$ on which the nose becomes unstable, the mean $v$ will determine the longer time behavior of interest.    Indeed, on the time scale of  one step,  $\st/v\approx \tsw/q$, the Levy exponent $\alpha$ only decreases from unity --- corresponding to a distribution $\delta(N-\Nt)$ --- to $1-\frac{1}{q}$:  thus the distribution of the non-oscillating component of $N$ hardly broadens at all.    In practice,  as noted earlier, the rapid oscillations at fixed speed are not large for reasonable ranges of parameters  as $v/\st^2$ in the multiple mutations regime of interest is unlikely to be  very small. 

As long as the approximation (discussed in \ref{DYNAPP}) that the effects of mutations on the dynamics of $\P$ and determination of $\Q$ are negligible except for the direct input from the saturated region to around $\P-\st$, is valid --- which it will be for moderate speeds ---  the heuristic result that 
$\log( \Nt\Q) \approx \frac{\Q^2}{2v}$
is good.

In the high speed limit, however, the transient $\phi_P$ will, like $\phix$, have an Airy regime just below the shoulder and this will contribute corrections to $\log \Nt$.  But these are small compared to the dominant term above. Thus in inverting $\Nt(v)$ to obtain $v$, they will only result in small corrections (down by $1/q$ times log-log factors).   Although more detailed analysis is possible in this regime, we will not carry it out here.

We  have found that the condition that relates $N$ to $v$ is that $N$ is, up to a numerical factor of order unity, the inverse of $\zetat(v)$ which is determined uniquely by the condition that the backwards-time dynamics  at speed $v$ starting from a constant $\phi$, $\phi(\chi,T)=\zetat$, converges to $\phix_v$ faster than $e^{-\eps_v (T-t)}$, i.e. with zero amplitude of the slowest eigenvector.   This is true very generally --- including with long-tailed $\mu(s)$ and with deleterious mutations --- as long as the fluctuations are not so large that non-linear effects change $v$ substantially: in the case analyzed here, they do not as we shall see in the next section.   Thus to very good accuracy, for the fixed population size model 
\be
v(N) \ \ \ {\rm is \ determined \ by} \ \ \ N(v)=\frac{1}{\zetat(v)} \ .
\ee

\subsection{``Typical" $\n(\chi)$ and weighted fixation probability}  

Thus far we have not explicitly studied the typical  fitness distribution of the population.   This will have structure on the scale of $\st$ reflecting the roughly periodic nature of the advance of the nose: we discuss this at the end of this subsection.  But  for understanding the longer time scale dynamics we are most interested in the envelope of fitness distribution which will be relatively smooth: call this $\nt(\chi)$.

 \subsubsection{Envelope of fitness distribution:} With the guess from the above analysis (and justified further below) that typical fluctuations in $\nt(\chi)$ are only by multiplicative factors of order unity and that these are driven almost entirely by the fluctuations of the nose, we could analyze the deterministic approximation with a cutoff at the nose representing the effects of the absence, typically, of individuals with fitness more than $\Q$ above the mean.  This is close to the original approach of Tsimring et al \cite{tsimring:etal:1996}.  The result  would be  
 \be
 \nt\approx \rhox_\Q
 \ee
  the formal fixed point of the deterministic dynamics truncated at $\Q$.  
 This gives as good an approximation for the  typical $\n$ as is possible given the looseness of the  definition of ``typical".    
 
 The distribution $\rhox_\Q$ can be analyzed similarly to the linear part of $\phix$ either using Laplace transforms, as in the Appendix, or approximately in terms of the  log-slope
\be
\xi(\chi)\equiv \frac{-d\log(\nt)}{d\chi} \ . 
\ee
This is defined with a minus sign to reflect the difference in the sign of the $v\pd{}{\chi}$ term between $\Lod$ and $\Lop$ which causes $\rhox$ to decrease with $\chi$ while $\phix$ increases. 
With the log-slope slowly varying on a scale of $\st$, we have
\be
\chi\approx v\xi-\Mz(\xi) \ .
\ee
This is exactly the same equation as for $\gamma=d\log(\phix)/d\chi$, but we are interested in the other branch of the solution: that with $\xi$ increasing from zero at $\chi=0$ to $\gamt$ at $\chit$ which is close to $\Q$.  Over all but the last part of this range of $\chi$, the mutational term is small: this yields a $\nt$ which is close to a simple gaussian (as assumed by \good and others). 

Near to $\chit$ (where the log-slope approximation breaks down) and beyond that to $\Q$, the behavior will change, similarly to $\phix$.  
In the high speed limit, $\nt$ in this regime will be an exponential $e^{-\gamt \chi}$ times the same  Airy function with $\Q$ just below the negative of its first zero.  For moderate speeds, there will be some additional structure from $\Q-\st$ to $\Q$ before cutting off $\nt$ at $\Q$. 
If 
\be
\nt(\Q) \sim \frac{1}{\Q\st} \ ,
\ee
 representing a just established population of size $1/\Q$ spread out over the scale of the typical fitness increments $\st$,  the normalization will be roughly correct with  
\be
 \int\nt(\chi)d\chi \sim \Nt
\ee
  With fixed population size rather than fixed $v$, this should give the correct relation between $N$ and the mean speed $v$, up to a multiplicative factor inside $\log(N\Q)$ of order unity (or perhaps logarithmically larger).  

\subsubsection{Affecting the future:}  We can now find the relative contributions of different parts of the fitness distribution to the future.  The probability of the individual whose descendants will take over the population having fitness $\chi$ above the mean  is roughly given by 
\be
p_{takeover}(\chi)\approx \frac{\phix(\chi) \nt(\chi)}{\nL}
\ee 
which, because both $\log \nt$ and $\log \phix$ curve downwards below $\Q$, is dominated by a small range near $\Q$. This dominant range has width of order $\st$ for moderate speeds --- in the staircase model simply the lead population ---  and of order the width of the Airy regime, $(v\st)^\frac{1}{3}$, for high speeds.

The reason that  at high speeds an individual that is a of multiple $\st$ behind the nose can takeover the population, is that mutations from lower fitness continue to contribute substantially to the growth of the fittest deterministic subpopulation even while this is producing enough mutants to establish the next subpopulation. But the same is true for the next-to-lead population being fed by the one below it and so on back of order $(v/\st^2)^\frac{1}{3}$ steps.  Note, however, that this does not imply that these populations are stochastic: they are not; it is only that their deterministic dynamics involves deterministic mutational input --- including to the lead population --- as well as deterministic growth.

\subsubsection{Fine structure of fitness distribution:}  In the staircase model, the fitness distribution is a series of equally separated delta-function peaks.
 as shown in \fref{FIG1}.  
One might expect that this would also be the case for short-tailed $\mu(s)$ albeit with each peak now a clump of subpopulations with slightly different fitnesses and the separations between these clumps varying somewhat around $\st$.  But the actual behavior is more complicated as illustrated in \fref{FIG3}.   

\begin{figure}
\includegraphics[width=6.0in]{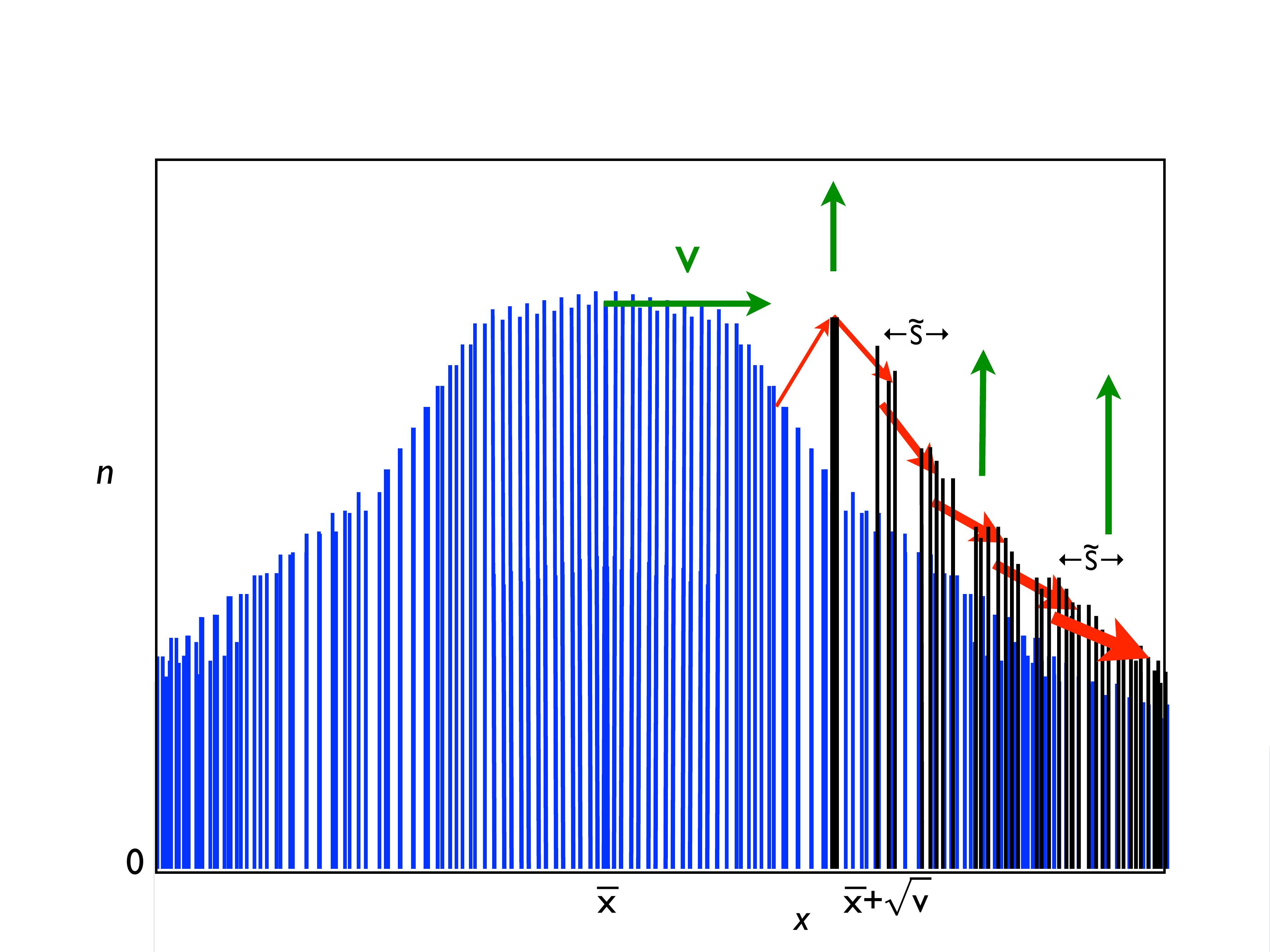}
\caption{Schematic of the fine structure of the fitness distribution for short-tailed mutation distributions, $\mu(s)$, in the huge population, high speed, regime.   The sizes, $n$, of the subpopulations are indicated on a linear scale as a function of their fitness, $x$.  Only the central portion of the distribution is shown.  Typically, the fitness distribution is a dense set of closely spaced subpopulations (blue) with an envelope that is close to gaussian with mean $\bar{x}$ and standard deviation $\sqrt{v}$.   Occasionally --- on average once per time $\tsw$ --- a lucky mutant is established anomalously early and its population (thick black bar) rises well above the others with similar fitness.  The leftmost red arrow indicates such a  lucky mutation (which occurred some time ago at the nose of the fitness distribution).  The descendants of this lucky mutant  are shown by the black bars, with the mutations that gave rise to them and advanced the nose indicated by the other red arrows.  The first set of these mutations has a narrow range of fitness increments around $\st$. This leads to a clump of subpopulations that stands out from the background of typical ones that arose at similar times.   After a few further sets of mutations, the clumps will have broadened, as shown, and the descendent subpopulations from the lucky mutant become a dense set with the roughly periodic structure no longer apparent.  

The bulk of the distribution is advancing at speed $v$, while all the lucky mutant populations are growing (vertical arrows) at rates equal to their fitness above the mean.  Soon after the time shown, the original lucky mutant subpopulation will 
become half the total population and the mean will jump rapidly to its fitness (at a speed a few times $v$). The mean will then stop increasing until the  descendent populations (black clumps) take over.  At that time  the mean will start increasing again at speed $v$ and the envelope of the distribution will become gaussian ---  until the next lucky mutation.   

\label{FIG3} }
\end{figure}

Consider a single lead population with a specific fitness and no  others close enough to contribute much to the future evolution.   The important mutations from this  will  have a range of $s$ around $\st$ that is narrower than $\st$: with width of  order  $\st/\sqrt{\ell}$ (assuming $d^2\Lambda/ds^2\sim \Lambda(\st)/\st^2$).  But in this range there will be a considerable number of independently established populations with somewhat different fitnesses and typically none of them will dominate.  A time $\test$ later, these will have collectively established a clump of new populations whose spread in fitness will be larger by a factor   of
$\sqrt{2}$, as the fitness increments are independent of the previous ones.   Similarly, after $h$ new sets of establishments, the width of the clump will be  $\sqrt{h}$ times as large.  After $h\sim \ell$ steps, one would expect the fitness distribution near the nose to  be spread out by as much as the typical spacing ($\st$) between the clumps and after that no periodic structure would remain.  But this is not correct: on average once in each $q$ steps, one of the new establishments will occur early enough to contribute a large fraction of the total population of the newly established clump.  Although the others will collectively contribute to the future, the lucky subpopulation will generate a roughly periodic set of clumps with the total population in each clump well above that in the  background of the  many small subpopulations  from the less-lucky ones. These clumps will again broaden out over  of order $\ell$ steps until the process repeats.   

For modest-sized populations, $\ell>q$ (this is equivalent to the condition for being in the modest-population regime).  Thus anomalously lucky mutations will occur often enough that the roughly periodic structure will persist until the next lucky one occurs: the fitness distribution will look similar to \fref{FIG1} but with each peak replaced by a clump of peaks with smaller peaks between the clumps.   But for huge populations, $\ell\ll q$ so that the periodic clumps will have broadened out and disappeared into the roughly uniform background before the next lucky mutation.  Thus over most of the width of  the fitness distribution, there will not be pronounced clumps.  But on average twice in the $2\Q$ wide distribution, there will be a sharp peak ---  a  delta-function with a single fitness ---  followed by a decaying and broadening series of clumps.  This behavior is illustrated schematically in \fref{FIG3}.

In the discussion above, we were sloppy about what constitutes a clump: what matters is the size of sub-populations relative to the envelope of the fitness distribution (for the same reason as the exponential weighting of higher values of $\chi$ reflected in the definition of $\nL$).     Near the mean of the distribution this distinction does not matter much, but near the nose it does.  In particular, if one considers mutations from  a single peak near the nose, far more will have small fitness increments than will have  $s\approx \st$.  Although the populations these establish will not grow as fast as those with larger $s$, they will nevertheless constiute for some time a higher population of new mutants near to the initial peak than further ahead.  Thus to observe the decaying clump structure one needs to normalize each subpopulation by the envelope of the typical fitness distribution.

\section{Feedback and Coupled Dynamics of Nose and Mean}

Armed with a general analysis of the stochastic effects of the fluctuations of the nose on the distribution of the population size at fixed speed, we can now analyze the dynamical adjustments of the speed required to keep $N$ fixed and the statistics of the fluctuations that arise from the coupled dynamics of the nose and the speed of the bulk of the population.  These give rise to an effective model on  time scales of order $\tsw$ which can be analyzed in detail.  But the fluctuations also require adjustments of the speed on intermediate time scales which leads to larger than expected jumps in the mean fitness. Although the typical fluctuations of the fitness distribution are modest, very rare large fluctuations dominate the average of the fitness distribution: these are analyzed in later subsections.

\subsection{Dynamics of feedback}

On time scales of order $\tsw$ on which the important feedback dynamics occurs, we can approximate the advance of the lead, $Q(t)$, by a differential equation with a deterministic part and a 
stochastic part that together will cause the unstable stochastic dynamics at fixed $v$. But we must also include  the effects of the mean advancing faster or slower than average in response to earlier fluctuations: these will decrease or increase $Q$, respectively.  It is easiest to work in the rest frame with $F(t)$ the position of the front and $\Up(t)$ the position of the mean (more precisely the value needed to keep $dN/dt$ zero).  We then have that 
\be
\frac{dF}{dt}\approx v + \eps [F-\Up-\Q]+ \eta(t)
\ee
for $F-\Up$ close to $\Q$.    The eigenvalue, $\eps$, of the instability is close to $1/\tsw$, and converges to it in the limit of high $v$. 
The stochastic part of the front speed, $\eta(t)$,  has  mean zero and consists, roughly, of a small negative constant part plus positive spikes with close-to-exponentially distributed magnitudes that give rise to the Levy distribution of the population size at later times. The effective noise, $\eta(t)$, has substantial correlations only over times of order single mutational steps which is smaller by a factor of $q$ than $\tsw$: this separation of time scales gives rise to simplifications which we have already made use of.  The variance  of the noise is related to the variance of $\log N$ that it would causes a time $\tsw$ later if $v$ were fixed: we will use this to compute the statistics of the noise, most simply its variance.   

A sub-population that was  in the lead at fitness $F$ at time $t$ , having just been established by reaching a size of order $1/\Q$,  will, at time $T$, have grown to a size  
\be
n(T;t) \approx \frac{1}{\Q} \exp\lb[F(t)(T-t) - \int_t^{T} dt' \Up(t')\rb] \ .
\ee
 (Note that we have here lumped together all the subpopulations in each peak of the fitness distribution discussed in the precious section.)
 
The total population at time $T$ is roughly the maximum over $t$ of these sub-populations:
\be
N\approx \max_t  n(T;t)  \ :
\ee
this maximum is achieved  at  some time $t_T$.   The derivative of  $\log N(T)$ with respect to $T$  must be zero.   This and the expressions for $n(T,t)$ and $N$  together imply   that 
\be
\Up(T)=\Up(t_T)+(T-t_T)\frac{dF}{dt} (t_T)=F(t_T)
\ee 
the equality of the first and last expression is simply that  the sub-population at any fixed fitness reaches its maximum when the mean catches up to it. 

 The  effects of the front advancing non-uniformly 
will, on average, be felt a time $\tsw$ later.  Thus we consider small fluctuations of the time delay $T-t_T$ around this, defining
\be
\dtau(T)\equiv T-t_T-\tsw \ , \label{TAUT}
\ee
and expand to linear order in $\dtau$.  We can then Fourier transform the dynamics in time, defining $\dtau(\omega)$ and $\eta(\omega)$, and obtain the variable part of  the speed,  $u(t)\equiv d\Up/dt-v$ with transform $u(\omega)$:
\be
u(\omega)=\eta(\omega)\frac{(-i\omega\tsw)^2}{(-i\omega\tsw-\eps\tsw)(e^{-i\omega\tsw}-1)-i\omega\tsw\eps\tsw}  \label{VELOMEGA}
\ee
the $e^{-i\omega\tsw}$ factor arising from the time delay between the fluctuations of the nose and its effect on the speed. 

In the absence of the feedback from the adjustment of the speed a time $\tsw$ after a fluctuation of the nose, response functions to $\eta$  would have had a simple pole $1/(-i\omega -\eps)$ reflecting the exponential instability of the nose that would then have existed.  This is stabilized by the feedback:  the only singularities in the frequency dependent response are a series of poles in the lower-half $\omega$ plane; nevertheless, oscillations, albeit strongly damped ones, will occur in response to a fluctuation of the nose. 
The fluctuations in the fitness of the nose, $F(\omega)$, are 
\be
\delta F(\omega) = \tsw\eta(\omega) \frac{e^{-i\omega\tsw}-1}{(-i\omega\tsw-\eps\tsw)(e^{-i\omega\tsw}-1)-i\omega\tsw\eps\tsw}  \ .
\ee

\subsubsection{Autocorrelations of the lead:} The fluctuations in the lead, $Q=F-\Up$, are 
\be
\delta Q(\omega) = \tsw\eta(\omega) \frac{e^{-i\omega\tsw}-1+i\omega\tsw }{(-i\omega\tsw-\eps\tsw)(e^{-i\omega\tsw}-1)-i\omega\tsw\eps\tsw}  \ . \label{QRESPONSE}
\ee
 With the feedback, 
the variance of the fluctuations in $Q$ are finite and at low frequencies, $\delta Q(\omega) \approx \tsw\eta(\omega)$. Surprisingly, for $\eps=1/\tsw$, the absolute value of the linear kernel relating $\delta Q$ to $\tsw\eta$ is unity at all frequencies of order $\tsw$.  This means that the covariance of $\delta Q$,  $\la \delta Q(t+\tau)\delta Q(t)\ra$, is proportional to the covariance of the stochastic part of the {\it speed} of the nose --- hardly to have been expected! --- and has correlations only on the short time scales of order $\tsw/q$ of individual mutational steps.    One would have expected that a larger than average $Q$ at one time, $t$,  would tend, via the higher speed of the nose that it implies, to cause  a larger than average $Q$ for times up to  at least $\tsw$ later since the feedback that limits further fluctuations will not occur until then.  But the rate of change of $Q(t)$ is  also affected by changes in the speed of the mean and these are anti-correlated, due to the feedback, with $\delta Q$ at earlier times.  The chance that $\delta F(t)$ is positive is positively correlated with $\delta Q$. But changes in $\Up(t)$ can be much sharper than changes in $F(t)$ and thus dominate short-time changes in $Q(t)$. For $\eps=1/\tsw$ these effects, taken together,  cancel on average.    More generally, for $\eps\tsw$ slightly different from unity, the residual correlations on time scales of order $\tsw$ will be present but small.   Note, however, that these results only apply to the covariance: higher order correlations of $\delta Q(t)$ should show structure on time scales of order $\tsw$; as the fluctuations in $\eta(t)$ are very asymmetric,  third order correlations should exhibit such structure. General higher order correlations are discussed below.

\subsubsection{Nose fluctuations and  fitness diffusion coefficient:}  At low frequencies  the fluctuations in $\frac{dF}{dt}$ and $\frac{d\Up}{dt}$ must be the same. These  are proportional to $\eta(\omega)$, the  fluctuations of the speed of nose, but larger by a  factor of $\frac{1}{1-\frac{1}{2}\eps\tsw}\approx 2$
since $\eps\approx \frac{1}{\tsw}$.  This factor   is due to the amplification of the short-time fluctuations over times of order $\tsw$. The nose and the mean fitness will thus advance at average speed $v$ plus a stochastic part that is diffusive at long times with diffusion coefficient, 
\be
D\approx \half \lb(\frac{1}{1-\frac{1}{2}\eps\tsw}\rb)^2 C_\eta \ \ \ {\rm with} \ \ \ C_\eta\equiv \int dt \la \eta(t)\eta(0)\ra \ .
\ee

The covariance of $\eta(t)$ can be obtained by using the results we have derived for constant speed, i.e. in the absence of feedback.  A fluctuation in $F$ results, at constant speed, in a fluctuation of $\log N$ a time $\tsw$ later of magnitude
\be
\delta\log N(t)\approx \frac{\Q}{v}\, \delta F(t-\tsw)  
\ee
with the coefficient $\frac{\Q}{v}=\tsw$.  After a further time, $\tau$, has elapsed, the distribution of $N/\Nt$ is given by the Levy distribution with parameter $\alpha(\tau)\approx e^{-\tau/\tsw}$.  As derived above, this implies a variance in $\log N$ equal to  $
\frac{\pi^2}{6}(\alpha^{-2}-1)$ which is $\sim\tau/\tsw$ for $\tau\ll \tsw$.  But we can also derive these fluctuations by integrating the effects of the fluctuations of $\eta$ exponentially amplified by the instability over a time $\tau$.    This yields
\be
{\rm var}[\log N(t+\tau)-\log N(t)]\approx \tsw^3 \half \lb[e^{2\tau/\tsw}-1\rb]  C_\eta
\ee
so that $C_\eta=\pi^2/3\tsw^3$: this is equivalent to the result derived in DFG in terms of the fluctuations in the establishment times.   Note that the correlations of $\eta$ are only over times of order $\st/v$ which is much shorter than $\tsw$ thus the integral over these correlations is all that matters (up to subtleties discussed below) for variances on time scales of order $\tsw$. 

By equating the results from the nose fluctuations to the phenomenological ones in terms of $\eta$,  the diffusion coefficient of the mean of the population at fixed $N$ is found to be
\be
D\approx \frac{2\pi^2}{3\tsw^3} 
\ee
a result that should become exact at asymptotically high speeds.  Note that because of the amplification caused by instability of the nose, this is larger by a factor of four than would have been expected from the short-time fluctuations of the speed of the nose.

This $D$ implies that over a time of order $\tsw$, the variations of the cumulative advance of the nose   are  typically only of order $1/\tsw$ which is much smaller (by $1/\log(\st/\Ub)$) than the typical fitness increment of $\st$, and even smaller, by $1/\log(N\st)$, relative to the cumulative fitness increase over $\tsw$ which is  $\Q$.
As the fitness of the lead population jumps by around $\st$ with the establishment of each new lead population, the fitness of the nose must be defined so as to smooth out these jumps.  At constant speed, we have seen that the appropriate weighting of the fitness distribution is by the fixation probability: $\nL \equiv  \int \phix \n \approx e^{-\delta Q \tsw}$  is the quantity that determines the speed a time $\tsw$ later.  This suggests defining $F$ at time $t$  in terms of $\n(x,t)$ in the rest frame to be the $F$ that makes 
\be
\int dx \phi_{F(t)}(x) \n(x,t) =1 \label{FDEF}
\ee
with the weighting function $\phi_F$ simply equal to $\phix$ shifted so that the position of the shoulder is $F$ rather than $Q$.  With this definition, $F$ will, indeed, have fluctuations  around its steady increase that are jumps forward by  of order $1/\tsw$ at an average rate of once per time $\tsw$, and backwards jumps times $\tsw$ later caused by the compensatory changes in $\Up(t)$ --- as we shall show.  The  fluctuations of $F$ are thus of the same order as the ambiguity in the position of the nose due to the width of the shoulder region: $\frac{v}{\Q}=\frac{1}{\tsw}$ (and the concomitant inaccuracies of the sharp shoulder approximation).  Changing the definition of $F$ by of order this ambiguity is equivalent to choosing in Eq.(\ref{FDEF}) another number rather than unity.  

While the position of the nose can be accurately defined, the nose population, $\nL$, is no longer a natural quantity because it is unclear what speed, or even what position of the 
mean, to use in defining the appropriate $\phix$ for the weighting of $\n$. But having used  $\nL$ to infer the statistics of fluctuations of the fitness, $F$, of the nose at constant speed and the implied adjustments of $\Up$ needed, we can now use $F$ instead. Note, however, that the fact that the fluctuations in $Q(t)$ do not accumulate over time scales substantially larger than $\tsw$, implies that,  if the speed were held constant for times of order $\tsw$, the fluctuations in the nose population would, as previously assumed, only be by factors of order unity (albeit with a long tail from the Levy distribution). Thus our approximations are self-consistent. 

  Although the fluctuations of $F$ are quite small, one has to, nevertheless, be careful as to which other quantities are considered: as we discuss below, $\Up(t)$ and $Q(t)$ typically have much larger fluctuations than $F(t)$ on time scales intermediate between $\test$ and $\tsw$: these are indicated by the presence of delta function components of the associated response functions, such as in Eq.(\ref{QRESPONSE}),  a time $\tsw$ after a jump in $F$.

\subsubsection{General analysis of dynamics on feedback time scales:}


Thus far we have only used the variance of the fluctuations of the speed of the nose.  But we can use the branching process analysis to do far better. We continue to focus  on time scales of order $\tsw$ in the asymptotic high speed limit. The crucial simplifications  are: that the statistics of the jumps in the position of the front are independent (except  on time scales much shorter than $\tsw$), that the fluctuations in $Q$ are small enough that the distribution of the sizes of the jumps is essentially independent of $Q$, and that the dynamics and effects of feedback changes in the speed are linear as analyzed above.    Moreover, the Levy process dynamics of the nose gives us the statistics of the jumps.    

Consider a general quantity (such as the advance in the front over some time interval) that is a linear function of the past dynamics of the front with some kernel $K_X(\tau)$:
$X(t)\approx\int_0^\infty d\tau\, K_X(\tau) \eta(t-\tau)$.   Since the $\eta$'s are independent, the generating function of $X$ can be immediately found from that of the $\eta$'s.  In the continuum limit (valid for large $q=\tsw/\test$), $\eta(t)$ is the derivative of a jump process: it  consists of a  small constant negative  part (to make its mean zero) and a series of positive delta-functions that occur as a Poisson process at rate $1/\tsw$ with independently distributed magnitudes.  The typical magnitude of the jumps is $\frac{v}{\Q}\approx \frac{1}{\tsw}$ so it is convenient to define their magnitudes in these units as $\frac{w}{\tsw}$ with the  generating function of the distribution  of the $w$'s  given by the formal limit of $\alpha\to 1$  of  the  result obtained from the  Levy process, Eq.(\ref{GAMMADIST}):
\be
\la e^{-zw}\ra = \lim_{\eps\to 0} \lb[ \frac{\Gamma(1+z(1+\eps))}{\eps \Gamma(1+z)}-1\rb] = z\psi(z) \label{WDIST}
\ee
with $\psi(z)\equiv \frac{d\Gamma}{dz}$ the digamma function. [There are subtleties associated with the asymptotically delta-function part for $w$ within of order $\eps$ of zero as indicated by the bad behavior in the large $z$ limit, but these will disappear with the integral over time, below.]    From Eq.(\ref{WDIST}) we obtain the generating function for the distribution of the $X$:
\be
\la e^{-zX}\ra = \exp\lb\{\frac{1}{\tsw}\int_0^\infty d\tau K(\tau)[z\psi(1+zK(\tau))+z\gamma_E]\rb\}
\ee
with the $\frac{1}{\tsw}$ the rate of jumps and $\gamma_E=-\psi(1)$ making $\la \eta \ra=0$ as it was defined to be.  For any bounded $K_X$, the distribution of $X$ falls off exponentially reflecting the exponential tail of the distribution of the $w$'s --- proportional to $e^{-w}$ --- which corresponds to the power law tail of the Levy distribution of the nose-population.    Note that for the dynamics of the nose population, a change in $\Up$ corresponds to {\it dividing} $\nL$ by a factor that is exponential of the
integrated effects of the change: the analysis in terms of the position of the nose (which is related to the logarithm of $\nL$) is exactly equivalent to combining this feedback effect with the Levy branching process.

There is one important caveat  to the analysis in terms of linear functions of $\eta$: these are  only meaningful if the kernel $K_X$ does not contain any delta-function parts: when it does, as for $\Up$ and $Q$, the linear relation between $X$ and $\eta$ must break down and  additional features appear as we  analyze in the next subsection.  
But for well behaved quantities, the full distribution can, at least in principle, be computed from the generating function.   A natural quantity to consider is the amount the nose advances in time $\tsw$, $Y(t)\equiv F(t)-F(t-\tsw)$.  The associated kernel  is well behaved at both high and low frequencies: $K_Y$ has no delta-functions and its integral is finite.    The finiteness of the integral of $K_Y$ --- i.e. its zero frequency limit  --- follows from the observation that once the effects of the shift in  the sweep time, Eq.(\ref{TAUT}), are incorporated, $Y$ is driven by the front fluctuations as 
\be
\frac{dY}{dt}=\frac{Y(t)-Y(t-\tsw)}{\tsw}+\eta(t)-2\eta(t-\tsw)+\eta(t-2\tsw) 
\ee  
with the discrete time differences acting like a second derivative of $\eta$ at low frequencies.

\subsection{Jumps in mean fitness and clustering of fixations}\label{JUMPS}

A jump in the position of the front would, a time $\tsw$ later,  cause a jump in the total population if $\Up$ did not adjust to compensate.  But since $\Up(t)$ only controls $\frac{dN}{dt}$ rather than $N$ directly, to compensate for a jump at the nose $\Up-vt$ must have a sharp spike.  In the linearized analysis, this is seen in the high frequency behavior of $\Up(\omega)$ which is proportional to $\eta(\omega)$, in contrast to the front position which advances as the integral of $\eta(t)$.  
[As $Q=F-\Up$, $Q(t)$  exhibits the same sharp spikes but with the opposite sign.]   Of course, these spikes, which are approximately delta-functions on the slow $\tsw$ time-scale, must be rounded out on shorter time scales: the natural expectation is on the scale of the steps of advance of the front, $\test$, but this is not correct. 

A jump in $F(t)$ can occur via an atypically early establishment, a larger than typical $s$, or a mutation from slightly closer to the nose than typical.  As long as  $\mu(s)$ falls-off faster than exponentially for large $s$, the effects of any of these is similar and all contribute to the distribution of jumps discussed above.   An increase of $Q$ from $\Q$ to $\Q+\delta Q$,  causes, a time $t$ later,  the resulting  subpopulation to be $n(t)\sim \exp[(\Q+\delta Q)t-\half vt^2]$ which reaches the total population size, $N\sim \exp(\frac{\Q^2}{2v})$ a time $\delta t \approx \frac{\sqrt{2\Q\delta Q}}{v}$ earlier than it would have starting from $\Q$.  To keep $N$ constant, this necessitates a jump in the mean fitness of $\delta \Up
=v\delta t \approx \sqrt{ 2\Q\delta Q}$ --- the square root appearing because of the quadratic maximum of $n(t)$. With $\delta Q = w\frac{v}{\Q}$, the amplitude $w$ is of order unity and 
\be
\delta \Up \approx w\sqrt{2v} 
\ee
with $w$ distributed as in Eq.(\ref{WDIST}) below. This distribution falls-off as $e^{-w}$ for large $w$ corresponding to a tail in the probability density of $\delta \Up$ of the form 
\be
\pr[d\delta\Up]\sim \delta\Up e^{-(\delta\Up)^2/2v}d\delta\Up.
\ee  
Thus the typical scale of a jump in the mean fitness is of order $\sqrt{v}$: far bigger than the typical jumps of $F(t)$ which are of size $\frac{v}{\Q}$.   

After the mean fitness jumps, it will not advance further until the effects of the continuing advance of the nose catch up: this takes a time $\delta\Up/v$ to occur.  
The  change in $\Up$ integrated over this time is simply $w$ which is the factor by which the jump of the nose causes the nose-population to increase and the weight of the delta-function in the dynamics of $\Up(t)$ on time scales of order $\tsw$.   One might worry that such strong non-linearities in the response to jumps of the front would invalidate the linearized results beyond this rounding of the delta-functions. But as jumps typically occur only once every time $\tsw$, the spikes will rarely overlap and do act almost independently as approximate delta-functions without compounding non-linear effects.

The jumps in the mean fitness are not really sharp because of the time it takes for a population fitter by of order $\sqrt{v}$  to take over from a less fit one: this time is of order $1/\sqrt{v}$.  Thus the  jumps in the mean fitness will be smeared out on the same time scale as the duration of their effects except for relatively rare ones with $w$ substantially larger than unity which will have more marked effects.  The speed of the mean of the population, $\frac{d\bar{x}}{dt}$, will change during the jumps by an amount of order the average speed: for jumps with $w$ substantially larger than unity,  $\frac{d\bar{x}}{dt}$ will increase to of order $\sqrt{w}\, v$ and then decrease to close to zero. Thus even in the huge-population regime in which there are not substantial variations in the speed associated with the successive establishments, the speed will still vary by factors of order unity. 

Note that although the jumps are quite smeared out, they  dominate fluctuations in the cumulative increase in $\Up$ for times considerably longer than $\tsw$  --- up to times of order $(\log N\st)\tsw$ --- implying that the result in the previous subsection for the diffusion coefficient of $F(t)$ and hence $\Up(t)$,  only applies for the latter  on very long time scales. 

In the modest-size  population regime, the jumps in the mean fitness from anomalously 
early mutations advancing the lead are in addition to the roughly regular jumps of sizes close to $\st$ from the typical steps of the nose.  The  anomalous jumps correspond to the mean taking several steps at a time, advancing quickly by those several $\st$. 

One might guess that the jumps in the mean fitness would be associated with fixations as  each jump arose from an anomalously lucky  mutation in the nose that by itself contributed a substantial fraction of the then-new lead population.  But during the time in which this lead population grew to {\it collectively} become the largest sub-population, the fraction of it that descended from the  lucky mutation did not change much. Thus for this mutation to takeover the whole population, its descendants must accumulate further beneficial mutations sufficiently rapidly.  For the original lucky mutation  to fix in the population --- or for a  less lucky mutation whose {\it descendants} are sufficiently lucky to do so ---  typically takes a few times $\tsw$.  This process was analyzed in reference \cite{DWF} where it was  shown that the time to fixation --- equal to the coalescence time of the whole population --- is $\tsw[\log\log N + \Oh(1)]$ (a seventh relevant time scale). But there is an interesting feature that arises from the same source as the jumps in the mean fitness: the mutations do not fix one after another but tend to fix in clusters of several at a time. Park et al \cite{park:etal:2010} found this in simulations and observed that the distribution of the number, $k$,  that fix together is close to an exponential distribution.   Analysis along the lines of this paper and ref. \cite{DWF} show that the tail of the distribution of $k$ is indeed generally exponential and that in the limit of large $q$,  the distribution is exactly exponential with mean simply $q$.  Thus $\Q$ is not only the additional fitness that the fittest member of a mutant lineage has accumulated by the time the lineage has risen to a significant fraction of the population,  it is also the average increase in the fitness associated with a fixation event.  Again, we see several difference fitness increments --- $\st$, $\sqrt{v}$, and $\Q$ --- here associated with different properties of  lucky mutants and their descendants.

\subsection{Average fitness distribution}\label{AVFITSEC}

The fluctuations in the nose, although limited by feedback from adjustments of the speed to keep the population size fixed, can, nevertheless, occasionally be so large as to  make the mean of the fitness distribution, $\la \n(\chi)\ra$, much larger than its typical form.  
The average fitness distribution is controlled by extremely rare large deviations of the position of the nose which are caused by very anomalously early establishments of a new lead population --- far earlier than the typical jumps analyzed above which occur on average once per time $\tsw$.  

To study the effect of anomalously large jumps of the nose, we can approximate the speed of the mean as roughly constant over the time between the jump and when it affects the mean.  As  analyzed above (and in DFG), the statistics of the jumps drive the distribution of the population size, $\nL$,  of the new nose, with a power law tail proportional to $d\nL/\nL^{1+\alpha}$ and the exponent after one step advance of the nose of $\alpha=1-\frac{1}{q}$.  (This corresponds to the exponential tail of the distribution of magnitudes of the jumps in $F(t)$ as discussed above).  As the earliest possible establishment time is just after the previous lead population has established, it is earlier than typical by $\la \test\ra\approx \ell/\Q$  and will result in a population at the new nose larger than typical by a factor of $\exp[\ell(1+\frac{1}{q})]$ (with $\ell\equiv \log(\st/\Ub)$).  This thus gives a cutoff for the distribution of $\nL$. As 
\be
\int^{\nu_{max}} \nu \frac{1}{\nu^{1+\alpha}} d\nu \sim \nu_{max}^{1-\alpha} \ ,
\ee
such rare events yield a contribution to $\la \nL\ra$ of order $\exp(\frac{2\ell}{q})$.  In the limit of large $L\equiv\log(N\st)$, $\ell/q$ is small so this is not a substantial increase in $\la \nL\ra$.  
But two effects have been left out of this simple estimate. First, the exponential instability of the nose causes subsequent establishments of new lead populations to be more likely to also occur earlier than typical. And second and more important,  a series of anomalously early establishments, while even rarer than a single one, can give much larger contributions to $\la \n(\chi)\ra$
near to the typical nose as well as between the nose and the peak of the fitness distribution, and, indeed,   further ahead than the typical nose.

Consider an anomalously     early establishment of $h$ steps in succession. If each of these is as early as possible --- occurring within a time around $1/\Q$ of the previous one --- the probability of this happening  is $(e^{-\ell})^h$ since each there is a factor of 
$\Ub$  for each mutation that occurs before any amplification of the lead population.  
As we are interested in the universal large $q$ limit it is convenient to define such a jump in the lead in units of the typical lead, $\Q$: $\Delta \equiv \frac{\delta Q}{\Q}= \frac{Q}{\Q} -1=w\frac{v}{Q}$ so that a jump in $Q$ of $h$ steps of size $\st$ corresponds to $\Delta=\frac{h}{q}$.  Since $q=\frac{2L}{\ell}$, this occurs with probability $e^{-2L\Delta}\sim \frac{1}{N^{2\Delta}}$:  simply a continuation of the exponential distribution of $w$ out to anomalously large values.   As the initial growth rate of the lead population is simply $Q$,  the anomalous lead population grows with rate $2L\Delta/\tsw$.   It is natural to rescale time and fitnesses, defining $T\equiv t/\tsw$ and $X\equiv \chi/\Q$.  Then the mutation rate, which appears only via $\ell$, drops out.  

From these scalings of the properties of rare jumps of the nose, we can guess the form of the averages of the fitness distribution that they result in.  As the logarithm of the probability of a fluctuation by $\Delta$ and the logarithm of $\n(\chi)$ are both proportional to $L$, we can anticipate that the averages of $\n(\chi)$ resulting from such fluctuations will scale as powers of $e^L \sim N$ (analogous to the power of $e^\ell$ that resulted from a single jump in $Q$ of size $\st$).   We conjecture a scaling form:
\be
\log[v\la\n(\chi)\ra]\approx L\,  \Xi\lb(\frac{\chi}{\Q}\rb) 
\ee
with $\Xi(X)$ a scaling function that is universal in the large $L$ limit (and the $v$ inserted to dedimensionalize $\n$). Therefore we expect that (slightly sloppily due to neglect of corrections in the exponent)
\be
\la \n(\chi) \ra \sim N^{\Xi\lb(\frac{\chi}{\Q}\rb)} 
\ee

Any class of rare process that yields a result of the above scaling  form yields a lower bound on $\Xi(X)$.  The simplest, of course, is the typical behavior which would yield $\Xi=1-X^2$ corresponding to the simple gaussian shape of the fitness distribution. But we can do much better.  

A jump of  of the nose by $\Delta_J$ is followed by further growth of the lead as $\Delta\approx \Delta_J e^T$ due to the instability of the nose. But if $\Delta_J$ is not small, this will result some (scaled) time $T_{\Delta_J}$ later in a large jump in $\Upsilon$ which will cause $\Delta$ to become negative for some period of time.   We conjecture that the dominant contributions to  the averages of $\n(X)$ occur at time $T_{\Delta_J}$ after such a jump.  For each $X$, we can then find the  $\Delta_J$ that gives the largest contribution to $\Xi$ at that $X$. This yields a lower bound for the function
 $\Xi(X)$. For small $X$ (for which small $\Delta_J$ dominates) 
\be 
\Xi \ge 1-X^2 + \frac{4}{27}X^3 \ ,
\ee
while at the average lead, the bound is
\be
\Xi(1)\ge \cong 0.29 
\ee
which  corresponds to a lower bound on the average nose population: 
\be
\la \nL\ra \ge N^{0.29} \ .
\ee
Beyond the typical nose, we find
\be
\Xi(X)>0 \ \ \ {\rm for} \ \ \   X<X_0 \ \ \ {\rm with} \ \ \  X_0\ge 1.15
\ee
corresponding to a large average population up to at least $15\%$ above the average lead of the nose.  Note that these results assume the fitness increments are all the same: if they are not, then other rare events involving larger than typical fitness increments could dominate.  But with a short-tailed $\mu(s)$,  simple considerations suggest that multiple small jumps will dominate over a single large one: this may well make the scaling function, $\Xi(X)$, universal.  

The above results mean that the --- at least implicit in many papers --- efforts to compute the mean shape of the fitness distribution from close-to-deterministic approximations are doomed to failure.  Of course, as the events that give rise to the large and skewed fitness distribution are very rare, what one should have been hoping for from the beginning was the typical rather than average behavior of the distribution. Nevertheless, the rare fluctuations resulting in anomalously large averages should, in principle, start showing up as a lack of convergence of the  inferred average as more and more separate evolutions (or separate times) are sampled.   Note, however, that near the peak of the distribution, the effects will be small as the  correction term to the gaussian is down by a factor of  order $1/\sqrt{L}$ at one standard deviation, $\sqrt{v}$, away from the peak. 

\section {Conditioning on Behavior at Infinite Time}

The above analysis is rather cumbersome with several levels of Ansatz's and perturbative or asymptotic approximations --- albeit ones that either have been or can be justified and done more systematically.   It would be nice to have a cleaner approach in which properties of interest could be computed directly, at worst involving numerical analysis of deterministic dynamical equations such as the backwards in time evolution of 
$\phi$.   

As fixing the population size exactly is both difficult and not necessarily realistic, and most of the properties depend only logarithmically on $N$,  one might guess that constraining  it  so that it does not fluctuate too wildly would give similar behavior to fixing it exactly. Hallatschek, \cite{OH}  has shown that by cleverly modifying the model to directly control the nose in a particular way, together with fixing the speed, one finds a well-behaved steady state for which one can obtain closed equations for the first moment of the population distribution.  

Here we discuss an alternate approach that has some advantages, in particular, it does result in everything being computable via deterministic backwards in time dynamics.  Its behavior  turns out to be quite similar to Hallatschek's model.   

The basic idea is quite general: for stochastic population dynamics for which at long times the total population size, $N$, will 
almost always go to either zero or infinity, one can analyze the behavior conditioned on it doing neither: i.e. on $N(T\to\infty)\ne 0 \ \rm or \ \infty$.  The hope is that this will keep the histories of populations  that stay finite but non-zero at long times close enough 
to the ``ridge" that separates explosion from extinction, that they fluctuate in some  well-behaved way around a typical population size characterizing the ridge.  
In the context of asexual evolution, Simon and Derrida \cite{DerridaCondl} have analyzed a model in which 
the speed is fixed and the condition that there be a single individual surviving at infinite times is imposed, although the selection is quite different  as discussed in Sec.(8).

We  consider the branching process at fixed speed (as analyzed above) but condition explicitly on the behavior at infinity.  
To obtain means, distributions or correlations of ``operators" at times in a range near $t$, say a quantity $R(t)$,  one computes 
\be
\langle R(t)\rangle_{(t_0,T,\beta_1,\beta_2)} = \frac{\langle R(t)e^{-\beta_1 N(T)} - R(t) e^{-\beta_2 N(T)}\rangle }{\langle e^{-\beta_1 N(T)} -  e^{-\beta_2 N(T)}\rangle}
\ee
where $t_0$ is the initial time with $t_0<t<T$. Then 
 $\langle R(t) \rangle_c = \lim_{t_0 \to -\infty,\ T\to\infty}\langle R(t)\rangle_{(t_0,T,\beta_1,\beta_2)}$ is a well behaved limit independent of $t_0$ and $T$.  [ A  simple example in which this can be seen  is a  single clone with a mean growth rate that is positive.] The limit ends up also independent of the initial state and of the parameters $\beta_i$, as long as both are positive.  Thus we can make the $\beta_i$ almost equal and replace the differences between them by a derivative with respect to $\beta$.  This corresponds to weighting the behavior at time $T$ by a factor proportional to $N(T)e^{-\beta N(T)}$  which suppresses both large and small $N(T)$.    
 
 
The key quantities are the fixed point of the branching process equation, $\Lod_v \phi -\phi^2=0$ and the lowest eigenfunction, $\psi(\chi)$, of the linearized operator about this fixed point which  has eigenvalue $-\epsilon$. 

 To compute any quantities of interest for times $t$ in some range far from the final and initial times, we must find the generating function running backwards from very long times $T$ starting with $\phi=\beta$, then adding in, at the appropriate times, the needed $\zeta's$ or $\theta's $  for the quantities of interest, and then running back to much earlier times.  Finally, we have to project onto the initial conditions, differentiate with respect to $\beta$ and then normalize (by doing the same thing with all the $\zeta's$ and $\theta's$ equal to zero). 

At a time $t$ long before the final time, all the transients will have become very small except the slowest one, so that $\phi\approx \phix + \bg(\beta) \psi e^{-\epsilon (T-t)}$  with the amplitude $\bg$ depending on $\beta$.  We now add in the desired ``fields" from the generating function: $-\dot{\phi}=\theta+\Lod_v \phi -\phi^2$ (if have $\zeta$ at time $t$, this is simply added to $\phi$ at that time).   If $\theta$ is small, such as to calculate, e.g., the mean of some quantity, then $\phi$ stays close to $\phix$.  But to get the conditioning on the behavior at $T$, we need to keep the dependence on $\beta$, thus the non-linear part is still crucial: the  term of order $\bg$ times $\theta$ is what is needed.   The non-linear dependence of $\bg$ on $\beta$ will in the end drop out with the normalization factor, thus all we need is the behavior linear in   $\bg$: derivatives with respect to $\bg$ can be used instead of with respect to $\beta$.   We now run the dynamics back to the initial  time, $t_0$, which, for $t-t_0$ large, will involve only the eigenfunction $\psi$ so that
\be
\phi\approx \phix + \ag \psi e^{-\epsilon (t-t_0)}  \ .
\ee  
 The coefficient $\ag$ is dominated by the $\{\theta\}$, but we are interested in its derivative with respect to $\bg$: this yields the conditioned generating function 
 \be
 \zc(\{\theta\})=\pd{\ag(\{\theta\},\bg)}{\bg}|_{B=0}
 \ee
Note that the normalization factor is now taken into account automatically, since in the absence of $\theta$, $\ag=\bg$.  The  probability that $N$ neither diverges nor goes to zero in time $T-t_0$ is, up to $\Oh(1)$ factors associated with behavior near the initial and final times, simply  $e^{-\epsilon (T-t_0)}$. Thus one loses a factor of order one for each sweep-time. This is because the fluctuations in the unconditioned, unconstrained $N(t)$ are by factors of order unity in time $\tsw$, as we have shown. 

The basic mathematical  structure of the effects of conditioning at infinity is similar to that of Simon and Derrida \cite{DerridaCondl}. 

\subsection{Exactly ``solvable" suppressed-fluctuation model}

Before analyzing the behavior of the population dynamics conditioned on infinite times, we observe that there is an exactly equivalent model that is local in time -- again similar to reference \cite{DerridaCondl}.   The local model can be obtained by, in the sense of dynamics in terms of time translation operators, commuting the operator at the late time $T$ used to condition, back through the operators of interest at intermediate times,  to the initial time.  The resulting stochastic model, which we call the suppressed-fluctuation model, is
\be
\pd{\n}{t} = \Lop_v \n + \sqrt{2\n}\, \eta_{BD} - 2\phix\n + \frac{2\psi\n}{\int dy \psi(y)\n(y)} \label{EQUIVMOD}
\ee
 with the first two  terms the same linear operator at constant speed and  stochastic noise as in the original model, and the two new terms preventing divergence and vanishing of $N$, respectively.   Both $\phix$ and $\psi$  are very small except near the nose.  Thus the effect of conditioning is equivalent to a particular way of preventing the 
nose from getting too far ahead or behind the mean speed by modifying the growth rate of populations in the nose, with one part depending inversely on a weighted  total population in the nose and the other strongly suppressing mutant populations that are substantially in front of the average lead. 

It is far from obvious that this non-linear model is exactly soluble --- but the generating function for the distributions of any quantities of interest can be computed from the backwards time dynamics as above. 

\subsection{Distribution of population size}

To compute the distribution of the population size we make use of the simplification derived earlier that, at fixed speed in the absence of conditioning, the population size is proportional to the population in the nose, $\nL$,  a time $\tsw$ earlier, as long as the fluctuations are not too large.   

The Levy distribution with time dependent exponent $\alpha(T)$ can be understood simply in the staircase model.  As discussed above and  in references \cite{DFG, DWF},  the feeding of the yet-to-be established population with fitness $\st$ larger than the lead population which is growing exponentially, results in an essentially Levy distribution of the new higher  fitness 
population once it is established with parameter $\alpha = \alpha_1 \approx 1- \frac{1}{q}=1-\frac{\st}{\Q}$.
If we consider each member of this population as giving rise, independently, to a population in the next higher fitness class with this distribution, then the dynamics becomes just like that of a population of individuals that have a Levy distribution of offspring.  This can be iterated straightforwardly with the generating function  at each successive step  found by replacing  the variable $z$ in $\la e^{-z\nL} \ra $ by $z^{\alpha_1}$.  In the limit of $1-\alpha_1 \ll 1$, this iteration results in $\alpha(T)=e^{-T/\tsw}$. The general analysis carried out above for a distribution of fitness effects yielded the same result.  
This implies that to a good approximation we can replace the full backwards in time dynamics of the generating function on time scales of order $\tsw$ and larger, by just keeping track of a single variable, $z(T-\tau)$ with simple dynamics: 
\be
\frac{dz}{d\tau}=z/\tsw
\ee
and adding in to $z$ a  $\zeta$ at time $t$ to get the distribution of $\nL(t)$ and thence that of $N(t+\tsw)$.    

We start with $z=\beta$ at time $T$, iterate  backwards to time $t$, add $\zeta$ to $z(t)$ and iterate back to time $t_0$.  We then have
\be
z(0)= [\zeta + \beta^{\alpha(T-t)}]^{\alpha(t)}   \label{ZETAAPPROX}
\ee

and the distribution of $\nL(t)$ can be computed by taking the derivative of $\exp[\nL(t_0)z(t_0)]$ with respect to $\beta$, normalizing by the same with $\zeta=0$, and taking the limit $T-t\to \infty$ and  $t_0\to -\infty$.  The initial conditions then drop out --- as they should in steady state --- and we get simply, after rescaling by $\Nt=1/\zetat$ to get the distribution of  $N$ rather than $\nL$, 
\be
\la e^{-\zeta N}\ra  \approx \frac{\zetat}{\zeta+\zetat}  \ee
so that 
\be
\pr[dN] \approx \frac{dN}{\Nt} e^{-N/\Nt}
\ee
and $\la N \ra \approx \Nt$.   

A more detailed  analysis that takes into account corrections to the linearized dynamics of $\P(\tau)$, etc., yields probability density proportional to $N^\delta e^{-N/\Nt}$ with very small $\delta\approx 1/[2 \log(\Nt\Q)]$.   The primary effect of this correction (which is only a factor of $1/e$  for $N\sim\sqrt{\Nt}$) is to make $\la \frac{1}{N}\ra$,  which for neutral drift is often considered as the inverse of an ``effective population size", finite: indeed only larger than $1/\Nt$ by a logarithmic factor.   [Note that this correction for small population sizes is associated with the failure of the approximation  of Eq.(\ref{PHIP}): the very small $N$ behavior is determined by the $\zeta\gg \zetat$ limit of the generating function which gives rise to backwards time dynamics that involves $\P(\tau)$ substantially smaller than $\Q$ so that the cutoff of $\phi_\P$ at $\P$ changes $\int \n \phi_\P$ substantially from the approximate form used earlier.]   

These results for the distribution of $N$ are strictly valid only in the huge population regime in which $v/\st^2$ is not small: if it is small, then there will be almost periodic oscillations in $N(t)$, with frequency $v\st$ and   ratio of maximum to minimum of order $\exp(\st^2/8v)$, superimposed on the longer time scale fluctuations.  

\subsubsection{Suppression of nose fluctuations:}

The effects of the $-2\phix\n$ term in the dynamics suppress the rare large fluctuations of the advance of the nose that give rise to the Levy distribution of the nose population, $\nL$.   In the fixed $N$ model, these fluctuations cause a  sub-population with a particular mutation that happens to establish anomalously early to have a substantial chance of taking over much of the population a time $\tsw$ later. As analyzed in reference \cite{DWF}, these relatively rare events control the coalescent properties.   In the conditioned model, on the other hand, these rare events are exactly what are suppressed: thus we expect the dynamics of fixation and coalescent properties to be somewhat different in this model.  This can be seen by computing the distribution of $\nL(t)$ given $\nL(0)$ which has generating function
\be
\la e^{-\zeta \nL(t)} | \nL(0) \ra \approx \frac{\exp\lb(\nL(0) \lb[(1+\zeta)^{\alpha(t)}-1\rb]\rb)}{(1+\zeta)^{1-\alpha(t)}}  \label{CONDNUL}
\ee
derivable from Eq.(\ref{ZETAAPPROX}).   This yields a  finite conditional mean of $\nL(t)$ that decays exponentially at rate $\eps$  from $\nL(0)$ to the mean $\nL$ which is unity.    
As will be shown elsewhere, this changes the coalescent properties from the universal ones analyzed in reference \cite{DWF}.

\subsection{Moments and correlations of $\n(\chi,t)$}

For the pure branching process model  at constant $v$, integer moments of the distribution are, as shown for the mean in Sec.(\ref{AVFITSEC}), very badly behaved, being dominated by extremely rare anomalously large events.  But in the conditioned model, moments  and correlations --- such as $\la \nL(t)\nL(0)\ra $ found from  Eq.(\ref{CONDNUL})  --- are well-behaved and quite characteristic of the typical behavior.   Beyond  the simple approximation, moments and correlations can be derived order by order in a perturbative expansion around the fixed point, $\phix$.  The eigenfunctions of the linear operator $\Lod_{v,c} \equiv\Lod_v -2\phix$ for deviations from $\phix$, and those of its adjoint $\Lop_{v,c}=\Lop_v-2\phix$, control the behavior.   The corresponding discrete sets of eigenfunctions we denote $\psi_j$ and $\nu_j$, respectively, with the orthonormality condition 
\be
\int d\chi \psi_j \nu_k = \delta_{jk} \ ,
\ee 
corresponding completeness relation 
\be
\sum_j \nu_j(\chi)\phi_j(\chi')=\delta (\chi-\chi') \ ,
\ee  
and eigenvalues $-\eps_j$.  The smallest eigenvalue, $\eps_0$, and corresponding $\psi_0$,  are what we have previously denoted simply $\eps$ and $\psi$.  Because the non-linearity in the backwards-time dynamics for $\phi$ is simply $\phi^2$, the non-linear matrix elements that enter are 
\be
H_{ij,l}\equiv \int \psi_i\psi_j \nu_k  \ .
 \ee

The mean of the  population density as a function of fitness, is
\be
\la \n(\chi) \ra = 2\sum_j \nu_j(\chi) \frac{H_{0j,0}}{\eps_j}
\ee
which is dominated by the lowest eigenfunction of $\Lop_{v,c}$.  Indeed, one can show that $\nu_0$ is a weighted average of $\n$:
\be
\la \frac{\n(\chi)}{\int \psi_0 \n} \ra = \nu_0(\chi)  \ ,
\ee
the weighting being just that that appears in the dynamical model Eq.(\ref{EQUIVMOD}), which  is equivalent to  conditioning at infinite time.     
One can also compute fluctuations of $\n$ and show that when $v/\st^2$ is not small, the relative fluctuations of $\n$, coarse-grained over a scale of order $\st$, are only modest.

\subsection{Relationship to Hallatschek model}

Hallatschek \cite{OH} has introduced a model that, while not obviously so, turns out to be quite closely related to our suppressed-fluctuation model.    He considers constraining other linear functionals of the fitness distribution rather than the total population size and shows that a particular choice leads to closed equations for the average steady state $\la \n(\chi) \ra$  at fixed speed.  The constraint is (in our notation) that 
\be
\int d\chi \phix(\chi)\n(\chi) =2
\ee
i.e. that the quantity related to the effective number of individuals in the lead population, $\nL$, is fixed to be equal to two.  The dynamic equations that $\la \n(\chi) \ra$  satisfies are qualitatively similar to our  suppressed-fluctuation model in that the nose is effectively prevented from either  running too far ahead or falling too far behind.  Although the details are different, we conjecture that --- in a sense that needs defining more carefully --- the two constrained models are in the same universality class for large population sizes. But it is a different class than fixed-population-size models: most  fluctuation-related properties are different.  In particular, the coalescent properties are quite different  than with fixed population size.  \cite{DWF} We leave an analysis of the similarities and differences between Hallatschek's model \cite{OH} and our suppressed-fluctuation model for future work.

\section{Field Theoretic Approaches}

Once one is using dynamic generating functions heavily, it is natural to ask whether one can enforce the ``mean-field" constraint on the total population size in a functional integral formulation and then use well-established field-theoretic methods to analyze the behavior, including performing perturbative expansions around mean-field in some of the small parameters.   Unfortunately, although it is straightforward to formulate the problem, analyzing it is far from easy and at this point it is not clear how much is gained.  

One can enforce the fixed population size  constraint by making $\Up$ depend explicitly on all the $\{\eta\}$ and $\{\n\}$ with the constraint enforced by an integral over an auxiliary field, $\K(t)$, that gives a delta function in $N$ at each time. One can then compute the dynamic generating function in terms of a conjugate field, $\phi(\chi,t)$: we use the same notation for future convenience although it is a somewhat different quantity than in the analyses thus far.   After shifting $\phi(x,t)$ by $\K(t)$,  integrating over the noise, and using dots for time derivatives, we have a generating function of the form 
\be
{\cal Z}=Tr_{(\n,\phi,\Up,\K)}(e^\A) 
\ee
 with ``action"  
\be
 \A=\int dt \int dx\  \phi[ \dot{\n} - \Lop_0 \n  +\Up(t)\n] + \phi^2 \n -\K \dot{\n} 
 \ee 
which is like that we could have obtained naively by putting in the fixed $N$ constraint and integrating over $\Up$: here we have  been careful in order to show that no Jacobian factors occur. 

We can now integrate out $\n$ to get a backwards  time equation for $\phi$ of the usual branching process form but with an extra piece from  the constraint that involves
\be
\k(t)\equiv \dot{\K} 
\ee
together with the adjoint selection and mutation operator:
\be
 -\dot{\phi}= [\Lod_0 -\Up(t)]\phi -\phi^2-\k(t) \ . \label{FTMODEL}
 \ee
Then we can go into the moving frame in terms of
$\chi\equiv x-\Up(t)$ with
\be
\Lod_{\dot{\Up}}\equiv\Lod_0 - \dot{\Up}\frac{\partial}{\partial \chi} \ .
\ee
We have thereby reduced the full set of fluctuations of the $\{\eta_{BD}(\chi,t)\}$ to fluctuations in just two quantities: $\Up(t)$ and $\k(t)$.   In principle, one could analyze this stochastic equation --- perhaps perturbatively around $\dot{\Up}=v=constant$ and small $\k$ --- and try to compute quantities of interest.  This does not appear to be straightforward for similar reasons to the difficulties described below.

Alternately, we could attempt to do a mean-field analysis by looking for a saddle point of the functional integrals by extremizing $\A$. 
Setting the variation with respect to   $\n$ to zero  is equivalent to integrating $\n$ out since $\A$ is linear in $\n$: this gives us the equation for $\dot{\phi}$. 
Expecting the saddle point solution to be uniform in time, i.e. constant speed, $\dot{\Up}=v$ and $\k$ constant, this has a well-behaved fixed point, $\phix_{\k,v}(\chi)$,  for any $\k<\kt$ with $\kt$ a velocity-dependent upper limit above which there is no well-behaved fixed point.  Note, however, that $\phix_{\k,v}$ does not have the interpretation of a fixation probability: indeed, for $\k>0$, it is negative for $\chi$ less than some value just below $\Q$. 

Formally, one can also extremize  over $\K$ to yield $\int d\chi \dot{\n}=0$, as expected,  and over $\Up$ to give $\int d\chi \phix_\k(\chi)\n(\chi)=0$.  Extremizing  over $\phi$ yields 
\be
\dot{\n}=[\Lop_v -2\phi(\chi)]\n 
\ee
which, with $\phi=\phix_\k$, is similar to 
the operator that appeared with conditioning on infinite times.   But we know that for $\k=0$ the eigenvalues of this linear operator will all be negative and thus there is no fixed point except $\n=0$ corresponding to extinction.  This will also be true for all but special values of $\k$ at which  the operator, or equivalently its adjoint $\Lod_v -2\phix_\k$ acting on $\phi$, has a zero eigenvalue.  There is one value of $\k$, $\kt>0$, at which this occurs: not surprisingly, this is the same value of $\k$  beyond which the solution $\phix_{\k,v}$ breaks down.  Thus the condition that a non-trivial saddle point exists  yeilds $\k=\kt(v)$. The condition for a zero eigenvalue can be shown to force the apparently extra condition $\int \phix \n =0$ and also yields $\int \chi \n =0$ as one would expect in  a naive ``mean-field" approximation where $\Up$ would  be exactly the time-dependent average fitness. 

The problem, however, is that we are missing a condition.  As the saddle point conditions are invariant under an overall rescaling of $\n$, nothing fixes the total $N$.  If we plug back into the action,  there is a boundary term $\K(T)\int\n(T)-\K(0)\int\n(0)$ in the action up to time $T$. With $\K=\k t$ this is $N\kt(v) T$.  We could now try to minimize over $v$ but, not surprisingly, this would push $v$ to infinity (for which $\kt\to 0$), the result for an infinite population.  Better is to keep $\k$ slightly smaller than $\kt(v)$ and then consider the linearized fluctuations around the (almost) saddle point.  Integrating over these --- which is rather complicated and we have not carried out in detail --- gives a term in the action which is not too singular for $\k$ near $\kt$, has only logarithmic dependence on $N$, and is larger for larger $v$ and $\k$ closer to $\kt$ because the lowest eigenvalue decreases in these limits.    Then minimizing the resulting approximation to the action --- the sum of the boundary term and the contribution from fluctuations (both proportional to $T$) --- over $v$ will yield $\kt$ of order $1/N\tsw$ with $\tsw$ only logarithmically dependent on $N$.   

For $\k_v <\kt$, the eigenvalue of $\Lop_v -2\phix_\k$  will be slightly negative and it would appear that the population is doomed to extinction. However the fluctuations are very asymmetric --- recall the statistics of the fluctuations in speed of the front, $\eta(t)$,
 that give rise to the Levy distribution --- and the non-;linear effects of these (which are included in the linearized dynamics of $\phi$ around $\phix$) will prevent  extinction.     

In principle, the quadratic fluctuations around the (almost) saddle point should yield the fluctuations in speed --- i.e. diffusion constant --- and how these are driven by the somewhat earlier fluctuations in the front: the value of $\chi$ at which the non-linear term $\phi^2$ becomes important and beyond which $\phix\approx \chi$.
It is not clear, however, that this procedure will yield a systematic asymptotic expansion rather than a (slightly) uncontrolled approximation. An important feature of the dynamics may well be problematic: even in the limit of arbitrarily large population sizes, the fluctuations in the population size at fixed $v$ are of order the typical $N$.  Although this only results in small fluctuations in $v$ on time scales of order $\tsw$, larger fluctuations occur on short time scales --- $\Oh(1/\sqrt{v}$) as shown in Sec.(\ref{JUMPS}) --- and can thus not be treated by a low frequency expansion that one might hope would be valid. Whether the non-linear aspects of these fluctuations will cause fundamental problems with perturbative field theoretic approaches is not clear.  

As discussed in section Sec.(8), perturbative field theoretic methods  may be much more useful for other models for which the time scales of the fluctuations of the nose and of the consequent  adjustments of $\Up$ are very different.

\subsection{Soft constraints on the population size}

An alternate approach is to allow the population to fluctuate but limit these fluctuations by an explicit coupling of the dynamics of $\Up(t)$ to the population size $N(t)$, its time derivative --- related to the difference between the instantaneous mean fitness and $\Up$ ---  or perhaps some convolution over past times of $N(t-\tau)$.   If the dynamical equations for $d\Up/dt$ are linear in $N$ (or more generally in $\n(\chi,t)$)   $\n$ can be integrated out and one is left with using the behavior of the $\Up(t)$ dependent backwards equation for $\phi$ to determine the dynamics of $\Up$.    This is analogous to conventional dynamical mean field theories, except that the fluctuations are far from gaussian.   Unfortunately, it  appears to be hard to find a dynamical model (even one for $d^2\Up/dt^2$ which is non-linear in $d\Up/dt$) that is stable at all frequencies without knowing details about the nature of the fluctuations.   The difficulties are associated with the fluctuations on  time scales of order  $1/{\sqrt v}$ caused by fluctuations in timing of advances of the nose,  as well as those on longer time-scales of order $\tsw$ (including the somewhat  oscillatory nature of the response of $\Up$ to the nose fluctuations as reflected in the complex poles in Eq.(\ref{VELOMEGA})), plus  the need for $\Up$ to  move rapidly backwards  when the population is shrinking to stop it going extinct.   

To prevent the total population size from fluctuating by factors more than of order unity, 
adjustments of $\Up(t)$ are needed on time scales of order $1/\sqrt{v}$ as discussed in Sec. (\ref{JUMPS}).  If one is willing to tolerate ignoring exponentially rare fluctuations that cause the population to go extinct suddenly, then a model of the form 
\be
\frac{d\Up}{dt}= v_0 + \frac{\alpha\sqrt{v}}{N_0} \frac{dN}{dt} + \beta v \lb[\frac{N}{N_0}-1\rb] \ ,
\ee
with $N_0$ of order a typical population size and $v_0$ the corresponding average speed, is stable for ranges of $\alpha$ and $\beta$ of order unity.  Writing this in a functional integral form as in the previous subsection and integrating out $\n(\chi,t)$ results in dynamics of $\phi$ of the same form as Eq.(\ref{FTMODEL})  with $k(t)=\alpha\frac{d\theta}{dt}-\beta\theta(t)$ in terms of the field, $\theta(t)$, that multiplies $\frac{d\Up}{dt}$ in the action.  This model can probably be analyzed in substantial detail, although how to systematically derive the slow-time-scale model of the coupling between the nose and mean --- not to mention corrections to this ---  using the backwards-time dynamics of $\phi$ is far from clear. 

\section{Summary and Extensions}

Understanding the dynamics of  the simplest asexual evolution in large populations has not been straightforward due to several factors, primarily  large fluctuations in some of the crucial quantities and a multiplicity of time scales.  Here we have developed various methods for analyzing the dynamics including the statistics of fluctuations.  The fundamental simplification is the separation between the stochastic dynamics of the nose of the fitness distribution --- treated by branching process methods --- and the non-linear dynamics of the bulk of the distribution that keeps the population size fixed.  One of the issues we have focussed on is typical versus average behavior and ways of computing the former from generating functions.   The analysis of the branching process aspects we have carried out with substantial generality and corrected some errors in previous analyses, \cite{GOOD} which do not, however, change most of the basic results. 

For the quantities that have been the focus of much of the previous work, the average speed of evolution as a function of population size (as well as  the fitness distribution), the primary new result is the condition for determining $N$ in terms of $v$ from the backwards-time dynamics: $N$ is the reciprocal of the value of $\zeta$, $\zetat$,   that from $\phi(\chi)=\zeta=const.$ converges most rapidly to the fixed point $\phix$ under the backwards time dynamics.  This condition should replace various results in the literature --- heuristic, somewhat ad-hoc, and/or cases in which $\n(\chi)$ is approximately gaussian.  It is applicable far more generally that the regimes analyzed in detail thus far, including when deleterious mutations play a substantial role.    For the model with only beneficial mutations analyzed here, we have  illustrated a quick method for obtaining the lead, $\Q(v)$, of the nose and hence other properties. And we have shown that for many purposes, the fluctuations of the fitness of the nose are small enough (or rare enough) to justify using typical quantities.

Beyond methodology, our focus has been primarily on the dynamics on time scales of order the nose sweep time, $\tsw$, the time for the lead population that contains the important new mutations to collectively dominate the population and thus the time scale on which the stochastic and non-linear dynamics are coupled. This is also  the time scale on which coalescence of genealogies occurs. \cite{DWF}  A simple stochastic model that couples the dynamics of the nose and the mean of the population is used to obtain the properties on this time scale, and its parameters derived from the underlying model.  

Note added: very recently, Neher and Shraiman, \cite{RATCHETSHRNEH}, have developed and analyzed an effective slow-time-scale model that couples the nose and mean of a population in which  deleterious mutations cause the mean fitness to stop increasing or actually decrease: Muller's ratchet. In this situation, the dynamics and the  methods for analyzing them are somewhat different but the basic effects of a time delay between the nose fluctuations and their consequent effects on the speed are qualitatively similar.

\subsection{Universality for huge  populations with short-tailed mutation distributions}

 In huge populations with short-tailed mutational distributions ---  $\mu(s)$ that falls off faster than exponentially for large fitness increments --- there is much universality.   The condition for the universal behavior is that $\log(N\st)\gg \log^2(\st/\Ub)$, with $\st$ and $\Ub$ the selective advantage and mutation rate, respectively, for the beneficial mutations that drive the evolution: these are determined mostly by $\mu(s)$.  
 
 Many of the universal features occur on the sweep time scale $\tsw$.   Distributions of the fluctuations of  certain quantities, in particular advances of the nose, are asymptotically universal and can be computed.  A crucial simplification is that for many properties  {\it only}  the sweep time scale  needs to be considered as the shorter time scales can be  `integrated out"  with asymptotic analysis of the branching process that controls the dynamics of the nose on time scales less than $\tsw$.  
Persistent variations in the mean of the fitness distribution, $\Up(t)$, over time scales of $\tsw$ are very small being only of order $1/\tsw$ --- the smallest of the important fitness scales. These give rise to diffusion of the fitness  (in addition to  its steady rate of increase at speed $v$)  with a diffusion coefficient which is a universal multiple of  $1/\tsw^3$.   

In the universal huge population regime, the average speed is large enough, $v\gg\st^2$, and concomitantly  the standard deviation of the fitness distribution, $\sigma\approx \sqrt{v}$, is large  enough that many subpopulations  contribute substantially to the total population. The speed  is then smooth on the time scale, $1/\st$, of the steps by which the nose advances.  But there are  transient fluctuations in $\Up(t)$ on intermediate time scales of order $1/\sqrt{v}$, the time in which the mean fitness increases by $\sigma$.  In the asymptotic  huge population limit, this time scale is much longer than  $1/\st$ but much shorter than $\tsw$. Occasional variations in the {\it timing} of the new beneficial mutations that advance the nose cause, a time $\tsw$ later,  transient increases in the mean fitness by of order $\sqrt{v}$  which  last for times of order $1/\sqrt{v}$, corresponding to  fluctuations in the speed by of order the average speed. The distribution of the amplitudes of these transient increases
has a gaussian tail.   These  (somewhat-smoothed) jumps in the mean fitness have important consequences for experiments. They should be readily apparent when two or more otherwise-identical populations with different colored fluorescent markers are mixed together and evolved, as in various experiments. \cite{hegreness:etal:2006,SHERLOCK}   Observations of increases in fitness of one of the sub-populations relative to the other have a natural interpretation in terms of new beneficial mutations with magnitude inferred from the rate of increase of the fraction of the total population with that color.  But  one must be very cautious in interpreting such changes: jumps in the difference in fitness between the populations can be --- indeed, for large populations are much more likely to be --- the amplified consequences of much smaller differences in the times of occurrences of new mutations that advance the nose of the fitness distribution of one of the colors.  

The fate of most driving mutations is that eventually they go extinct because they are out-competed by the few that are lucky enough for their lineages to accumulate multiple subsequent mutations atypically early. \cite{DWF}  For those few that do eventually fix, they have typically accumulated several mutations early enough to result in jumps in the mean fitness.  Although they fix only several sweep times later, the consequences of the fluctuations that cause these jumps is that mutations will fix in clusters:  on average $q=\Q/\st$ driving mutations  at a time, with an exponential distribution of the size of the clusters. \cite{park:etal:2010} 
 
 The statistics  of the fitness distribution  are also universal for huge populations with short-tailed mutation distributions.  But the fluctuations in the advance of the nose cause a power law tail out to large sub-population sizes  which is only partially cutoff by the non-linear feedback that keeps the total population size constant. Indeed,  the average fitness distribution is dominated by very rare fluctuations which make the average far larger than the typical envelope of the fitness distribution: we conjecture  that the average is also universal.
 
 Much of the universality comes simply from the nature of the approach to the backwards-time fixed point, $\phix$, associated with the branching process: one slow direction governs the eventual approach with the controlling eigenvalue much smaller than all the others.  More generally,  the backwards time dynamics converges rapidly to a one parameter family of functions, $\phi_\P(\chi)$,  that obtain along this slow approach. And the essential features of this family are (asymptotically) independent of the mutation distribution and details of the short time dynamics of the nose. 
  
On shorter time scales than either  $\tsw$ or $1/\sqrt{v}$,  there is substantial universality associated with details of  the stochastic dynamics of the nose: new mutations that advance the nose and contribute substantially to the future --- the driving mutations --- have fitness increments that vary little.  As studied earlier, \cite{DFG,GOOD} 
with short-tailed mutation distributions there is a narrow range of fitness increments around a ``predominant" value $\st$,  that drive the dynamics. Although there are small variations in the timing and fitness increment of these driving mutations, only a small fraction of them --- on average one per time $\tsw$ ---  are anomalous enough to result in the jumps in the mean fitness discussed above.   

In the huge population size limit, there is an additional fitness scale,  $(v\st^2)^\frac{1}{3}\gg \st$,  that emerges: this is the range of fitness of  almost all the individuals that are close enough to the nose to have a chance of taking over the population.  Mutations from subpopulations in this range are important for the dynamics of the nose.

Asymptotically, there are thus many important fitness scales (as shown in \fref{FIG1}): from smallest to largest, these are $1/\tsw$, the typical fluctuations in the fitness increase of the nose over a sweep time; $\st$, the typical selective advantage of the mutations that drive the dynamics; $(v\st^2)^\frac{1}{3}$, the range of fitness behind the nose from which an individual whose descendants takeover the population are likely to come; $\sqrt{v}$, the standard deviation of the fitness distribution and the characteristic magnitude of transient jumps in the mean fitness that compensate for the fluctuations in the nose;   and $\Q$, the lead of the nose over the peak of the fitness distribution and the average total fitness increment associated with clusters of mutations that fix at the same time.   Of course, as the ratios between these scales involve powers of logarithms, they will, in practice, never be far apart. Nevertheless, processes that occur on each of these time scales are important for understanding various aspects of the dynamics --- even for this simplest model of evolution in large populations.   The reduction of the complexities of the dynamics to the simple effective model coupling the nose and the mean, only strictly works for $\tsw$ much longer than the other time scales. But many of the qualitative --- and semiquantitative --- properties will be correctly captured by the effective model and the jumps in the mean fitness that it produces on  intermediate time scales.

 We have not discussed coalescent properties here nor statistics of individual mutations (beyond the clustering of their fixations) or of genomes: these have been analyzed recently in this and related contexts,  \cite{DWF,DerridaExactLong}  and our branching process results can be used to show their universality.  
But an important observation concerning the statistical properties of alleles is in order.  As the results for the models considered here rely heavily on the Levy distribution of the fluctuations of the lead population and how these fluctuations grow with time, one might think that the cutoff of these fluctuations by the constant population size constraints would alter the coalescent and related properties.  
But if one is concerned with the {\it relative frequencies} of different mutant lineages, the rescalings of the effective lead population, $\nL$, that result from adjustments of the mean, cancel out. Thus we expect that  the statistics of the frequencies should still, up to corrections that are small for large $q$, be given correctly by the branching process analysis and the resulting Levy distributions.    Thus, for example,  the variations in the relative sizes of differently colored sub-populations (as discussed above) should reflect the statistics of the iterated Levy distributions.   
 
 \subsection{Beyond the universal limit}
 
 \subsubsection{Modest-sized populations with short-tailed mutation distributions:}
 
Microbial laboratory populations can easily have $N\ll \Ub\gg 1$,  but are unlikely to have $\log(N\st)$ larger than $\log^2(\st/\Ub)$. [Note that as the magnitude, $\st$, of the mutations that  drive the evolution and  the mutation rate, $\Ub$, for these must be determined self-consistently,  $\Ub$ can be much smaller than the total beneficial mutation rate making it even harder to get to the huge population regime.]
In this ``modest-sized" population regime,  the average speed, $v$, is  smaller than $\st^2$. 
This implies that the population will almost always be dominated by a peak in the fitness distribution that is narrower than $\st$.  
And the mean of the fitness distribution will  usually advance by steps of size roughly $\st$ following the close-to-step-wise advance of the nose. 
Nevertheless,   anomalously early (or large) steps in the advance of the nose  will sometimes give rise to jumps of the mean fitness of a few times  $\st$, similar to the $\Oh(\sqrt{v})$ sized jumps in the huge population regime. But these jumps do  {\it not} reflect similar sized steps of the nose: the mutations in the nose that cause these jumps  are still quite close to $\st$.    

When $v/\st^2$ is small enough that the mean fitness is not smooth even without the anomalously large jumps, the detailed analysis we have carried is not strictly correct. But, as shown by DFG, the qualitative behavior is essentially the same and much of the quantitative aspects  are only changed slightly, including the condition that determines the predominant $s$, $\st$. The one notable exception is the range of fitness of subpopulations that have a reasonable chance of taking over the population in the future: with $v<\st^2$,  some member of  the  population within $\st$ of the nose will almost always win (and the scale $[v\st]^{1/3}$ --- which is here less than $\st$ ---  no longer plays a role).   Other than this, the behavior is --- up to corrections that are likely to have smaller consequences than the approximations that gave rise to the model --- very well captured by the analysis of the universal huge-population regime.

 \subsubsection{Long-tailed mutation distributions:}
 As discussed by several authors, \cite{DFG,sniegowski:gerrish:2010,OH,GOOD}  mutation distributions, $\mu(s)$, that decay more slowly than exponentially behave differently because individuals far from the nose can give rise to large effect mutations that contribute to advancing the nose: indeed, these, and the clonal interference competition between them, \cite{gerrish:lenski:1998,sniegowski:gerrish:2010} can dominate the dynamics.  Although the methods used here can again be used to analyze the long-tail cases, there is substantially less universality: this is reflected in $q$ (the number of mutations by which  the nose leads to the mean) never becoming  large even in asymptotically large populations.  What happens instead is that the typical size, $\st$,  of the mutations that drive the evolution grows with the population size as a power of   $\log N$. \cite{DFG,fogle:etal:2008,park:etal:2010}  This causes the speed to grow faster than linearly in $\log N$ for large populations.  Concomitantly, large oscillations in the speed of the mean fitness on time scales of order $\test$ persist for arbitrary large populations since $\frac{v}{\st^2} \propto \frac{1}{\log(N\st)}$.  Nevertheless, the width of the distribution of selective strengths that drive the evolution can still be narrow compared to $\st$ --- although this needs analyzing in more detail to understand whether or not each new large-fitness-increment establishment also involves smaller effect mutations. 

If $\mu(s)$ falls off more slowly than $\exp(-C s^{\beta_2})$, with some borderline value of $\beta_2$ less than unity, the behavior for asymptotically large populations is different. The multiple mutation regime does not obtain: instead --- at least in a loose sense --- $q$ sticks at two.     DFG concluded that $\beta_2=\frac{2}{3}$, while more detailed considerations suggest that 
$\beta_2$ is somewhat larger.  Nevertheless, for more-modest population sizes, $q$ depends on the mutation rate as well and can still be considerably larger than two with long-tailed $\mu$ as seen in simulations. \cite{park:etal:2010}  We thus conjecture that  $q$ will  be non-monotonic in $N$ for sufficiently long-tailed $\mu$. 
In the intermediate regime when $q$ is considerably  larger than two,  aspects of the dynamics for modest population sizes are expected to be similar to modest-sized populations with short-tailed $\mu$.   It may be, however, that effects of  fluctuations in the speed will be considerably larger in the long-tailed case because of the lack of a narrow predominant range of $s$  and this might make the fixation probability approximation much worse than for short-tailed $\mu$ with similar $q$.   
Simulations at extremely large $N$ and further analysis of the fluctuations would be needed to fully understand the long-tailed case. 

 An oft-studied  case is a simple exponential distribution of beneficial mutations. \cite{orr:2002,orr:2003,joyce:etal:2008,GOOD} This  is marginal in that $q$ increases without bound as the population size increases --- albeit very slowly. \cite{DFG} Its marginality can also be seen from how the clonal interference approximation breaks down. \cite{park:etal:2010} For an  exponential mutational distribution, the branching-process fixed point equation becomes  simply a second order differential equation.  \cite{GOOD} The required absence of one of its two independent solutions in the linear regime sets the condition for $\Q$.  Again, this can be analyzed efficiently via Eq.(\ref{SOLVCOND}) as  in \ref{SHOULDAPP}.

\subsubsection{Applicability of model:} 
 In general, as discussed in the Introduction,
the condition under which the model studied here is applicable is that the statistical properties   of the fitness landscape do not change much as the population evolves. This will certainly hold if the distribution of  available mutations, $\mu(s)$, is the same for any genome in the region of the fitness landscape that can be accessed.  But the discreteness of  the set of beneficial mutations available means that the approximation that $\mu(s)$ is continuous may not be good.   If there are a large number of available mutations in the relatively narrow predominant range around $\st$ that dominates the dynamics of the nose --- the driving mutations ---  then the continuous approximation is valid. Furthermore,  the exact values of the available $s$'s in this range can vary as mutations are acquired.    In this case, it is also reasonable to assume that the predominant-range mutations are not be depleted rapidly ---  as long as they occur in many genes whose effects  are not too closely linked.   

 In fact, the needed conditions are somewhat less restrictive: the behavior of the model will not change much even if the number of mutations in the predominant range does vary  substantially. This is because the newly established population that advances the nose by $\st$, consists of a mixture of many sub-populations with different new mutations.   If some of these  new mutants have fewer available further mutations, their lineages  are less likely to survive. But as long as the new mutants on average have  similar number of predominant-range further mutations available, and there are enough mutant lineages in the nose that the fluctuations around this average are not too large, then the behavior will be essentially the same.  The exact condition on the statistics of the  ``evolvability" of the new mutants ---  the full collection of possible multiple-mutation uphill directions from each --- is not clear: it is an interesting avenue for future research.  
 
But their is an intrinsic problem with long-tailed distributions:  for large populations, the predominant range is very far out in the tail of $\mu(s)$ and continues to move further  out as $N$ increases so that  the mutation rate of the predominant mutations, $\Ub$, continues to decrease --- in the clonal-interference approximation, $\Ub \sim \frac{1}{N}$.  \cite{sniegowski:gerrish:2010} This means that three of the conditions for validity of the assumptions of the model are likely to break down:  first, that there are a large number of mutations sufficiently far out in the tail, second, that such large effect mutations will 
not be depleted, and third, that even if enough new mutations are made available by the previous mutations to stop this depletion,  their spectrum will be sufficiently independent of which mutations occurred first.  Thus, while large effect mutations with a collectively low target size may in many circumstances drive the initial stages of the evolution, it seems unlikely that the further  evolution will be well described by an even-close-to-steady long-tailed mutation distribution.   A possible exception is discussed in the next subsection.

 
 \subsubsection{Very high beneficial mutation rate:} 
All our analysis  has been carried out in the small mutation rate limit, $\Ub\ll\st$: indeed, 
$1/\log(\st\Ub)$ has been used as a small parameter.  Rouzine and collaborators \cite{rouzine:etal:2003,rouzine:etal:2008}  have studied the opposite limit, $\Ub\gg\st$, possibly relevant for some viruses. From their results, one can anticipate that  at high speeds the lead is sufficiently large that the time scale for growth of the lead population becomes fast compared to other time scales. Correspondingly, one can show (most simply by the methods used in the present paper) that  the shoulder in $\phix$ becomes sharp in this limit.   The condition that determines $\Q$ in terms of $v$, Eq.(\ref{RHOQCOND}), can then be used to analyze this case.  Although there are several regimes, in the limit of very large populations the dynamics of the nose becomes similar to that in the small mutation rate limit that we and many others have studied.   The universality thus extends to this regime although some of the time scales will  change.  [The intermediate population size regimes can  be analyzed similarly although the behavior is somewhat different than that of the  modest-population-size limit with small mutation rate.]

 \subsubsection{Suppression of nose fluctuations:}   
  
  Hallatschek has introduced a model in which the nose of the distribution is constrained instead of the overall population size and he was able to compute various properties.  \cite{OH} Here we have introduced a related model for which the computation of the generating function for any linear functional of the fitness distribution  is reduced   to analyzing deterministic dynamical equations.  In both these models, the mean speed is held fixed and the population size adjusts accordingly.   Although this is very similar to what has, in effect, usually been done to analyze the steady-state of the fixed-population-size model, the nature of the fluctuations in the  fixed-speed constrained-nose models  is quite different. These constrained-nose models are in a different universality class.   In particular, the fluctuations are far smaller, with only exponential rather than power-law tails in the size of sub-populations. This difference will also modify the coalescent properties analyzed in reference \cite{DWF}.

\subsection{Extensions}

Most work has focussed on the statistical steady state of evolving populations or how the speed changes on long time scales due to the depletion of the supply of beneficial mutations. \cite{rouzine:etal:2003,RATCHETNEW,RATCHETSHRNEH} But the initial dynamics starting from a clonal population is also of interest, especially for understanding laboratory evolution experiments (e.g. \cite{hegreness:etal:2006,SHERLOCK}). The backwards-time methods can readily be generalized to studying the initial transients --- indeed until the mean fitness of the population starts to increase substantially, the branching process approximation is very good. But the time-dependent speed of the mean, $v(t)$, will have to be worked out self-consistently: this should again be doable via deterministic, rather than  stochastic, computations although integrating the backwards time dynamics of $\phi$ may require numerics for  realistic (non-asymptotic)  parameters.   Such questions as to whether individual mutations or multiple mutations will first takeover a population --- especially with a long-tailed $\mu(s)$ --- and variations  between different initially identical populations  (or sub-populations) are of particular interest and readily analyzable.  

The models discussed thus far all assume linear increase in growth rate as a function of fitness, $\chi$,  above the mean.  But in many artificial and natural contexts, this is a poor assumption, especially when the fitness differences become large.   An alternative is to select, for example, the best half of the population to divide and to kill the rest.  \cite{DerridaSelLett,DerridaExactLong,DerridaCondl}  While more general forms of the growth rate as a function of $\chi$ can be studied, the dynamics of very large populations is mostly determined by the large $\chi$ limit as this is where the nose will be. For constant growth advantage for large $\chi$ --- such as the ``best-half" model --- there is an absolute maximum speed (unless $\mu(s)$ has an exponential or slower tail \cite{DerridaExactLong}).  Corrections to this speed limit have been analyzed in several contexts, (especially by Derrida and collaborators,  reference  \cite{DerridaExactLong} and references therein) and some analyses of fluctuations carried out.  To obtain many of the basic results, the  methods of the present paper can be used directly: indeed the analysis is somewhat easier due to the nature of the linear operators involved.  
There is a substantial simplification that occurs in the constant-growth-advantage model: the time scales on which the nose is unstable is much longer than the sweep time, $\tsw$. A similar separation of time scales  occurs for any sub-linear  (for large $\chi$) increase in the growth rate.   This should enable methods that take advantage of the separation of time scales, such as those of Sec.(5), to be developed. 

 Deleterious mutations, which we have thus far neglected, can be important in populations with high mutations rates (such as mutator strains of microbes or in viruses) and in very well adapted populations for which there are few beneficial mutations available. The combined effects of deleterious and beneficial mutations can be analyzed by combinations of branching process and effective stochastic models  as we have done here. For asymptotic analysis via matching of different regimes as done in Sec.(3), deleterious mutations do not add major complications --- at least as long as the speed is still positive  which has been proved rigorously for asymptotically large $N$ with any distribution of deleterious and beneficial mutations.  \cite{ETHERIIDGERIG}   However the quick method for obtaining $\Q$ from the sharp shoulder approximation --- and thereby the approximation to the fitness distribution, $\rhox_\Q$ --- no longer works because the deleterious mutations from $\chi>\Q$ to $\chi<\Q$ ruin the analytic properties of the linear part of $\phix$ (although Wiener-Hopf methods could potentially be used to handle this).   For modest $N$, deleterious mutations can cause  the speed to change sign --- Muller's ratchet --- in which case
additional features occur. \cite{ORR2000,rouzine:etal:2003,RATCHETNEW,RATCHETSHRNEH,SCHIFFELS}

A potentially important effect, especially when there are few beneficial mutations available with substantial fitness advantages, are  double (or multiple) mutations. Individually, these are highly improbable, but  if there are enough of them available, they could collectively drive the evolution. Such processes would involve one mutation that is either deleterious or whose benefit is sufficiently small that it cannot establish a new population that advances the nose, but which enables a spectrum of previously-unavailable further mutations some of which have larger benefit than the predominant single mutations.  As the number of such two-hit processes is potentially very large --- and they are, arguably, less likely to have been ``found" by earlier  evolution in different conditions --- they could perhaps yield a long tail to $\mu(s)$ that drives the evolution and  includes enough different processes that the tail  is sustained as the evolution progresses.  The general issue of mutational processes that do not go steadily uphill --- and the balance between individual rareness and large numbers of possibilities ---  is an interesting direction for future study. \cite{WEISSMAN} 

In large sexual populations with low rates of recombination ---  important at least for microbes ---  the evolutionary dynamics is again controlled by the nose of the fitness distribution.  In this case, problems associated with fluctuations in the nose invalidate the methods of analysis of reference \cite{neher:etal:2010} for  low recombination rates. Thus one would like better means to study these important fluctuations. Some of the approaches in the present paper --- for example analyzing the rare events that dominate averages --- may be useful for such populations in which the diversity and anomalously high fitness individuals are generated by a mixture of approximately-asexual accumulation of new mutations occasionally brought together or broken apart by recombination.  

\bigskip
\noindent
{\bf Acknowledgements}

The author wishes to thank Michael Desai, Oskar Hallatschek, and Benjamin Good for useful discussions, for encouraging the working out and inclusion of various parts of this paper, and for comments on the manuscript. 
This work was supported in part by the NSF via grant DMS-1120699.

\newpage

\bibliographystyle{unsrt}

\bibliography{jstat-dsf-refs.bib}

\newpage

\appendix



\section{Average fitness distribution at fixed speed}\label{AVDYNSAPP}

The dynamics of the average fitness distribution at fixed speed, $\na(\chi,t)$, is given by the linear deterministic equation:
\be
\pd{\na}{t}=\Lop_v \na
\ee
which can be solved via Laplace transforms.  Defining 
\be
\nl(\lambda,t)\equiv \int d\chi \, \exp(\lambda\chi)\na(\chi,t) \ ,
\ee
--- noting the sign convention for $\lambda$ ---
one obtains
\be
\nl(\lambda,t)=\nl(\lambda+t,0)\exp\lb[ -v\lambda t  -\half v t^2 + \int_{\lambda}^{\lambda+t} d\lambda'\, \Mz(\lambda')    \rb] \label{RHOHAT}
\ee
with the total population $N(t)=\nl(\lambda=0,t)$ and 
\be 
\Mz(\lambda)\equiv  \int ds\, \mu(s)\lb [e^{\lambda s} -1\rb] \ . \nonumber
\ee  
It is convenient  to also define derivatives and an integral of the transform of the mutation spectrum:
\bea
&\Mo(\lambda) \equiv & \frac{d\Mz}{d\lambda} = \int ds\, \mu(s) s \, e^{\lambda s} \\
&\Mt(\lambda) \equiv & \frac{d^2\Mz}{d\lambda^2} = \int ds\, \mu(s) s^2 e^{\lambda s}  \\
&\Mmo(\lambda) \equiv &  \int_0^\lambda \Mz(\lambda') d\lambda' = \int ds\, \mu(s) \lb[ \frac{e^{\lambda s}-1}{s} -\lambda \rb ]  \ .
\eea

If the initial distribution has maximum fitness above the mean of $Q$, the long time behavior will be dominated by the factor in the exponential of Eq.(\ref{RHOHAT})
\be
\int_{\lambda}^{\lambda+t} d\lambda'\, \Mz(\lambda')  =\Mmo(\lambda+t)-\Mmo(\lambda) , 
\ee
which grows rapidly  for $t+\lambda > \log[1/\mu(\tilde{s}))]/\tilde{s}$  with  $\tilde{s}$ the $s$ that dominates this integral  at the time  at which the integral becomes of order unity. Beyond time $  \log[1/\mu(\tilde{s}))]/\tilde{s}$,  the position of the peak  of  the average distribution, $\na(\chi)$, grows exponentially. But the way in which the dynamics of the linearized equation for the average $\n(\chi,t)$ breaks down provides a useful bound on the actual behavior with fluctuations. 

Starting from a roughly gaussian fitness distribution cutoff at $Q$, 
 the lead $Q$ 
must be sufficiently large that with the correct speed, $v$,  the average population  --- as it is an overestimate of the typical population --- does not drop below its initial value, which it could only do before the new mutations start to dominate.  For times longer than $Q/v$, the original population near $Q$ and mutations that arise from this dominate and  $d\log \la N\ra/dt \approx Q-vt + \Mz(t)$ which must thus be positive. The minimum of this expression,   which occurs when $d\Mz/dt =v$, must thus either be positive or must occur before  time $Q/v$.   
The value of $\lambda$ at  which the nose-dominated contribution to $d\log \la N\ra/dt$ is minimum we have denoted  $\gamt$, with
$\Mo(\gamt)= v $.
The argument above implies that if $Q/v < \gamt$,  then $Q-v\gamt + \Mz(\gamt)$ must be positive, thus generally, 
\be
Q>\chit \equiv \gamt v-\Mz(\gamt)  \nonumber
\ee
the bound quoted in the text.  

\subsection{Formal steady state  fitness ``distribution"}\label{FORMALAPP}

The average fitness distribution will run away for any non-negative initial conditions.  But formally, there exists a fixed point, $\rhox(\chi)$, of  the average equation, although it is not a distribution as it becomes negative.  As the fixed point satisfies $\Lop_v\rhox=0$, we have
\be
\rhox(\chi)=\int_{\cal C} \frac{d\lambda}{2\pi i} e^{-\chi\lambda-{\cal E}(\lambda)}
\ee
with 
\be
{\cal E}(\lambda)+\chi\lambda= \chi\lambda -v\lambda^2/2 + \Mmo(\lambda) \ .
\ee 
The behavior of $\rhox(\chi)$ can be analyzed in different regimes via  the inverse Laplace transform (or other means) to show that it is  positive and 
well-behaved for $\chi$ less than some value and  beyond that it oscillates around zero with decaying amplitude as  $\chi$ increases further.

\section{Well-behaved solution in linear regime of $\phix$} \label{WELLBEHAPP}

The general solution to the homogeneous linear equation, $\Lod_v\phi=0$, can be found by Laplace transforms, as above.  We have
\be
\lb[-v\lambda -\pd{}{\lambda} + \Mz(\lambda)\rb]\hat{\phi}(\lambda)=0
\ee
which can be integrated and inverse Laplace transformed to obtain the formal solution
\be
\phi(\chi)=\int_{\cal C} \frac{d\lambda}{2\pi i} e^{\chi\lambda+{\cal E}(\lambda)}  \ .
\ee
The integral must be done along an appropriately chosen contour, ${\cal C}$.   There are saddle points at values  $\hat{\lambda}(\chi)$ that are solutions to  
\be
v\lambda-\Mz(\lambda)=\chi \ . 
\ee

The saddle point approximation is closely-related to the log-slope WKB approximation discussed in the main text (a related approximation is used in reference \cite{rouzine:etal:2003}.)
It can be justified over most of the range of $\chi$, and how it breaks down near $\Q$ also seen, by rescaling $\chi$ by $\Q$, and $\lambda$ by $\tsw$ --- although this is somewhat unnatural as it makes fitness scales and time scales no longer reciprocals.  With the rescaling $X=\chi/\Q$ and $\Lambda=\lambda/\tsw$, the exponent ${\cal E}+\chi \lambda$ is proportional to $\frac{\Q^2}{2v}\approx \log(Ns)$ times a function of $X$ and $\Lambda$ except for   he mutational term which has a coefficient that becomes of the same order only when $\Lambda$ is within $\st\tsw$ of unity. This thus yields the value of $\chi$ at which mutations become important and in this regime, the rapid dependence on $\Lambda$ changes the large parameter that would justify the saddle point approximation from $\frac{\Q^2}{2v}$ to $\frac{v}{\st^2}$: i.e. it is only valid for $\chi$ near to $\Q$ in the high speed limit in which $\frac{v}{\st^2}$ is large and even then, it eventually breaks down as discussed below.

For large negative $\chi$, there is a saddle at $\hat{\lambda}\approx \chi/v$ for which the mutational term is negligible, hence a contour that goes over this saddle would yield $\phi\sim \exp(\chi^2/2v)$ for large negative $\chi$: this corresponds to  the badly behaved, but generic solution as this saddle is the highest.   But there is also a saddle with large positive real $\lambda$  at  $\Mz(\lambda)\approx -\chi $.  This corresponds to a solution that goes to zero rapidly as $\chi$ decreases.   We thus want to choose a contour that goes over this saddle but  not over the negative $\lambda$ saddle (or its continuation for positive $\chi$).   There are also complex $\lambda$ saddles with $\Re(\lambda)$ large and positive: in general there could be contributions from the linearly independent solutions defined by contours that go over each one of these saddles, but the coefficients of these must be small enough that their oscillatory behavior does not cause $\phix(\chi)$ to go negative --- which it cannot.  The complex-saddle solutions go to zero faster than this real-saddle solution for large negative $\chi$, thus their contributions will be negligible in this limit. We thus have a single condition that is needed to give well-behaved behavior in the linear regime for $\Q-\chi$ large: that there is no contribution to the single badly behaved solution.  This single condition is sufficient to determine $\Q$. 
[Note that the behavior is qualitatively similar to defining Airy functions via a contour integral, which is essentially an inverse  Laplace transform: to get the well-behaved solution, Ai, requires not going over a particular saddle point, otherwise the exponentially growing solution, Bi, is mixed in.]

The asymptotic behavior of $\phix$  for large negative $\chi$ is 
\be
-\log(\phix)\approx (-\chi)\hat{\lambda}(\chi)
\ee
with $\hat{\lambda}$ the saddle point, which is at, e.g., $\hat{\lambda}\approx \frac{1}{s} \log(-\chi/\Ub)$ for the constant $s$ model, and $\hat{\lambda}\approx\frac{1}{\sigma}\sqrt{2 \log(-\chi/\Ub)}$ for a half-gaussian mutational spectrum 
of width $\sigma$.   Note that $\phix$ is much smaller than the badly behaved upside-down gaussian solution. 

What happens to the linear solution as $\chi$ increases?   At a critical value of $\chi$, $\chit=v\gamt-\Mz(\gamt)$, the saddle merges with the badly behaved saddle, as signaled  by $d\hat{\lambda}/d\chi=\infty$: this occurs at $\lambda=\gamt$ and $\chi=\chit$ with $\Mo(\gamt)=v$ --- the same special value that appeared earlier. Beyond $\chit$, the saddle splits into two complex saddles and for slightly larger $\chi$,  $\phi$ would start oscillating.  This suggests that the breakdown of the linear approximation must occur near $\chit$ as guessed from the analysis of the deterministic approximation to $\n(\chi,t)$ discussed above.  As the relevant linear operators are adjoints of each other and their Laplace transforms simply related, this is hardly surprising.    

The condition for the validity of the saddle-point or WKB approximations for the smooth part of $ \log\phi$, is that the curvature at the saddle point, $\Mo(\lambda)$, times the square of the scale at which the non-quadratic terms come in is large. Near to $\gamt$ we can expand the exponent, ${\cal E}+\chi\lambda$, in the contour integral:
\be
{\cal E}+\chi\lambda \approx  (\chi-\chit)(\lambda-\gamt) + \frac{1}{6} \Mt(\gamt)(\lambda-\gamt)^3
\ee
and infer that the scale on which the saddle-point approximation breaks down is when
\be
\chit-\chi \sim [\Mt(\gamt)]^{\frac{1}{3}} \approx (v\st)^{\frac{1}{3}} \ .
\ee
When $v\gg \st^2$, this is a distance of more than $\st$ from $\chit$ and higher order terms will have smaller effect.  But for smaller $v$, the scale $\st$ is less than $(v\st)^{\frac{1}{3}}$  and higher order terms will make the saddle point approximation break down for $\chit-\chi$ of order $\st$.

\subsection{Moderate speeds, $v\ll \st^2$}\label{MODAPP} 

To find the fixed point, $\phix$, we must match the linear solution(s) to the non-linear solution just below $\Q$.   For small $v/\st^2$, the behavior is relatively simple.    For  $\Q-\chi$ substantially less than  $\st$, the  mutations primarily come from  the non-linear region which gives  $\Q\int_{\Q-\chi}^\infty \mu(s)ds$ in the fixed point equation. This is much smaller than the other contributions which are both of order $\Q^2 \exp[-(\Q-\chi)\Q/v]$ until values of $s$ near the peak of $\exp(s\Q/v)\mu(s)$ start to enter: this occurs for $\chi$ near $\Q-\st$ for $\Q/v \approx \gamt$. For larger $\chi$, i.e.  in the range $\Q-\st < \chi < \Q-v/\Q$, 
\be
\phix\approx \Q \exp[-(\Q-\chi)\Q/v +(\Q-\chi)^2/2v] \ . 
\ee

We pull out, as in the main text,  the rapidly varying factor, $e^{\gamt\chi}$ and use 
$f\equiv \phi e^{-\gamt(\Q-\chi)}\Q$
which matches onto the shoulder solution near $\Q$ and  obeys, in the regime of interest for $\chi<\Q-\st$,  the linearized fixed point equation
\be
-v \pd{f}{\chi} \approx (v\gamt-\chi)f -\int ds \mu(s)e^{\gamt s}f(\chi+s)
\ee
where the  $-v\pd{\phi}{\chi}+\chi\phi$ has become $-v \pd{f}{\chi} -(v\gamt-\chi)f$ and the convolution over $\mu(s)\phi(\chi+s)$  a convolution over $\mu(s)\exp(\gamt s)f(\chi+s)$ with $\int ds \mu(s)\exp(\gamt s) = \Mz(\gamt)\approx \frac{v}{\st}$. [ Note that in the main text we added and subtracted this times $f(\chi)$ to yield $f(\chi+s)-f(\chi)$ in the convolution together with    
a term $(\chit-\chi)f(\chi)$.]  

Near $\Q-\st$, the mutational input to $f$  must play a significant role. If it did not, the balance of the other terms would make $f$ start to increase as $\chi$ decreases implying that the log-slope of $\phix$ would continue to get  more and more negative, making the mutational contributions more negligible and leading to the badly behaved $\exp(\chi^2/2v)$ solution for $\phix$.  But if the mutational term were too large, it would tend to drive $f$ negative.   
Thus the mutational input at $\chi-\st$,  roughly  $v/2\st$, needs to be comparable to $(v\gamt-Q+\st) f(\Q-\st)\approx (v\gamt-Q+\st)\Q\exp[-\st(\Q/v -\gamt)+\st^2/2v]$. 
With $\st^2/v$ relatively large, this condition implies that the terms in the exponential roughly cancel yielding the result in the main text:
\be
\Q\approx v\gamt+\frac{1}{2}\st +  {\cal O}\lb[\frac{v}{\st}\log(\st^2/v)\rb] \approx \chit + \frac{1}{2}\st + {\cal O}\lb[\frac{v}{\st}\log(\st^2/v)\rb] \nonumber
\ee
since the $\Mz(\gamt)$ contribution to $\chit$ is $-v/\st$ which is smaller than the terms above in this moderate speed regime.  The behavior of $\phix$ can be constructed step by step with the dominant balance being between the $(v\gamt-\chi)f$ and the mutational input terms.  The steps will be   of size close to $\st$ as long as $\gamma$ stays sufficiently close to $\gamt$ that $s\approx \st$ still dominates the mutational input, which it will for multiple steps below $\Q$.  On top of the parabolic dips of $\log f$ between each integer multiple of $\st$ below $\Q$ that are caused by those in the first segment, 
the envelope can be seen to be roughly 
\be
-\log f \approx  \frac{\Q-\chi}{\st}\lb[\log\lb(\frac{\Q-\chi}{e\st}\rb) + \log(\st^2/v)\rb]
\ee      
which  corresponds to a log-slope of $\phi$ of
\be
\gamma(\chi) -\gamt \approx \frac{1}{\st}\log\lb[\frac{\st(\chit-\chi+\st/2)}{v}\rb] \ .
\ee
In this regime, $\Mz(\gamma)\approx \exp[(\gamma-\gamt)\st]v/\st$ thus the WKB approximation yields
\be
\gamma(\chi)-\gamt \approx  \frac{1}{\st} \log\lb[\frac{\st(v\gamt-\chi)}{v}\rb] \ ,
\ee
which, with correction factors inside the log, is almost the same. 

\subsection{High speeds, $v\gg\st^2$}\label{FASTAPP} 

When $v\gg \st^2$, the saddle point approximation breaks down much further  below $\Q$ than in the moderate speed case. But the behavior just below $\Q$ is much smoother as the $\exp[(\Q-\chi)^2/2v)$ factor in $\phix(\chi)$  between $\Q$ and $\Q-\st$ is close to unity.   These both suggest expanding in derivatives.  Again, this is best to do in terms of the rescaled function, $f$.  
We have 
\be
0 = \frac{1}{2}\Mt(\gamt)\pd{^2f}{\chi^2} +(\chi-\chit)f + {\cal O}\lb[v\st^2 \pd{^3f}{\chi^3}\rb] \nonumber
\ee
with $\Mt(\gamt)\approx v\st$.   The scale of $\chit-\chi$ that enters we  denoted $b\equiv[\Mt(\gamt)/2]^\frac{1}{3}\approx (v\st/2)^\frac{1}{3}$.   This suggests that the neglected terms will be smaller by a factor of $(\st^2/v)^\frac{1}{3}$.  The dominant terms are simply Airy's equation, $\pd{^2f}{y^2}=yf$  in the variable 
$y\equiv\frac{\chit-\chi}{b}$
and we are interested in the $f\propto {\rm Ai}(y)$ solution that decays rapidly as $y\to \infty$: it can be seen that this matches on to the WKB approximation in this regime.  
For negative $y$, the solution oscillates, first passing through zero at a value $-a_1\cong -2.34$.   Thus we expect $\Q<\chit + a_1b$. Very near to $\Q$ the logarithmic derivative of $f$ should be (from the previous subsection)   $\Q/v-\gamt=-\Mz(\gamt)/v+(\Q-\chit)/v\approx -1/\st +(\Q-\chit)/v$ which, with $\Q-\chit<a_1b$ is close to $-1/\st$.   This is much larger in maginitude  than  the typical logarithmic derivative of $f$ from the Airy solution which is of order $(v\st)^{-\frac{1}{3}}$. 
But near to its zeroes, the derivative of $ \log {\rm Ai}$ diverges, in particular as $1/(y+a_1)$ near to its first negative zero.   Thus near to $\Q$, the log-slope of $\phix$ is
\be
\gamma-\gamt\approx \frac{-1}{\chit+a_1b-\chi} \ .
\ee
In the interval $(\Q-\st,\Q)$, the mutational input is very different than at lower $\chi$ and the Airy approximation is not valid. Thus it is easiest to match the solutions near to $\Q-\st$ at which the log-slope is still close to $1/\st$: this means that $\Q$ should be closer to the Airy zero than ${\cal O}(\st)$ therefore
\bea
\Q &\approx& \chit +a_1 (v\st/2)^\frac{1}{3} + o(\st) \\ &\approx& \gamt v - \frac{v}{\st} + a_1 (v\st/2)^\frac{1}{3}  \\ &\approx& \gamt v -\Mz(\gamt) + a_1 \lb[\Mt(\gamt)\rb]^{\frac{1}{3}} 
\eea
with the last  expression in terms of the ${\cal M}_k(\gamt)$ being somewhat more general. 
A more detailed calculation, outlined below, yields a correction term in $\Q$ of order  $-1/\gamt$ which is indeed much less than $\st$.   

\section{Sharp shoulder approximation}\label{SHOULDAPP} 

In the main text we introduced the well-justified approximation of treating the shoulder as a sharp boundary at $\Q$ with the linearized equation for $\chi<\Q$ and the effective boundary condition that $\phix(\Q)=\Q$  The approximate fixed point equation is 
\be
\Lod_v \phix + \Q \int_{\Q-\chi}^\infty \mu(s)ds =0 \nonumber
\ee
with the second term the input from $\chi>\Q$.  This is again solved  by Laplace transforms with the input term and a boundary term $-v\Q e^{-\Q\lambda}$ (from the motion of the shoulder at speed $v$ in the rest frame) combining to yield an inhomogeneous equation of the form 
\be
\hat{\Lod}_v \hat{\phix} + I(\lambda)=0 
\ee
with 
\be
I(\lambda)=\Q e^{-\Q\lambda}\lb[\frac{\Mz(\lambda)}{\lambda} -v\rb]
\ee
The $-\pd{}{\lambda}$ in $\hat{\Lod}_v$ that comes from the $\chi$ term in $\Lod_v$ means that the solution is of the form
\be
\hat{\phix}(\lambda)=e^{{\cal E}(\lambda)}\int^\lambda d\lambda' e^{-{\cal E}(\lambda')}I(\lambda') \ .
\ee
with the exponent ${\cal E}(\lambda)=-v\lambda^2/2+\Mmo(\lambda)$ as defined earlier.     A multiple of the homogeneous solution, $e^{{\cal E}(\lambda)}$, is included in the general solution by virtue of the undetermined lower bound on the integral over $\lambda'$.   
Unlike for the formal linear solution over $(-\infty,\infty)$, the Laplace transform of $\phix$ on $(-\infty,\Q)$ should be well behaved.   In the left half-plane $\hat{\phix}$ should be  
$e^{-\lambda \Q}$ times an analytic function since the  solution should vanish for $\chi>\Q$.    The difficulty is that the integral over a contour parallel to the imaginary axis that is required for the inverse Laplace transform cannot be deformed into the left half-plane to make $\phix(\chi>\Q)$ vanish because of the $e^{-v\lambda^2/2}$ term in $e^{{\cal E}(\lambda)}$ which diverges badly at $\pm i \infty$.  Thus for general $\Q$, there is no well-behaved solution. This should not be surprising as we know that the general linear solution grows as $e^{\chi^2/2v}$ for large negative $\chi$ while the solution that is well-behaved for large negative $\chi$   will not obey the boundary condition at $\Q$ that makes it vanish for $\chi>\Q$.   The only way the solution can be well-behaved in both ways is if the integral over $\lambda'$ almost exactly cancels the divergence at $\lambda \to \pm i \infty$.   With the lower bound of the integral at $\lambda' = -i\infty$, the cancellation occurs near this  end but, in general, not near the $+i\infty$ end --- unless the integral over the full range vanishes. Thus the condition for a well-behaved solution is that 
\be
\int_{-i\infty}^{i\infty} e^{-{\cal E}(\lambda')}I(\lambda')d\lambda' =0 \label{SOLVCOND}
\ee
which will occur only at  discrete values of $\Q$, only one of which --- if there are more than one --- will yield an everywhere-positive $\phix$.   (Note that other values of $\Q$, which have negative $\phi$ for some $\chi<\Q$, can yield the eigenfunctions and eigenvalues of the linear operator that determines the convergence to $\phix$)

The above condition that determines $\Q$ is a solvability condition enforcing the condition that the projection onto the badly behaved solution vanishes.   The function $e^{{-\cal E}(\lambda)}$ in which it is expressed, is  the transform of the formal steady state solution, $\rhox$, of the linear dynamics of $\n$ which is the zero eigenvector of the adjoint operator, $\Lop_v$.   Transforming back to $\chi$ space, we obtain  the condition quoted in the main text:
\be
\int d\chi \, \rhox_{\Q} (\chi) \Omega(\chi) = v\rhox_{\Q}(\Q) \nonumber
\ee
with $\Omega(\chi)=\int_{\Q-\chi}^\infty ds \mu(s)$.

For the high velocity limit, $v\gg\st^2$, $\Q$ is in the regime in which the Airy approximation is valid and the  solvability condition becomes  essentially
\be
{\rm Ai}[(\Q-\chit)/b]\approx \frac{1}{\gamt\st}{\rm Ai}[(\Q-\chit+\st)/b]
\ee
with the left-hand side arising from the boundary term and the right from the mutational input term.  This yields, as Ai goes to zero linearly, the result that $\Q$ is below the position of the  first zero of the Airy function by $1/\gamt$. But  this is of order the width of the shoulder region so that the sharp wall approximation breaks down.  Nevertheless, the scale of the correction should still be of order $1/\gamt$.  It will turn out that the fluctuations in speed at fixed population size correspond to fluctuations in the nose by of order $1/\tsw\approx 1/\gamt$ over times of order $\tsw$.  Thus it is not possible to define $\Q$ to a precision of order $1/\gamt$ without taking into account these fluctuations.  But even to do that well, one would need to redefine $\Q$ in a way that is useful in the presence of the much larger fluctuations in the position of the mean discussed in Sec. \ref{JUMPS}.  [Note that the contributions from the complex saddle points with $\Im(\lambda')\approx 2\pi k/\st$ for non-zero-integer $k$, which the contour also goes over, are strongly suppressed for $v\gg\st^2$ corresponding to the smallness of  the deviation of  $\exp[-(\Q-\chi)^2/2v$ from unity in the interval $(\Q-\st,\Q)$.]

In the moderate velocity limit,  $v\ll \st^2$, the result found that 
\be
\Q\approx \chit +\frac{1}{2}\st+ \frac{v}{\st}[\log(\gamt\st)+1] \approx v\gamt +\frac{1}{2}\st+ \frac{v}{\st}\log(\gamt\st) \nonumber
 \ee
 can  be derived simply by doing the integral in the solvability condition along the line $\Re(\lambda')=\gamt$. This expression for $\Q$ only differs by a small correction from the heuristic result and is close to that of \good.   Note that in this limit one can neglect the mutational term in ${\cal E}(\lambda')$ in deriving $\Q$: this also justifies the neglect of all the mutational effects except for the input from the saturation regime, which is equivalent to neglecting the effects of mutations for all but the lead population as it is being established.  If $\rhox_\Q$ is replaced by its gaussian approximation, then the condition Eq.(\ref{RHOQCOND}), weights the $s$ that contribute by a factor $\mu(s)\exp(\frac{s\Q}{v} -\frac{s^2}{2v}$ which is the same weighting factor that appears in \good.

\section{Dynamics of $\phi$}\label{DYNAPP}

Here we derive some of the results summarized in the main text about  the backwards-time dynamics of $\phi$.  With the definition 
 $\tau\equiv T-t$,  $
\pd{\phi}{\tau}=\Lod_v \phi -\phi^2 $
with ``initial" condition 
$\phi(\chi,\tau=0)=\zeta(\chi)$: $\zeta=const.$ being of particular interest. 
We again first use Laplace transforms in $\chi$ for the linear part: 
\be
\pd{\phih}{\tau}+\pd{\phih}{\lambda}= [-v\lambda + \Mz(\lambda)]\phih(\lambda,\tau)+\ \ {\rm nonlinear\ terms}
\ee
so that functions of $\lambda-\tau$ will appear.  With initial condition, $\phih_0(\lambda)\equiv\phih(\lambda,0)$, the linearized evolution yields
\be
\phih(\lambda,\tau)=\phih_0(\lambda-\tau)e^{E(\lambda,\tau)}
\ee
with 
\be
E= -v\lambda \tau +\frac{v\tau^2}{2} +\int_{\lambda-\tau}^\lambda d\gamma \Mz(\gamma) \ .
\ee
For constant initial condition, $\phi(\chi,0)=\zeta$, so that only $\lambda=0$ exists initially,  this yields simply 
\be
\phi\approx \zeta \exp[\chi\tau-\half v\tau^2 +\Mmo(\tau)] \ .
\ee
 Formally, we can proceed to handle the non-linearity in the sharp boundary approximation used for the fixed point.  The boundary part in the equation for $\phih$ must again be supplemented by the mutational input from the saturation region above $\P(\tau')$ for all $\tau'<\tau$, and the effects of these integrated over $\tau'$ from zero to $\tau$.  The condition that $\phih(\chi,\tau)=0$ for $\chi>\P(\tau)$ for all $\tau$ can in principle be used to determine $\P(\tau)$ but the analysis is messy and we have not carried it out in detail.

A heuristic approach again works well.  We can do an approximate analysis  in terms of the log-slope, $\gamma(\chi,\tau)$, which obeys, if sufficiently slowly varying that the approximation $\phi(\chi+s)\approx \phi(\chi)\exp[\gamma(\chi) s)] $ is valid,
\be
\pd{\gamma}{\tau}=1 -  \pd{\gamma}{\chi}[v-\Mo(\gamma)]
\ee
which implies that $\gamma$ grows linearly in time while being advected at speed $v-\Mo(\gamma)$. This hints at the range of validity of the linearized expression for $\phi$: it should be valid provided the path of advection from the initial condition to the point of interest does not pass through the boundary of the linear region which is the curve $\P(\tau)$.  It also shows how the effects of the shoulder region just below $\P(\tau)$ propagate to smaller $\chi$.  Take an advection curve that passes through a point $(\chi,\tau)$ of interest which at an earlier time $\tau'$  hit $\P(\tau')$: with $\gamma'\equiv\gamma(\P(\tau'),\tau')$, this occured when $\chi=\P(\tau')+v(\tau-\tau')-\Mz(\gamma'+\tau-\tau')+\Mz(\gamma')$.  The linear growth of $\gamma$ as it is being advected then yields  $\gamma(\chi,\tau)\approx\gamma' +\tau-\tau'$.   This will only be applicable when the log-slope at $\P$ has reached roughly $\gamt$ which is the value at which the advection speed become negative.  

For $\tau$ less than $\gamt$, the linearized approximation ignoring the input from the saturation region will be good up to $\P(\tau)$  which can   be found from the self-consistency condition 
\be
\P(\tau)\approx \zeta \exp[\tau \P(\tau) -v\tau^2/2 -\Mmo(\tau)]
\ee
indicating only modest corrections to the simple result ignoring mutations: Eq.(\ref{PAPPROX}).
Beyond $\tau\approx \gamt$, the behavior will be determined, as for the fixed point, by $\phi$ in a small region just below $\P(\tau)$ and the mutational input to it from just above $\P(\tau)$. The remaining effects of the original $\zeta$ will quickly become negligible near the shoulder as, without mutational input from higher $\chi$,  they would give a rapidly dropping $\phi$ leading to an increasing  $\P(\tau)$ as  seen in the solution ignoring mutations, $\P\approx \frac{1}{\tau}\log(\P/\zeta)+v\tau/2$ which has a minimum near $\tau \approx \sqrt{2\log(\P/\zeta)/v}$   of $\P_{min}\approx \sqrt{2v\log(\sqrt{v}/\zeta)}$.  Note that if $\zeta\approx \zetat \approx 1/\Nt$, $\P_{min}$ is close to $\Q$.  

In the modest speed limit with $v$ of order $\st^2$ or smaller, we can again make the approximation that the only mutational effects that matter for determining $\P(\tau)$ are those from the saturation region.   Input at an earlier time $\tau'$ from just above $\P(\tau')$ to $\chi\approx \P(\tau')-s$ and its effects move to higher $\chi$  at speed $v$ so that if $\tau'=\tau-t_s$ with $t_s$ such that
\be
\P(\tau-t_s)-s + vt_s=\P(\tau)\ ,
\ee
the effects of this input will reach $\P(\tau)$ at $\tau$.  The contribution to $\phi(\P(\tau),\tau)$ from this input is, to logarithmic accuracy, roughly
\be
\P(\tau)\exp\lb[-\Lambda(s)+t_s(\P(\tau-t_s)-s)+\frac{v}{2}t_s^2\rb]  \ .
\ee
To get $\phi(\P(\tau),\tau)\approx \P(\tau)$ which is required for self-consistency, the integral over $s$ of the exponential factors must be approximately unity. Anticipating that this integral will be sharply peaked, we can approximate the integral by the maximum of the integrand so that 
\be
\max_s \lb[-\Lambda(s)+(t_s)(\P(\tau-t_s)-s)+\frac{v}{2}t_s^2\rb] \approx 0 \label{MAXSP}
\ee
If $\P(\tau)$ is slowly varying, we can expand $\P(\tau-t_s)\approx \P(\tau)-\frac{dP}{d\tau}t_s$  and obtain
\be
t_s\approx \frac{s}{v-\frac{d\P}{d\tau}}
\ee
with the dominant $s$ determined by the maximization condition. Substituting in and keeping terms to firs order in $\frac{d\P}{d\tau}/v$, we obtain the two conditions 
\bea
\P-s &\approx& \lb(v-\frac{d\P}{d\tau}\rb)\frac{d\Lambda}{ds} \\
\P-\frac{1}{2}s &\approx& \lb(v-\frac{d\P}{d\tau}\rb)\frac{\Lambda}{s} \ .
\eea
The fixed point obtained from $\frac{d\P}{d\tau}=0$ is at $\P^*=\Q$ with $\Q$ essentially the same as obtained earlier.  

Note that $\st$ is determined, up to small corrections, by $d\Lambda/ds=\Lambda/s=\gamt$ so that the $s$ that dominates the input near the shoulder is slightly larger than $\st$.  But this effect cancels in determining $\Q\approx v\gamt +\frac{1}{2}\st$ which agrees with the above for $v\ll \st^2$ as it should.  The static condition, $\P=\Q$, has  $t_s=s/v$ so that there is a $-s^2/2v$ term in  Eq. (\ref{MAXSP}).  This condition for determining the pre-dominant $s$ (and also $\Q$) is essentially the same as that from Eq.(\ref{RHOQCOND}) with the gaussian approximation for $\rhox_\Q$ and also that of \good, as discussed earlier.  


As long as $\frac{dP}{d\tau}/v$ is small, the neglect of higher order terms and derivatives can be justified and we obtain the simple result
\be
 \frac{d\P}{d\tau} \approx \frac{v}{\Q} (\Q-\P)
 \ee
 so that the fixed point is approached exponentially with eigenvalue 
 $\eps\approx/\Q\approx1/\tsw$.
 The stability backwards in time corresponds to  the instability of the nose so it is  controlled by the same eigenvalue $\eps$.  

In the high speed limit with $v\gg \st^2$, the analysis is somewhat more involved because the effects of mutations that cascade down to $\P-\Oh[(v\st)^\frac{1}{3}]$ and are carried back up to $\P$ are all important.  However the basic result for the exponential approach to the fixed point with eigenvalue close to $v/\Q$  remains the same.

When starting from a constant value of $\phi=\zeta$ at time $T$, until $T-\gamt\approx T-\tsw$ the dynamics is mostly linear --- corresponding to deterministic for $\n(\chi,t)$ --- in the regime of interest, and $\P(\tau)$ changes rapidly as given by Eq.(\ref{PAPPROX}).  During  a narrow time range around $T-\gamt$, $\P$ does not change much 
but a form of $\phi$ that is very similar to that at the fixed point is established below $\P$.   After this time, the dynamics is exponential relaxation of $\P(\tau)$ to the fixed point $\Q$ with the shape of $\phi(\chi,\tau)$ close to its fixed point form but shifted by $\P(\tau)-\Q$.  
The results in the main text follow.

\end{document}